\pdfoutput=1

\documentclass[11pt,twoside,a4paper,cmspaper,final,collab]{cms-tdr}
\def\svnVersion{164910}\def\cmsCernNo{2012-095}\def\cmsCernDate{\today}\def\cmsMessage{Submitted to Physical Review C}

\begin{document}\cmsNoteHeader{HIN-10-002}

\hyphenation{had-ron-i-za-tion}
\hyphenation{cal-or-i-me-ter}
\hyphenation{de-vices}
\RCS$Revision: 164909 $
\RCS$HeadURL: svn+ssh://alverson@svn.cern.ch/reps/tdr2/papers/HIN-10-002/trunk/HIN-10-002.tex $
\RCS$Id: HIN-10-002.tex 164909 2013-01-15 17:31:04Z alverson $
\newlength\cmsFigWidth
\ifthenelse{\boolean{cms@external}}{\setlength\cmsFigWidth{\columnwidth}}{\setlength\cmsFigWidth{0.85\textwidth}}
\cmsNoteHeader{HIN-10-002} 
\title{Measurement of the elliptic anisotropy of charged particles produced in PbPb collisions at \texorpdfstring{$\sqrt{s_{NN}}$ = 2.76\TeV}{nucleon-nucleon center-of-mass energy = 2.76 TeV}}

\date{\today}

\abstract{
The anisotropy of the azimuthal distributions of charged particles produced in $\sqrt{s_{NN}}=2.76$\TeV PbPb
collisions is studied with the CMS experiment at the LHC. The elliptic anisotropy parameter,
$v_{2}$, defined as the second coefficient in a Fourier expansion of the particle invariant yields, is
extracted using the event-plane method, two- and four-particle
cumulants, and Lee--Yang zeros.
The anisotropy is presented as a function of transverse momentum ($\pt$), pseudorapidity ($\eta$)
over a broad kinematic range: ${0.3<\pt<20}$\GeVc, $|\eta| < 2.4$, and in 12 classes of collision centrality from
0 to 80\%. The results are compared to those obtained at lower center-of-mass
energies, and various scaling behaviors are examined.
When scaled by the geometric eccentricity of the collision zone, the elliptic anisotropy is found to obey a universal scaling
with the transverse particle density for different collision systems and center-of-mass energies.

}
\hypersetup{%
pdfauthor={CMS Collaboration},%
pdftitle={Measurement of the elliptic anisotropy of charged particles produced in PbPb collisions at nucleon-nucleon center-of-mass energy = 2.76 TeV},%
pdfsubject={CMS},%
pdfkeywords={CMS,quark-gluon plasma, heavy ions, elliptic flow}}

\maketitle 

\section{Introduction}
\label{sec:Introduction}
\begin{figure}[hbtp]
  \begin{center}
    \includegraphics[width=0.5\textwidth]{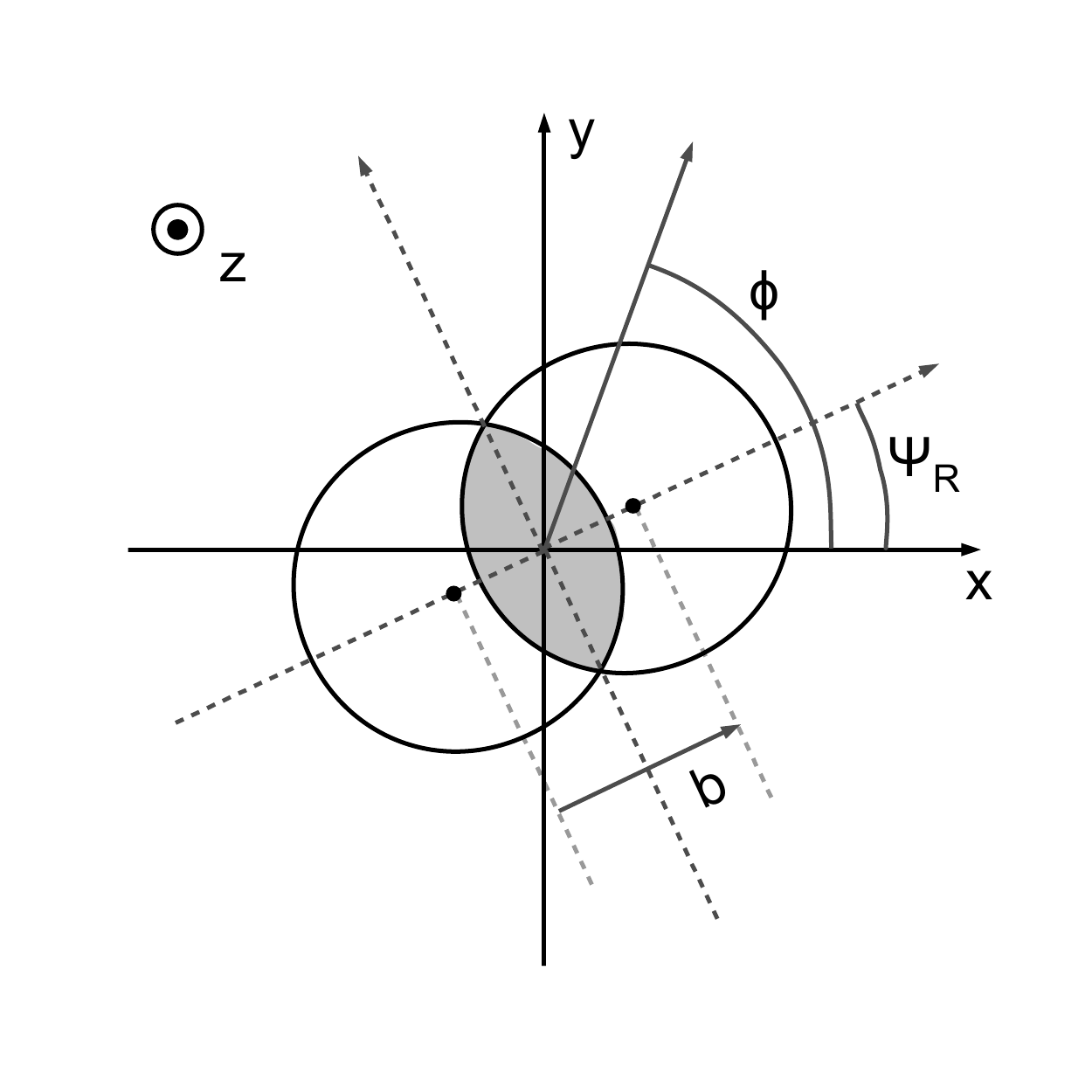}
    \caption{A schematic diagram of a noncentral nucleus-nucleus collision
viewed in the plane orthogonal to the beam.
The azimuthal angle $\varphi$, the impact parameter vector \textbf{b}, and the reaction-plane angle $\Psi_{\textrm{R}}$ are shown. The
event-plane angle $\Psi$, with respect to which the flow is measured, fluctuates around the reaction-plane angle.
}
    \label{fig:almond}
\end{center}
\end{figure}

The azimuthal anisotropy of emitted charged particles is an
important feature of the hot, dense medium produced in heavy-ion collisions,
and has contributed to the suggestion of a strongly coupled quark-gluon plasma
(sQGP) being produced in nucleus-nucleus collisions at
RHIC~\cite{PHENIX,STAR,PHOBOS,Shuryak:2004cy,Gyulassy:2004zy}. In noncentral collisions,
the beam direction and the impact parameter vector define a reaction plane for
each event. If the nucleon density within the nuclei is continuous, the initial
nuclear overlap region is spatially asymmetric with an ``almond-like'' shape. In this approximation, the impact
parameter determines uniquely the initial geometry of the collision, as illustrated in  Fig.~\ref{fig:almond}.
In a more realistic description, where the position of the individual nucleons that participate in inelastic interactions
is considered, the overlap region has a more irregular shape and the event-by-event orientation of the almond fluctuates around
the reaction plane. Experimentally, the azimuthal distribution of the particles detected in the final state can be used to
determine the ``event plane''  that contains both the beam direction and the azimuthal direction of maximum particle density.
Strong rescattering of the partons in the initial state may lead to local thermal equilibrium and the build up
of anisotropic pressure gradients, which drive a collective anisotropic expansion.  The acceleration is greatest
in the direction of the largest pressure gradient, i.e., along the short axis of the almond. This results in an
anisotropic azimuthal distribution of the final-state hadrons. The anisotropy is quantified in terms of a Fourier
expansion of the observed particle yields relative to the event-by-event orientation of the event plane~\cite{Poskanzer:1998yz}:

\ifthenelse{\boolean{cms@external}}
{
\begin{multline}
\label{eq:1}
E\frac{{\rd^3 N}}{{\rd^3 p}} = \frac{{\rd^3 N}}{{\pt\, \rd\pt\, \rd{}y \rd\varphi}} = \frac{1}{{2\pi }}\frac{{\rd^2 N}}{{\pt\, \rd\pt\, \rd{}y}}\\
\times\left( {1 + \sum\limits_{n = 1}^\infty  {2v_n(\pt,y) \cos \left[ {n\left( {\varphi  - \Psi} \right)} \right]} } \right),
\end{multline}
}
{
\begin{equation}
\label{eq:1}
E\frac{{\rd^3 N}}{{\rd^3 p}} = \frac{{\rd^3 N}}{{\pt\, \rd\pt\, \rd{}y\, \rd\varphi}} = \frac{1}{{2\pi }}\frac{{\rd^2 N}}{{\pt\, \rd\pt\, \rd{}y}}\left( {1 + \sum\limits_{n = 1}^\infty  {2v_n(\pt,y) \cos \left[ {n\left( {\varphi  - \Psi} \right)} \right]} } \right),
\end{equation}
}

where  $\varphi$, $E$, $y$, and  \pt are the particle's azimuthal angle, energy, rapidity,
and transverse momentum, respectively, and  $\Psi$ is the event-plane angle.
The second coefficient of the expansion,
often referred to as the ``elliptic flow'' strength, carries information about
the early collision dynamics \cite{Ollit, Sorge}. The coefficients in the Fourier expansion may depend on
\pt, rapidity, and impact parameter. Typically, the measurements are obtained for a particular class of events based on
the centrality of the collisions, defined as a fraction of the total inelastic nucleus-nucleus cross section,
with 0\% denoting the most central
collisions with impact parameter \text{b} = 0, and 100\% the most peripheral collisions.
Expressions similar to Eq.~(\ref{eq:1}) can be written for the Fourier expansion of the yield
integrated over \pt or rapidity:

\begin{equation}
\label{eq:2}
\frac{{\rd^2 N}}{{\rd{}y\, \rd\varphi}} = \frac{1}{{2\pi }}\frac{{\rd{}N}}{{\rd{}y}}\left( {1 + \sum\limits_{n = 1}^\infty  {2v_n(y) \cos \left[ {n\left( {\varphi  - \Psi} \right)} \right]} } \right),
\end{equation}
and

\ifthenelse{\boolean{cms@external}}
{
\begin{multline}
\label{eq:3}
\frac{{\rd^2 N}}{{\pt \rd\pt\, \rd\varphi}} = \frac{1}{{2\pi }}\frac{{\rd{}N}}{{\pt\, \rd\pt }}\\
\times\left( {1 + \sum\limits_{n = 1}^\infty  {2v_n(\pt) \cos \left[ {n\left( {\varphi  - \Psi} \right)} \right]} } \right).
\end{multline}
}
{
\begin{equation}
\label{eq:3}
\frac{{\rd^2 N}}{{\pt\, \rd\pt\, \rd\varphi}} = \frac{1}{{2\pi }}\frac{{\rd{}N}}{{\pt\, \rd\pt }}\left( {1 + \sum\limits_{n = 1}^\infty  {2v_n(\pt) \cos \left[ {n\left( {\varphi  - \Psi } \right)} \right]} } \right).
\end{equation}
}

The Fourier coefficients $v_n(y)$ and  $v_n(\pt)$ can be obtained by directly analyzing the yields integrated over \pt or rapidity,
or from the coefficients of the triple-differential invariant yield in  Eq.~(\ref{eq:1}), $v_n(\pt, y)$, by performing the following
yield-weighted average over the ranges of transverse momentum $\Delta \pt$ and rapidity $\Delta y$, from which the particles are taken:

\begin{equation}
\label{eq:4}
v_n(y) = \frac{ \int_{\Delta \pt} \pt\, \rd\pt \frac{{\rd^2 N}}{{\pt\, \rd\pt\, \rd{}y}}v_n(\pt,y)}{\int_{\Delta \pt} \pt\, \rd\pt \frac{{\rd^2 N}}{{\pt\, \rd\pt \,\rd{}y}}}
\end{equation}

or

\begin{equation}
\label{eq:5}
v_n(\pt) = \frac{ \int_{\Delta y} \rd{}y \frac{{\rd^2 N}}{{\pt\, \rd\pt\, \rd{}y}}v_n(\pt,y)} {\int_{\Delta y} \rd{}y \frac{{\rd^2 N}}{{\pt\, \rd\pt \,\rd{}y}}}.
\end{equation}

The coefficients of the  Fourier expansion of the particles' invariant yield integrated over a broad rapidity and \pt window,
are often referred to as ``integrated flow'':

\begin{equation}
\label{eq:6}
\frac{{\rd{}N}}{{\rd\varphi}} = \frac{1}{{2\pi }}N\left( {1 + \sum\limits_{n = 1}^\infty  {2v_n \cos \left[ {n\left( {\varphi  - \Psi } \right)} \right]} } \right),
\end{equation}
where

\begin{equation}
\label{eq:7}
v_n = \frac{\int_{\Delta y} \rd{}y \int_{\Delta \pt} \pt\, \rd\pt \frac{{\rd^2 N}}{{\pt\, \rd\pt\, \rd{}y}}v_n(\pt,y)}{\int_{\Delta y} \rd{}y \int_{\Delta \pt} \pt\, \rd\pt \frac{{\rd^2 N}}{{\pt \,\rd\pt\, \rd{}y}}}.
\end{equation}

In obtaining the Fourier coefficients, the absolute normalization in the particle yields is not important, as long as the
particle detection efficiency is constant over the chosen transverse momentum and rapidity range. However, if the efficiency varies,
the appropriate efficiency corrections need to be applied. This often leads to a two-step procedure in which first  $v_n(\pt,y)$ is obtained in
a narrow phase-space window where the efficiency is constant, and then a yield-weighted average is performed using  Eqs.~(\ref{eq:4}),~(\ref{eq:5}),
or (\ref{eq:7}), folding in the efficiency-corrected particle spectra. When the particle mass is not
determined in the measurement, the pseudorapidity $\eta=-\ln[\tan(\theta/2)]$, with $\theta$ being the polar angle, is used
instead of the rapidity.

Elliptic flow has been measured at the AGS, SPS, and RHIC.
A notable feature in these measurements is that the elliptic flow measured as a function of
transverse momentum, $v_2(\pt)$, increases with the nucleon-nucleon center-of-mass energy ($\sqrt{s_{NN}}$)
up to about 22\GeV~\cite{Adler:2004cj,Agakishiev:2011id}, and then saturates at
a value compatible with predictions from ideal hydrodynamics~\cite{Heinz:2001xi,Kolb:2003dz}. The most extensive
experimental studies have been performed at the highest RHIC energy of $\sqrt{s_{NN}} = 200$\GeV in AuAu
collisions~\cite{PHENIX,STAR,PHOBOS}. First results from  PbPb collisions at $\sqrt{s_{NN}}= 2.76 $\TeV from the Large Hadron Collider
(LHC)~\cite{ALICE:PhysRevLett.105.252302,:2011yk} indicate that there is little or no change in the transverse
momentum dependence of the elliptic flow measured at the highest RHIC energy and the LHC, despite the approximately
14-fold increase in the
center-of-mass energy. Recent theoretical studies of elliptic flow have focused on
quantifying the ratio of the shear viscosity to the entropy density  of the produced medium assuming viscous
hydrodynamics~\cite{Teaney:review,Romatschke:2009im}, and taking into account a variety of possible initial conditions.

Based on experimental
results and the corresponding theoretical descriptions of the data, the
underlying physics processes that generate the elliptic anisotropy are thought
to vary for different kinematic regions:
\begin{itemize}
\item{\textbf{Elliptic flow in the bulk system}}

Hadrons produced in soft processes, carrying low transverse momentum  ($\pt\lesssim2$\GeVc for
mesons, and  $\pt\lesssim3$\GeVc for baryons~\cite{Adler:2003kt,Adams:2003am,Afanasiev:2007tv,Abelev:2007qg}),
exhibit azimuthal anisotropies that can be attributed to collective flow driven by the pressure gradients in the system.
The description of elliptic flow is amenable to hydrodynamic calculations~\cite{Kolb:2003dz,Huovinen:2006jp,Heinz:2009xj,Teaney:review}.
Comparisons to theory indicate that the flow is primarily generated during an early stage of the system evolution.
\item{\textbf{Recombination region}}

At intermediate transverse particle momentum  ($2\lesssim\pt\lesssim4$\GeVc), the RHIC data show that the elliptic anisotropies
for various identified hadron species approximately follow a common behavior
when both the $v_2(\pt)$ value and the \pt of the particle are
divided by the  number of valence quarks in the
hadron \cite{Adler:2003kt,Adams:2003am,PHENIX_scaling}.
This behavior is successfully reproduced by models invoking
quark recombination as the dominant hadronization mechanism in this momentum
range~\cite{Greco:PhysRevC.68.034904,Fries:PhysRevC.68.044902,Hwa:PhysRevC.70.024904}.
The quark-number scaling of the elliptic flow has been interpreted as evidence
that quark degrees of freedom dominate in the early stages of heavy-ion collisions,
when the collective flow develops~\cite{Muller200584}. The quark recombination may involve both thermally produced quarks
and quarks originating in jet fragmentation. Therefore, the elliptic anisotropy in the recombination region results from
 an interplay between the bulk flow of the system and the azimuthal anisotropy in hadron production
induced by jet quenching.

\item{\textbf{Jet-fragmentation region}}

At intermediate and high transverse momentum  ($\pt\gtrsim 3$\GeVc), where fragments from increasingly harder partons begin
to contribute to the particle spectra, anisotropy in the azimuthal distributions may be generated
from the stronger jet quenching in the direction of the long axis of
the almond-shaped reaction zone~\cite{gyulassy-2001-86,PhysRevC.80.054907,Adare:2010sp,Bass:2008rv}. It is expected that
this mechanism will dominate the elliptic anisotropy of hadrons with momenta in excess of 8\GeVc. Thus, measurements
extending beyond this  \pt range carry information about the path-length dependence of energy loss in the produced medium.
\end{itemize}

The measurements presented in this paper span the transverse momentum range of $0.3<\pt<20\GeVc$ and provide
the basis for comparisons to theoretical descriptions of the bulk properties of the system and the quenching of jets. Such comparisons
may give insight in determining the transport properties of sQGP, namely: the shear viscosity, and the opacity of the plasma.
The theoretical interpretation of the elliptic anisotropy in the recombination region requires identified-hadron measurements
that are not included in this analysis.

In ideal hydrodynamics, the integrated elliptic flow is directly proportional to the initial spatial eccentricity of the
overlap zone~\cite{Kolb:2000sd}. There are many factors that can change this behavior, including viscosity in the
sQGP and the hadronic stages of the system evolution, incomplete thermalization in peripheral collisions, and variations
in the equation-of-state. Event-by-event fluctuations in the
overlap zone~\cite{PhysRevLett.104.142301,PHOBOSeccPART,Bhalerao:2006tp,Voloshin:2007pc,Ollitrault:2009ie,Qiu:2011iv}
could also influence the experimental results, depending on the method that is used to extract the $v_2$ signal. Invoking multiple
methods with different sensitivities to the initial-state fluctuations is important for disentangling the variety of
physics processes that affect the centrality dependence of the elliptic flow. Comparisons to results from
lower-energy measurements and explorations of empirical scaling behaviors could provide additional insights into the nature of
the matter produced in high-energy heavy-ion collisions.

The pseudorapidity dependence of the elliptic flow, $v_2(\eta)$, provides information on the initial state and the early-time
development of the collision, constraining the theoretical models and the description of the longitudinal dynamics in the
collisions~\cite{Hirano:2005xf,Busza:2011pc}. Longitudinal scaling in $v_2(\eta)$ extending over several units of
pseudorapidity (extended longitudinal scaling) has been reported at RHIC~\cite{Back:2004zg} for a broad range of
collision energies ($\sqrt{s_{NN}} = 19.6$--200\GeV). Studies of the evolution of
 $v_2(\eta)$ from RHIC to LHC energies may have implications for the unified description of sQGP in both energy regimes.

The paper is organized as  follows: Section~\ref{sec:Experimental Method} gives
the experimental details of the measurement including the Compact Muon Solenoid (CMS) detector,
triggering and event selection, centrality determination, Glauber-model calculations, reconstruction of the
charged-particle transverse momentum spectra,  methods of measuring the elliptic anisotropy,
and the studies of the systematic uncertainties. In Section~\ref{sec:Results} we present results of $v_2$ as a function of
transverse momentum, centrality, and pseudorapidity, and the measurement of the charged-particle transverse momentum spectra.
The elliptic flow results are obtained with
the event-plane method~\cite{Poskanzer:1998yz}, two- and four-particle cumulants~\cite{Borghini:2001vi},
and Lee--Yang zeros~\cite{Bhalerao:2003xf, Borghini:2004ke}. The analysis is
performed in 12 classes of collision centrality  covering the most central 80\% of inelastic collisions.
We study the eccentricity
scaling of $v_2$, and investigate the differences in the
results obtained from different methods, taking into account their sensitivity to initial-state fluctuations.
Section~\ref{sec:Discussion} is devoted to detailed comparisons with results obtained by other experiments at lower energies
and the exploration of different scaling behaviors of the elliptic flow. The results of our studies are summarized in
Section~\ref{sec:Summary}.

\section{Experimental method}
\label{sec:Experimental Method}
\subsection{CMS detector}
\label{sec:CMS}
The central feature of the CMS apparatus is
a superconducting solenoid of 6\unit{m} internal diameter, providing a 3.8\unit{T} field.
Within the field volume are a silicon tracker, a crystal electromagnetic
calorimeter, and a brass/scintillator hadron calorimeter.
Muons are measured in gas-ionization detectors embedded in the steel return
yoke. In addition to these detectors, CMS has extensive
forward calorimetry. The inner tracker measures charged particles
within the pseudorapidity range  $|\eta| < 2.5$, and consists of
silicon pixel and
silicon strip detector modules.
The beam scintillation
counters (BSC) are a series of scintillator tiles
 which are sensitive to almost the full PbPb interaction cross section.
These tiles are placed along the beam line
at a distance of $\pm 10.9$\unit{m} and $\pm 14.4$\unit{m} from the interaction point,
and can be used to provide minimum-bias triggers.
The forward
hadron calorimeter (HF) consists of a steel absorber structure that is composed of grooved
plates with quartz fibers inserted into these grooves.
The HF calorimeters have a cylindrical shape
and are placed at a distance of 11.2\unit{m} from the interaction point, covering
the pseudorapidity range of $2.9<|\eta|<5.2$.
A more detailed description of
the CMS detector can be found elsewhere~\cite{JINST}.

The detector coordinate system has the origin centered at the
nominal collision point inside the experiment, with the $y$ axis pointing vertically
upward, the $x$ axis pointing radially inward towards the center of the LHC ring, and the
$z$ axis pointing along the counterclockwise beam direction.

\subsection{Event selection}
\label{sec:Event Selection}
The measurements presented are performed by analyzing PbPb collision events recorded by the CMS detector in 2010.
From these data, the minimum-bias event sample is collected using coincidences
between the trigger signals from both the $+z$ and $-z$ sides of either the
BSC or HF. The minimum-bias trigger used for this
analysis is required to be in coincidence with the presence of both colliding ion bunches in the
interaction region. This requirement suppresses noncollision-related noise, cosmic rays,
and beam backgrounds (beam halo and beam-gas collisions).
The total hadronic collision rate varied
between 1 and 210\unit{Hz}, depending on the number of colliding bunches and the bunch intensity.

In order to obtain a pure sample of inelastic hadronic collisions, several offline selections
are applied to the triggered event sample. These selections remove contamination from
noncollision beam backgrounds and from ultraperipheral collisions (UPC) that lead to
an electromagnetic breakup of one or both of the Pb nuclei~\cite{Djuvsland:2010qs}.
First, beam-halo events are vetoed based on the BSC timing.
Then, to remove UPC and beam-gas events, an offline HF coincidence of at least three towers
on each side of the interaction point is required, with a total deposited energy of at least
3\GeV in each tower. A reconstructed primary vertex made of at least two tracks
and consistent with the nominal interaction point position is required.
To further reject beam-gas and beam-halo events, the pixel clusters are required to have a length
along the beam direction compatible with being produced by particles originating
from the primary vertex, as for the study in~\cite{Khachatryan:2010us}.
Additionally, a small number of  noisy events with uncharacteristic
hadron calorimeter responses are removed.

The acceptance of the silicon tracker for $|\eta| > 0.8$ is found to vary significantly with respect to
the longitudinal position of the collision point relative to the geometric center of the detector.
This event-by-event variance in the tracking efficiency contributes to a systematic
bias of the elliptic flow measurements at forward pseudorapidity. In order to remove this bias,
events in this analysis are required to have a longitudinal vertex position
within 10\unit{cm} of the geometric center of the detector.
After all selections, 22.6  million minimum-bias events, corresponding to an integrated luminosity of approximately
3\mubinv, remain in the final sample.

\subsection{Centrality determination and Glauber model calculations}
\label{sec:Centrality Determination}
In this analysis, the observable used to determine centrality is the total energy
in both  HF calorimeters. The distribution of this total energy is used to divide the event sample
into 40 centrality bins, each representing 2.5\% of the total nucleus-nucleus interaction
cross section. The events are then regrouped to form 12 centrality
classes used in the analysis: 0--5\% (most central), 5--10\%, 10--15\%, 15--20\%, 20--25\%,
25--30\%, 30--35\%, 35--40\%,
40--50\%, 50--60\%, 60--70\%, and 70--80\% (see Table~\ref{tbl:glauber}).
Using Monte Carlo (MC) simulations, it is estimated that the minimum-bias trigger and the event selections include  $(97\pm3)$\% of
the total inelastic cross section. For the events included in this analysis (0--80\% centrality), the trigger is fully efficient.

\begin{figure}[htb]
\begin{center}
\includegraphics[width=0.5\textwidth]{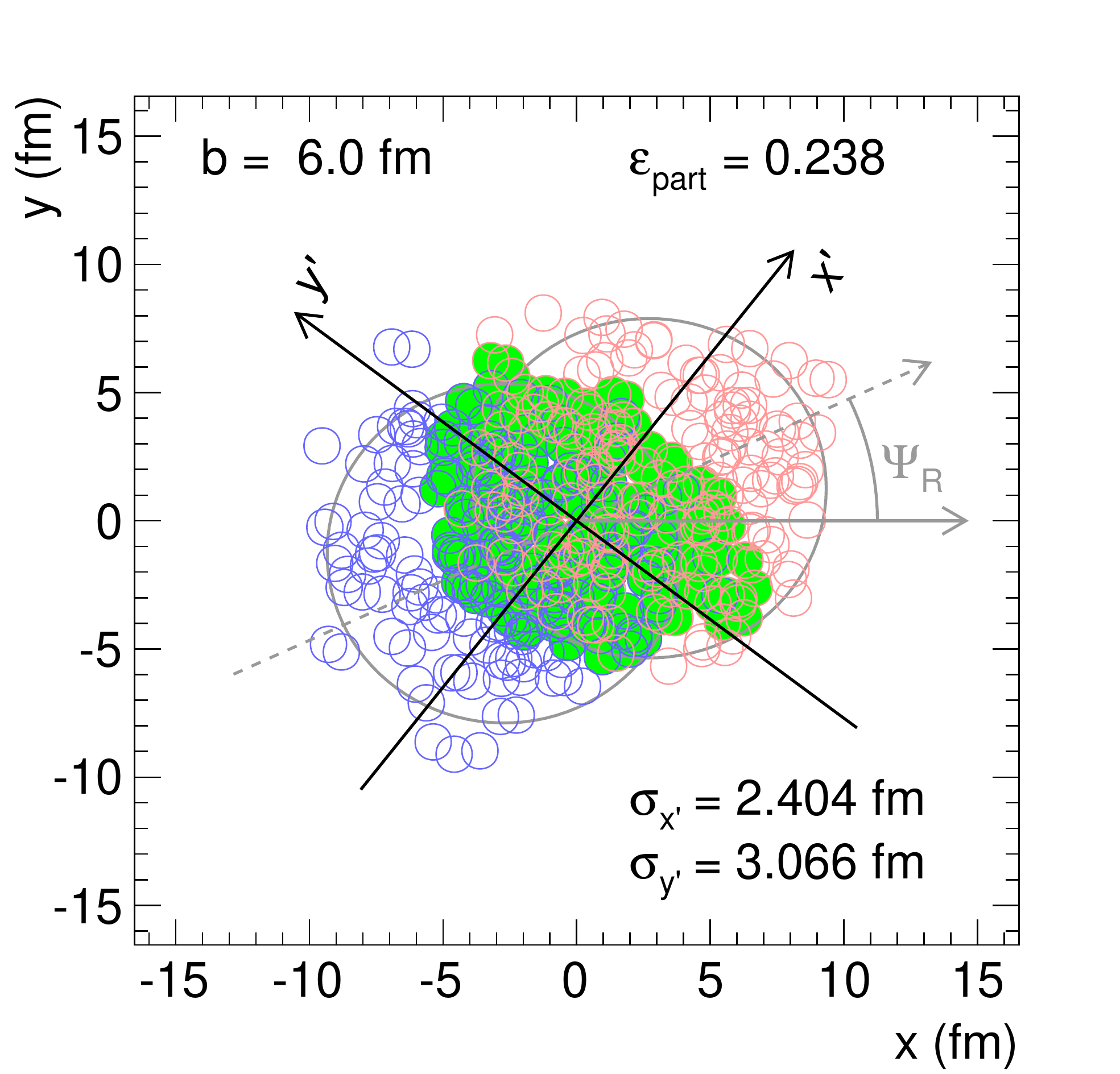}
\caption{\label{fig:glauber}
(Color online) A schematic view  of a PbPb collision with an impact
parameter $ b = 6$\unit{fm} as obtained from the Glauber model. The nucleons that participate in inelastic interactions
are marked with filled circles. The x and y coordinates represent the laboratory frame, while x' and y' represent the
frame that is aligned with the axes of the ellipse in the participant zone. The participant eccentricity
$\epsilon_{\text{part}}$ and the standard deviations of the participant spatial
distribution $\sigma_{y'}$ and $\sigma_{x'}$ from which the transverse overlap area of the
two nuclei is calculated are also shown. The angle $\Psi_{\textrm{R}}$ denotes the
orientation of the reaction plane.}
\end{center}
\end{figure}

For each group of events that comprises a centrality class, we evaluate a set of quantities that characterize the initial geometry
of the collisions using a MC Glauber model. The Glauber model is a multiple-collision model that treats a nucleus-nucleus
collision as an independent sequence of nucleon-nucleon collisions  (see Ref.~\cite{Miller}
and references therein).   The nucleons that participate in inelastic interactions are called ``participants''.
A schematic view of a PbPb collision with an impact parameter $ \text{b} = 6$\unit{fm}, as obtained from the Glauber model, is
shown in Fig.~\ref{fig:glauber}. The direction and the magnitude of the impact parameter vector and the corresponding reaction-plane angle $\Psi_{R}$, are the same as in Fig.~\ref{fig:almond}. However, the initial interaction zone
as determined by the spatial distribution of the participants (filled circles) is no longer regular in shape and is not necessarily
symmetric with respect to the reaction plane. In each event one can evaluate the variances $\sigma_{x}^{2}$ and $\sigma_{y}^{2}$, and the
covariance  $\sigma_{xy}= \langle xy \rangle -  \langle x \rangle \langle y \rangle$ of the
participant distributions projected on the $x$ and $y$ axes. One can then find a frame $x'$-$y'$ that minimizes  $\sigma_{x'}$, and define a
``participant plane'' using the beam direction and the $x'$ axis~\cite{PHOBOSeccPART,PhysRevLett.98.242302}. In this frame, the covariance $\sigma_{x'y'}$
of the participant spatial distribution vanishes. To characterize the geometry of the initial state of the collision,  we
define~\cite{PHOBOSeccPART,PhysRevLett.98.242302} the eccentricity of the participant zone  $\epsilon_\mathrm{part}$, its cumulant moments
$\epsilon\{2\}$ and $\epsilon\{4\}$, and the  transverse overlap area of the two nuclei $S$:

\begin{equation}
\label{eq:8}
\epsilon_{\text{part}} \equiv \frac{\sigma_{y'}^{2} - \sigma_{x'}^{2}}{ \sigma_{y'}^{2} + \sigma_{x'}^{2} } = \frac{ \sqrt { \left( \sigma_{y}^{2} - \sigma_{x}^{2} \right)^{2} + 4\sigma_{xy}^{2} } }{ \sigma_{y}^{2} + \sigma_{x}^{2} },
\end{equation}
\begin{equation}
\label{eq:9}
\epsilon\{2\}^{2} \equiv \langle \epsilon_\text{part}^{2} \rangle ,
\end{equation}
\begin{equation}
\label{eq:10}
\epsilon\{4\}^{4} \equiv  2 \langle \epsilon_\text{part}^{2} \rangle^{2} - \langle \epsilon_\text{part}^{4} \rangle, ~\text{and}
\end{equation}

\begin{equation}
\label{eq:11}
S \equiv \pi \sigma_{x'}\sigma_{y'} = \pi \sqrt{\sigma_{x}^{2}\sigma_{y}^{2}-\sigma_{xy}^{2}}.
\end{equation}

In Eqs.~(\ref{eq:9}) and ~(\ref{eq:10}), the average is taken over many events from the same centrality interval.

\begin{table*}[htb]
\begin{center}
\caption{For the centrality bins used in the analysis, the average values of the number of participating nucleons,
transverse overlap area of the two nuclei, participant eccentricity, and cumulant moments of the participant eccentricity,
along with their systematic uncertainties from the Glauber
model.
\label{tbl:glauber}}
\begin{tabular}{ c c c c c c c}
\hline
Centrality & $\langle N_{\text{part}} \rangle$ &
$\langle S \rangle [\mathrm{fm}^{2}]$ &
$\langle \epsilon_{\text{part}} \rangle$ &
$\epsilon\{2\}$   &   $\epsilon\{4\}$  \\
\hline
0--5\% & 381$\pm$2 & 29.4$\pm$1.2  &0.074$\pm$0.003 & 0.084$\pm$0.003 &0.053$\pm$0.002      \\
5--10\% & 329$\pm$3 & 26.6$\pm$1.1  &0.111$\pm$0.005 & 0.128$\pm$0.005 &0.060$\pm$0.003     \\
10--15\% & 283$\pm$3 & 24.0$\pm$1.0  &0.154$\pm$0.007 & 0.175$\pm$0.007 &0.122$\pm$0.005      \\
15--20\% &240$\pm$3 & 21.6$\pm$1.0  &0.198$\pm$0.009 & 0.219$\pm$0.009 &0.171$\pm$0.007      \\
20--25\% &204$\pm$3 & 19.5$\pm$0.9  &0.238$\pm$0.009 & 0.262$\pm$0.010 &0.214$\pm$0.008     \\
25--30\% & 171$\pm$3 & 17.5$\pm$0.8  &0.276$\pm$0.010 & 0.302$\pm$0.012 &0.253$\pm$0.010      \\
30--35\% &143$\pm$3 & 15.7$\pm$0.8  &0.312$\pm$0.011 & 0.339$\pm$0.012 &0.288$\pm$0.010   \\
35--40\% & 118$\pm$3 & 14.1$\pm$0.7  &0.346$\pm$0.010 & 0.375$\pm$0.011 &0.322$\pm$0.009     \\
40--50\% & 86.2$\pm$2.8 & 12.0$\pm$0.6  &0.395$\pm$0.010 & 0.429$\pm$0.011 &0.370$\pm$0.010   \\
50--60\% &53.5$\pm$2.5 & 9.4$\pm$0.5  &0.465$\pm$0.008 & 0.501$\pm$0.009 &0.437$\pm$0.007    \\
60--70\% & 30.5$\pm$1.8 & 7.1$\pm$0.4  &0.543$\pm$0.011 & 0.581$\pm$0.012 &0.514$\pm$0.010     \\
70--80\% & 15.7$\pm$1.1 & 4.8$\pm$0.3  &0.630$\pm$0.016 & 0.662$\pm$0.017 &0.598$\pm$0.015     \\
\hline
\end{tabular}
\end{center}
\end{table*}

To implement the Glauber model for PbPb collisions at $\sqrt{s_{NN}} = 2.76$\TeV,
we utilize the foundation of
a published Glauber MC software package \textsc{TGlauberMC} \cite{PublishedPHOBOSGlauberMC},
which was developed for the PHOBOS Collaboration at RHIC.
Standard parameters of the Woods-Saxon function used for modeling the
distribution of nucleons in the Pb nuclei are taken from Ref.~\cite{DeJager:1987qc}.
The nucleon-nucleon inelastic cross section, which is used to determine how
close the nucleon trajectories need to be in order for an interaction to
occur, is taken as $64\pm5$\unit{mb}, based on a fit of the existing data for total and
elastic cross sections in proton-proton and proton-antiproton collisions~\cite{PDBook}.
The uncertainties in the parameters involved in these calculations contribute to
the systematic uncertainty in $N_\text{part}$, $S$, and $\epsilon_\text{part}$
for a given centrality bin. The connection between the experimentally defined centrality classes using the HF energy distribution
and $N_\text{part}$ from the Glauber model is obtained~\cite{Chatrchyan:2011sx} from fully simulated and reconstructed MC events generated with
the AMPT~\cite{Lin:2004en} event generator. The calculated Glauber model variables for each centrality class are shown in
 Table~\ref{tbl:glauber}.

\subsection{Reconstruction of the charged-particle transverse momentum distributions and  the mean
transverse momentum}
\label{sec:Tracking}

To determine the transverse momentum distributions of the charged particles produced in the collisions, we first need to reconstruct
the  particles' trajectories (``tracks'') through the 3.8\unit{T} solenoidal magnetic field. The tracks are reconstructed by starting
with a ``seed'' comprising two or three reconstructed signals (``hits'') in the inner layers of the silicon strip and pixel detectors
that are compatible with a helical trajectory of some minimum \pt and a selected region around the reconstructed primary
vertex or nominal interaction point. This seed is then propagated outward through subsequent layers
using a combinatorial Kalman-filter algorithm.
Tracking is generally performed in multiple iterations, varying the layers used in the seeding
and the parameters used in the pattern recognition, and removing duplicate tracks between iterations.
This algorithm is described in detail in Ref.~\cite{CMSTracking}. The algorithm
used in most of the CMS proton-proton analyses, as well as the tracking detector performance
for the 2010 run, are described in Ref.~\cite{Tracking2010}.

The six-iteration process used in proton-proton collisions is computationally not feasible in
the high-multiplicity environment of very central PbPb collisions. In place of this, a
simple two-iteration process is used. The first iteration builds seeds from hits in some combination
of three layers in the barrel and endcap pixel detectors compatible with a trajectory of $\pt > 0.9\GeVc$
and a distance of closest approach to the reconstructed vertex of no more than 0.1\unit{cm} in the transverse
plane and 0.2\unit{cm} longitudinally.
These tracks are then filtered using selection criteria based on a minimum number of reconstructed hits,
vertex compatibility along the longitudinal direction and in the transverse plane, and low relative uncertainty on the
reconstructed momentum.

In the second iteration, seeding is also performed using three layers of the pixel detector, but the
minimum transverse momentum requirement is relaxed to $\pt > 0.2\GeVc$. These tracks are not propagated
through the silicon-strip detector, but simply refitted using the transverse position of the
beam spot as an additional constraint.
These pixel-only tracks are then filtered using selection criteria of vertex compatibility
along the longitudinal axis and statistical goodness of fit.

\begin{figure*}[hbtp]  \begin{center}
    \includegraphics[width=0.9\textwidth]{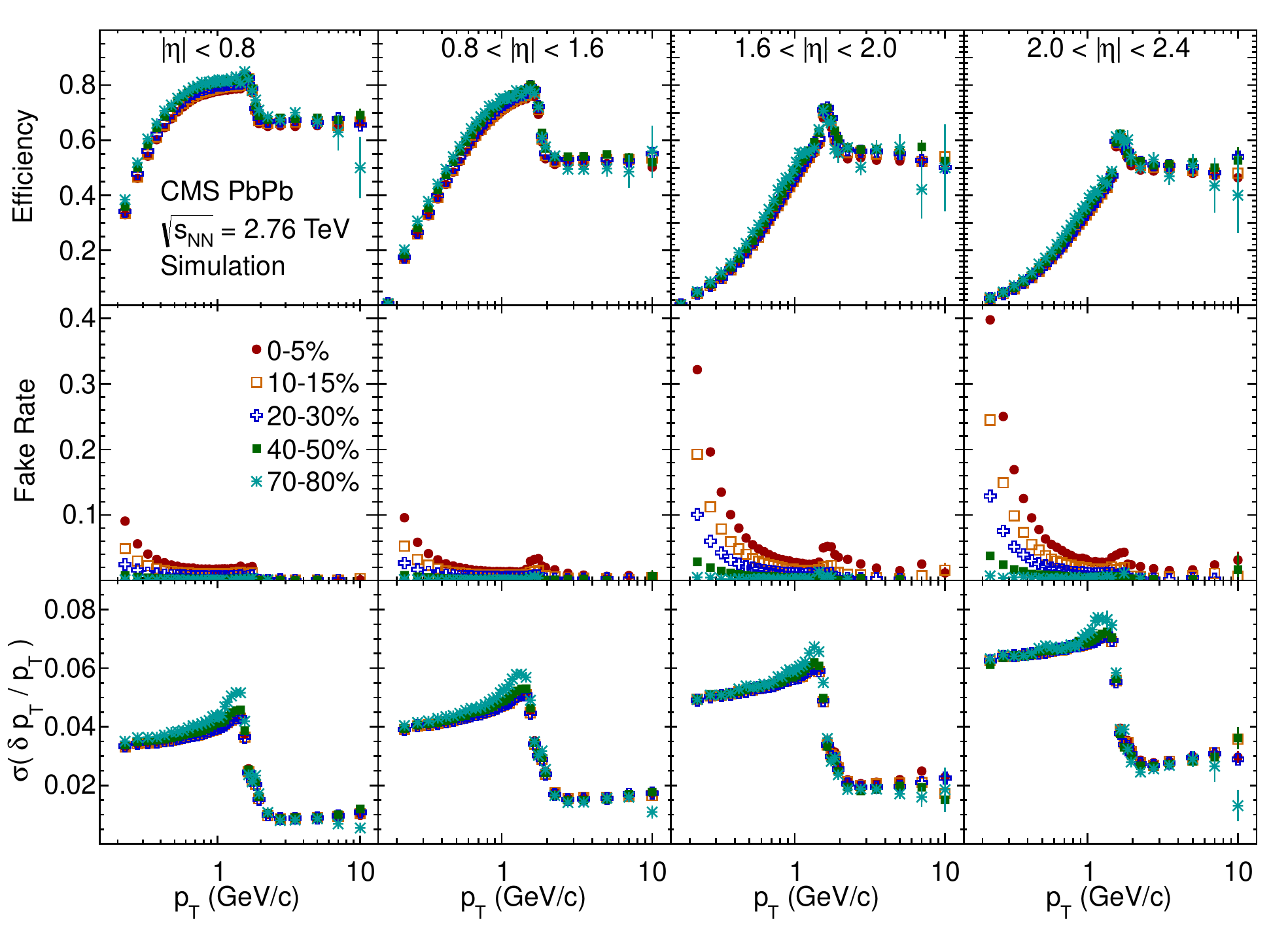}
    \caption{(Color online)
Efficiency (top), fake rate (middle), and momentum resolution (bottom) of charged tracks obtained from \textsc{hydjet}
simulated events in four pseudorapidity regions: $|\eta| < 0.8$, $0.8 < |\eta| < 1.6$,
$1.6 < |\eta| < 2.0$, and $2.0 < |\eta| < 2.4$ displayed from left to right, and for
the five centrality classes given in the legend.
}
    \label{fig:tracking_technical}
  \end{center}
\end{figure*}

The tracks from both collections are checked for duplicate tracks using the number of
hits in common between the two tracks, and duplicates are
removed giving preference to the first-iteration tracks.
The tracking algorithm may sometimes misidentify tracks by combining silicon detector signals that
do not originate from the same charged particle. It is important to keep the proportion of misidentified tracks
(referred to as ``fake tracks''), or fake rate, as low as possible.
To create the final collection, first-iteration tracks with $\pt > 1.5\GeVc$ are combined with second-iteration
pixel-only tracks with $\pt < 1.8\GeVc$.
These \pt limits were chosen to exclude kinematic regions where a given iteration has a high fake rate.
 The efficiency and fake rate
of this modified tracking collection are found both by using a full
MC simulation of PbPb
collisions based on the \textsc{hydjet} event generator~\cite{Lokhtin:2005px} and by embedding simulated
charged pions into PbPb data events.
The efficiency, fake rate, and momentum resolution
of this final collection determined by the simulated events from the \textsc{hydjet} event generator are shown in Fig.~\ref{fig:tracking_technical}
for events of five different centrality classes. The abrupt change in efficiency, fake rate, and momentum resolution
seen in the figure occurs at the transverse momentum where the two track collections are
merged.

The charged-particle transverse momentum distributions are corrected for the loss of acceptance, efficiency, and the contributions from
fake tracks. Each detected track is weighted by a factor $w_{tr}$ according to its centrality, transverse
momentum, and pseudorapidity:
\begin{equation}
w_{tr} (\text{centrality},\pt, \eta) = \frac{1-f}{e}.
\end{equation}

Here $f$ is the fraction of fake tracks that do not correspond to any charged
particle, and $e$ is the absolute tracking efficiency accounting
for both geometric detector acceptance and algorithmic tracking efficiency.
The proportion of multiply reconstructed particles and reconstructed particles
from secondary decays is negligible and is not included in the correction factor.
The fully corrected transverse momentum distributions are measured in 12 centrality classes
over the pseudorapidity range $|\eta| < 2.4$  in bins of $\Delta \eta = 0.4$, as discussed in Section~\ref{s:spectra}.
These distributions are used in obtaining integrated $v_2$ values. We also study the evolution of $\langle\pt\rangle$ with
pseudorapidity and centrality (Section~\ref{s:spectra}), and center-of-mass energy (Section~\ref{sec:Discussion}).

To evaluate $\langle\pt\rangle$, the spectra need to be extrapolated down to $\pt = 0\GeVc$. The
extrapolation is performed using a Tsallis distribution~\cite{Tsallis:1987eu,Wilk:2008ue,Biro:2008hz}:

\begin{equation}
 E \frac{\rd^3N_{\mathrm{ch}}}{\rd p^3} =
\frac{1}{2\pi p_T} \frac{E}{p} \frac{\rd^2N_{\mathrm{ch}}}{\rd\eta \rd p_T} =
C\left(1 + \frac{E_T}{nT}\right)^{-n} ,
 \label{eq:tsallis}
\end{equation}

where $\ET = \sqrt{m^2 + \pt^2} - m$, and $m$ is taken to be the charged pion mass.
The measured spectra are fitted in the range $0.3<\pt<3.0\GeVc$ , and the fit parameters $C$, $n$,
and $T$ are determined. The mean transverse momentum is then evaluated using the fit function in the
extrapolation region of $0\le\pt<0.3 \GeVc$, and the data from the range $0.3\le\pt\le6.0 \GeVc$. This method has been
previously applied in CMS~\cite{Khachatryan:2010us} in the measurement of $\langle\pt\rangle$ in pp collisions at $\sqrt{s}=7\TeV$.

\subsection{Methods for measuring the anisotropy parameter \texorpdfstring{$v_{2}$}{v2}}
\label{sec:methods}

Anisotropic flow aims to  measure the azimuthal correlations of the particles produced in heavy-ion collisions in relation
to the initial geometry of the collisions. Originally, the flow was defined as a correlation of the particle emission angles with
the reaction plane~\cite{Ollit,Poskanzer:1998yz}. More recently, it was recognized~\cite{PHOBOSeccPART,PhysRevLett.98.242302}
that the initial geometry is better characterized by the positions of the individual nucleons that participate in inelastic
interactions, and thus define a participant plane that fluctuates around the reaction plane on an event-by-event basis.
Neither the reaction plane nor the participant plane are directly measurable experimentally. Instead, there are several experimental methods that
have been developed to evaluate the anisotropic flow based on the final-state particle distributions.  In the present analysis, we use the
event-plane method, two-particle and four-particle cumulants~\cite{Borghini:2001vi}, and the Lee--Yang zeros method~\cite{Bhalerao:2003xf,
Borghini:2004ke} the last of which is based on correlations involving all the particles in the event.

The anisotropic flow measurements are affected by fluctuations that come from several sources. Statistical fluctuations arise due to the fact
that a finite number of particles is used to determine a reference plane for the flow and the multiplicity fluctuations within the chosen
centrality interval. The effect of these fluctuations is to reduce the measured flow signal and is largely compensated for by the resolution
corrections described below. Any remaining effects from statistical fluctuations on our measurements are included in the systematic
uncertainties (see Section~\ref{sec:Systematic Studies}). Another more important source of fluctuations comes from the event-by-event fluctuations
in the participant eccentricity that are present even at fixed $N_{\text{part}}$. These dynamical fluctuations have been
shown~\cite{Bhalerao:2006tp,Voloshin:2007pc,Ollitrault:2009ie,PHOBOSeccPART,Qiu:2011iv} to affect the various methods for estimating the flow
differently, since each of them is  based on a different moment of the final-particle momentum distribution. For example, the two-particle
cumulant method measures an r.m.s.\ value of the flow that is higher than the mean value. Conversely, the four-particle cumulant
and other multiparticle correlation methods return a value that is lower than the mean value. For the event-plane method, the results vary
between the mean and the r.m.s.\ value depending on the event-plane resolution, which varies in each centrality
interval.

In addition, there exist other sources of correlations in azimuth, such as those
from resonance decays, jets, and Bose--Einstein correlations between identical particles. These correlations,
which are not related to the participant plane, are called nonflow correlations. The various methods proposed to
estimate the magnitude of anisotropic flow have different sensitivities to the nonflow correlations, thus allowing systematic
checks on the flow measurements.
The multiparticle correlation methods are least affected by nonflow correlations, but they do not work reliably
when either
the flow anisotropy ($v_2$) or the multiplicity in the selected phase-space window is small.
This happens in the most central and in the most peripheral events~\cite{Borghini:2001vi,Bhalerao:2003xf}.
Thus, the two-particle cumulant and the event-plane methods
provide an extended centrality range, albeit with a larger nonflow contribution.

All four methods used here have been extensively studied and applied in different experiments. Thus, we
limit our description in the following subsections
only to the features that are specific for our implementations of these methods.

\subsubsection{Event-plane method}
The event-plane method estimates the magnitude of anisotropic flow by reconstructing an
``event-plane'' containing both the beam direction and the direction of the maximal flow determined from the azimuthal
distributions of the final-state particles. Under the assumption that the flow is driven by the
initial-state asymmetry in the nuclear overlap zone, and that there are no other sources of azimuthal correlations in the
final-state particles, the event-plane is expected to coincide with the participant plane~\cite{PHOBOSeccPART} defined in Fig.~\ref{fig:glauber}.
Recent theoretical calculations~\cite{Holopainen:2010gz,Petersen:2010cw,Qiu:2011iv,Gardim:2011xv} confirm that the event plane and the
participant plane are strongly correlated event by event.
Since the event plane is determined using a finite number of particles, and detected with finite angular resolution,
the measured event-plane angle fluctuates about its true value. As a
result, the observed particle azimuthal anisotropy with respect to the event-plane is smeared compared to its true value.
The true elliptic flow coefficient $v_2$ in the event-plane method is evaluated by dividing the observed $v^{\mathrm{obs}}_2$
value by a resolution correction factor, $R$, which accounts for the event-plane resolution.

To determine the event-plane resolution correction, a technique that sorts the particles from each event into three subevents based on their
pseudorapidity values~\cite{Poskanzer:1998yz} is used.
For subevents $A$, $B$, $C$ in
three different pseudorapidity windows, the event-plane resolution correction
factor $R_A$ for subevent \rm{A} is found as:

\begin{equation}
\label{f.EP_7} R_A=\sqrt{\frac{\langle \cos[2 (\Psi^A-\Psi^B)]\rangle
\langle\cos[2 (\Psi^A-\Psi^C)]\rangle}{\langle\cos[2
(\Psi^B-\Psi^C)]\rangle}},
\end{equation}
where $\Psi^A,\Psi^B$, and $\Psi^C$ are the event-plane angles determined from the corresponding subevents,
and the average is over all events in a selected centrality class used in the $v_{2}$ analysis.

In our implementation of the method, the event-plane angle determined for the
subevent furthest in $\eta$ from the track being used in the
elliptic flow analysis is used, and the corresponding resolution
correction is employed.
This selection minimizes the contributions of auto-correlations
and other nonflow effects  that arise if the particles used in the event-plane determination and those used in
the flow analysis are close in phase space.

\begin{figure}[hbtp]
  \begin{center}
    \includegraphics[width=0.5\textwidth]{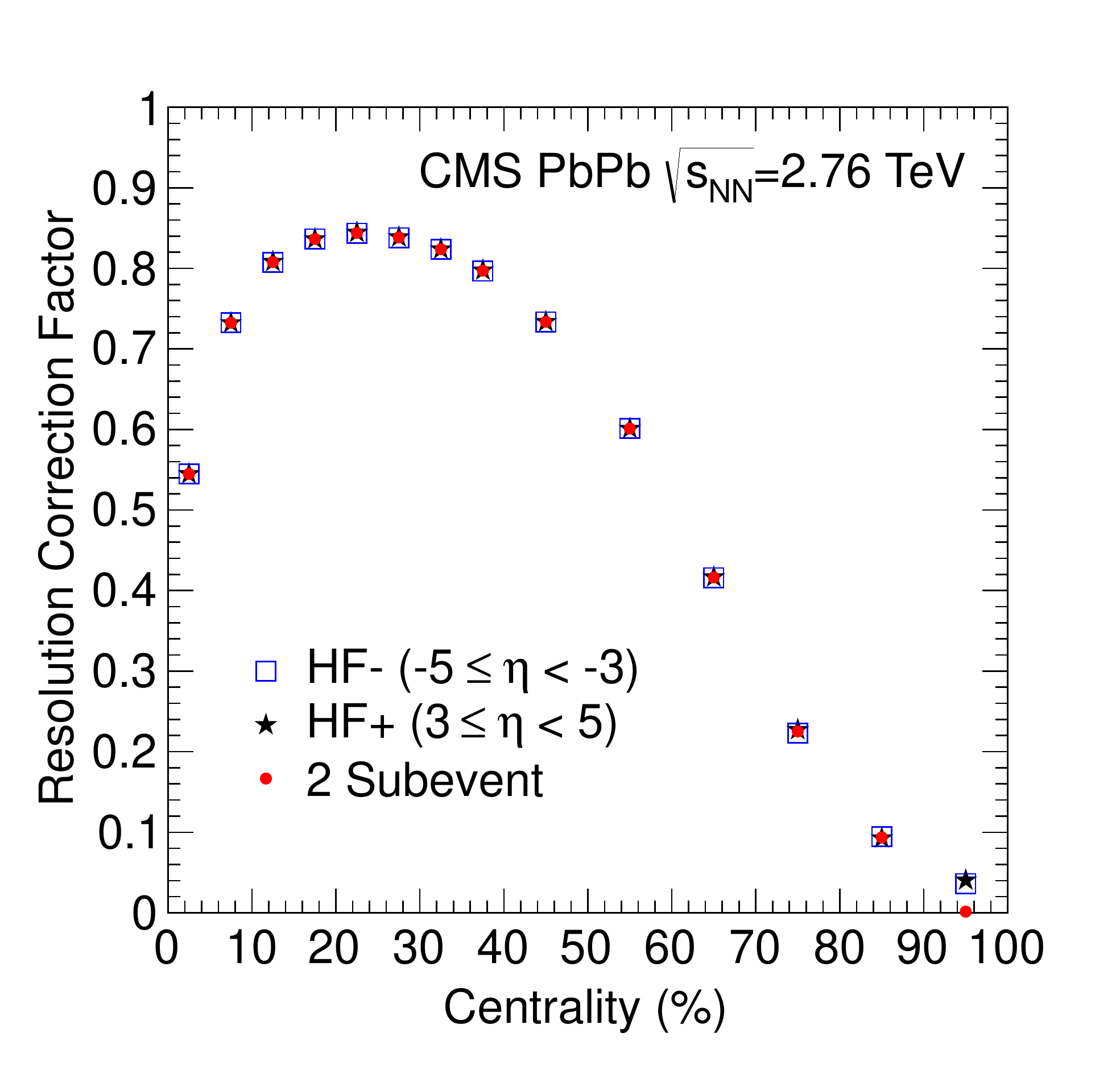}
    \caption{(Color online) Event-plane resolution correction factors as a function of centrality for the two event-planes
(\textit{HF-} and \textit{HF+}) used in determining the elliptic anisotropy parameter $v_2$. The corrections determined
with the three-subevent method used in the analysis are shown as open squares and star symbols.
The results from a two-subevent method used in evaluating
the systematic uncertainties are shown as filled circles,
though they overlap the other points in all but the most peripheral bin.
}
    \label{fig:Rescor}
\end{center}
\end{figure}

To achieve the largest pseudorapidity gap possible, two event planes are
defined, with calorimeter data  covering the pseudorapidity
ranges of $-5 <\eta<-3$ and $3<\eta<5$,  labeled ``\textit{HF-}'' and ``\textit{HF+}'', respectively.
These pseudorapidity ranges are primarily within the coverage of the HF
calorimeters.  A third event plane, found using charged particles detected in the tracker in the pseudorapidity range $-0.8<\eta<0.8$,
is also defined and used in the three-subevent technique for determining the resolution
corrections for \textit{HF-} and \textit{HF+}. The resulting resolution corrections are presented in Fig.~\ref{fig:Rescor}.
Particles detected in the tracker with $\eta>0$ are then correlated with the \textit{HF-} event plane, and those with
$\eta<0$ with the \textit{HF+} plane.  In this manner, the minimum pseudorapidity
gap between particles used in the event-plane determination and those for
which the $v_2$ signal is  measured is 3 units. A two-subevent technique based on \textit{HF-} and \textit{HF+}, with
a resolution parameter defined as
$ R_{A/B}=\sqrt{\langle \cos[2 (\Psi^A-\Psi^B)]\rangle}$, is also implemented and used for systematic studies. The values of $ R_{A/B}$ are shown for comparison in Fig.~\ref{fig:Rescor}.

A standard flattening procedure~\cite{Barrette:1997pt} using a Fourier decomposition of the distribution of the event-plane
angles to $21^{st}$ order is used to shift the event-by-event
plane angle to correct for asymmetries in the event-plane distribution that arise from the detector acceptance
and other instrumental effects. Although most of these effects are already accounted for with a correction involving just the first four coefficients
in the expansion, the larger order was used for data quality monitoring purposes. For each centrality class in the analysis, the flattening parameters are
calculated by grouping events according to the longitudinal location of their primary collision vertex in 5\unit{cm}-wide bins.

\subsubsection{Cumulant method}
The cumulant method measures flow utilizing a cumulant expansion of multiparticle
azimuthal correlations, without determining the orientation of the event plane. The idea is that if the
particles are correlated with the event-plane orientation, then there also exist correlations between them.
In our analysis, we utilize two- and four-particle correlations.
To calculate the cumulants of these correlations, from which the flow coefficient is
extracted, we use a generating function of the multiparticle correlations
in a complex plane~\cite{Borghini:2001vi}. First, we evaluate the ``integrated'', or reference, flow by
constructing the corresponding  generating function including all particles from a broad ($\pt$, $\eta$) window,
and averaging over the events in a given centrality class. The reference flow may not be corrected for tracking efficiency and
should not be equated with the fully corrected integrated flow of the events. Then, the differential flow, i.e.,
the flow in a narrower phase-space window, either in \pt or $\eta$, is measured with respect to the
reference flow. In the cumulant and Lee--Yang zeros methods, the reference flow serves the same purpose as the
determination of the event-plane angle and the resolution correction factors in the event-plane method. In our
analysis of
 $v_{2}(\pt)$, the \pt and $\eta$ ranges
for the reference flow are $0.3<\pt <3$\GeVc and $|\eta|<0.8$, respectively. In the analysis of  $v_{2}(\eta)$, the
reference flow is obtained for the range $0.3<\pt <3$\GeVc and $|\eta|<2.4$ in order to maximize the
resolution parameter, which improves as the charged
hadron multiplicity $M$ in the selected  phase-space window increases.  The transverse momentum restriction of
$\pt <3$\GeVc is imposed to limit the contributions from hadrons originating in jets, and thus reduce the
nonflow correlations
contributing to the measured elliptic anisotropy parameter. To avoid auto-correlations, the particles used for determining
differential flow are not included in evaluating the reference flow. The generating function for the reference flow is
calculated at several different points in the complex plane, and we then interpolate between these points. We use three values for the
radius parameter, $r_0$, and seven values for the polar angle, as described in Ref.~\cite{Borghini:2001vi}. The radius parameters
are determined according to the detected charged particle multiplicity and the number of events analyzed in each centrality class.
Each particle in the differential \pt or $\eta$ bin is correlated to the particles used for the reference flow
through a differential generating function. To account for the fact that the track reconstruction efficiency may
vary across the chosen bin, we implement an efficiency correction that is applied as a track-by-track weight
in the construction of the differential generating function.

\subsubsection{ Lee--Yang zeros method}
\begin{figure}[hbpt]
\begin{center}
    \includegraphics[width=0.5\textwidth]{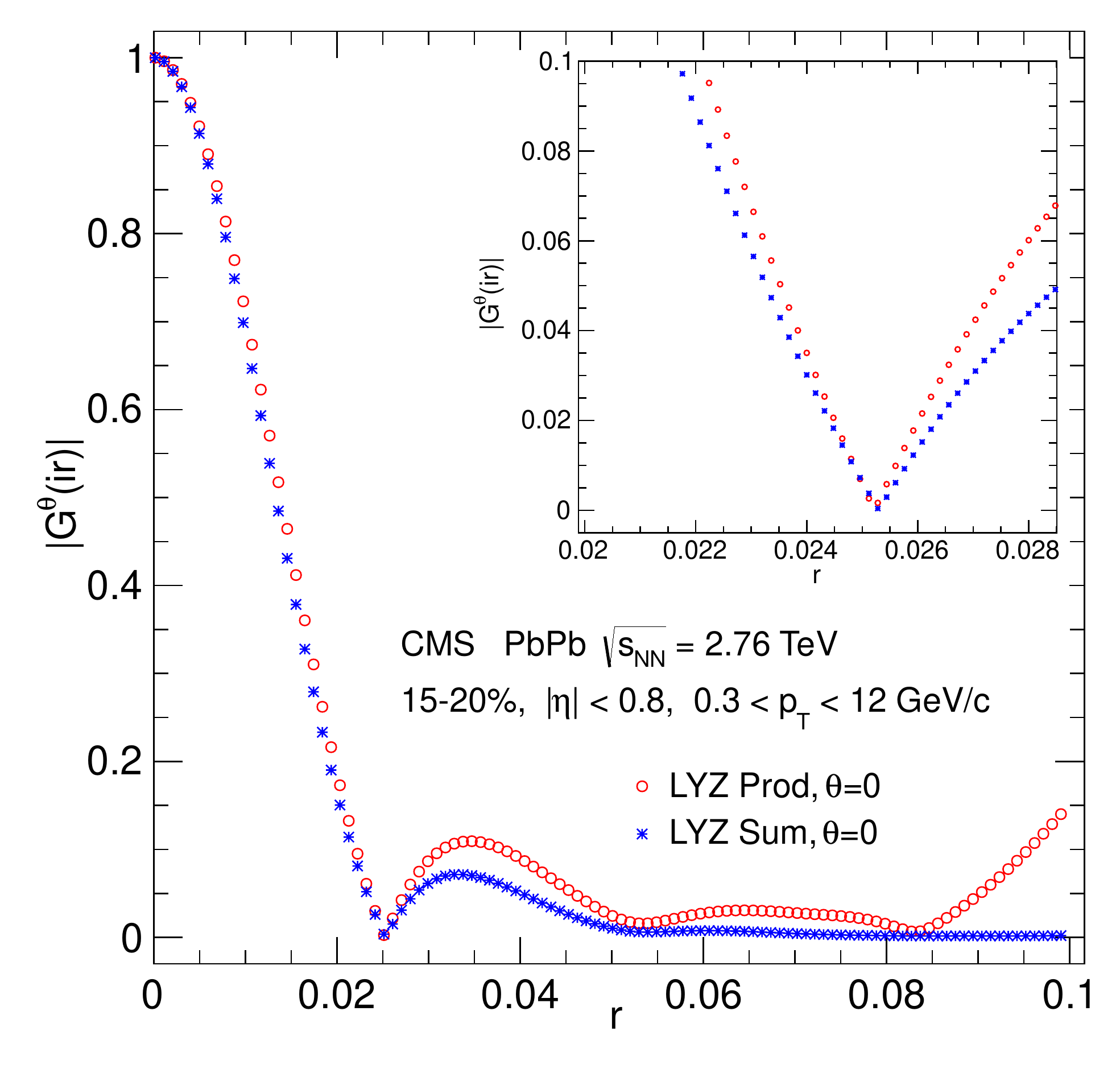}
\caption{\label{fig:LYZ_Gir_r}(Color online)
An example of the modulus of the second harmonic Lee--Yang zero generating function
$G^{\theta}(ir)$ as a function of the
imaginary axis coordinate $r$ for $\theta$ = 0. Both the sum and product generating functions are shown,
calculated from events with centrality $15$--$20\%$, $|\eta| < 0.8$, and $0.3 < \pt < 12$\GeVc.
An enlargement of $|G^{\theta}(ir)|$ around its first minimum is shown in the inset. }
\end{center}
\end{figure}

The Lee--Yang zeros (LYZ) method~\cite{Bhalerao:2003xf, Borghini:2004ke} for directly measuring the flow is based on
multiparticle correlations involving all particles in the event. It uses the asymptotic behavior of the cumulant
expansion to relate the location of the zeros of a complex function to the magnitude of the
integrated flow in the system.  For a detector with a uniform detection efficiency for all particles in the chosen  $\eta$
and \pt window, it is thus possible to obtain the integrated flow in a simple one-step procedure. Since this is not the case
for the CMS measurements presented here, we replace the term ``integrated flow'' that is used in the literature describing the
method~\cite{Bhalerao:2003xf, Borghini:2004ke} with ``reference flow''.
In the generating function (i.e., $G^{\theta}(ir)$, where $r$ is the imaginary axis coordinate), the flow vector,
constructed from all particles in
the event, is projected onto a direction that makes an arbitrary angle  $\theta$ with respect to the $x$ axis.
We use five different projection angles and then average the results over events from the same centrality class to reduce
the statistical uncertainties. Subsequently, the minimum of the generating function is found,
and the differential flow is determined with respect to the reference flow. There are different ways to
define the generating function that involve either a sum or a product of the individual particle contributions.
An example is given in Fig.~\ref{fig:LYZ_Gir_r},
showing how the minimum of the generating function is determined using both definitions.
The results presented here are based on the product generating function. In our
analysis of  $v_{2}(\pt)$, the \pt and $\eta$ ranges for the reference flow are $0.3<\pt <12$\GeVc
and $|\eta|<0.8$, respectively. Since the Lee--Yang zeros method is less sensitive to jet-induced charged-particle correlations than
the two-particle cumulant method, the \pt range of the tracks included in the determination of the reference flow is not
restricted to low \pt. In the analysis of $v_{2}(\eta)$, the pseudorapidity range for the particles included
in the reference flow is extended to $|\eta|<2.4$.

The Lee--Yang zeros method is sensitive to multiplicity fluctuations. To evaluate the effects of these fluctuations we perform
a toy-model MC study. Event ensembles are generated sampling the multiplicity for each event from Gaussian distributions
with mean and r.m.s. values comparable to the ones measured in the centrality bins used in the measurement.
For each particle, the corresponding \pt, $\eta$, and $\phi$  values are sampled from a realistic input $v_{2}(\pt,\eta)$ distribution.
These events are then analyzed using the same procedure as in the data and the resulting  $v_{2}(\pt,\eta)$ values are compared to the input.
We find that the  Lee--Yang zeros method tends to underestimate the results if the r.m.s. of the multiplicity distribution is more than
about 14\% of the mean. In order to keep the systematic uncertainties below 2\%,
the data are analyzed in 5\%-wide centrality classes and then averaged and weighted with the charged-particle yield, to obtain results in
wider centrality intervals needed for comparisons with other methods. Efficiency corrections are implemented as a
track-by-track weight in the differential generating function.

\subsection{Corrections to the anisotropy parameter \texorpdfstring{$v_{2}$}{v2}}
\label{sec:corrections}
In determining the $v_2(\pt)$ distributions for particles detected in the
tracker, it is necessary to correct for the influence of misidentified (i.e., ``fake'') tracks on
the measurements.   As shown in Section~\ref{sec:Tracking}, the fake-track contribution is
particularly significant at low-\pt values and for pseudorapidities
$|\eta|>1.6$.  Of particular concern is the observation that the fake tracks
can carry a $v_2$ signal at low \pt similar to that of properly reconstructed (i.e., ``real'') tracks at
a higher \pt. Since the true  $v_2$ signal is very small at low \pt, but increases at higher  \pt, the fake tracks
may contribute a significant fraction of the measured  $v_2$ signal at low \pt.

Studies using a full MC CMS simulation of PbPb
collisions based on the \textsc{hydjet} event generator~\cite{Lokhtin:2005px} indicate that the component of the $v_2$ signal
due to fake tracks is relatively constant for $\pt < 0.8\GeVc$, where the
fraction of fake tracks is largest. For higher \pt, where the fraction of fake tracks is
quite small, the value of $v_2$ is consistent with the measured value
from correctly reconstructed tracks. This suggests the following simple correction scheme.
Let $N_{\mathrm{det}}(\pt)$ be the number of reconstructed tracks in a given \pt bin,
$f$ the fraction of these tracks that are ``fake,''
$N_{\mathrm{true}}$ the number of ``true'' tracks in the \pt bin, and $e$  the
efficiency for reconstructing a true track in the bin.  Then
$N_{\mathrm{det}}-fN_{\mathrm{det}} = e N_{\mathrm{true}}$. The $fN_{\mathrm{det}}$ fake
tracks are characterized by a constant $v_2$ value given by $v_2^{\mathrm{fake}}$.
The $N_{\mathrm{det}}-fN_{\mathrm{det}}$ real tracks are characterized by  $v_2^{\mathrm{real}}$.   Then the observed value $v_{2}^{\mathrm{obs}}$ of $v_{2}$ will be

\begin{equation}
v_2^{\mathrm{obs}}=(1-f)v_2^{\mathrm{real}}+fv_2^{\mathrm{fake}}
\end{equation}
and so

\begin{equation}
v_2^{\mathrm{real}} = \frac{{v_2^{\mathrm{obs}} - fv_2^{\mathrm{fake}}}}{{1 - f}}.
\end{equation}
This correction for the fake-track signal is only significant for
\pt  values less than $\approx$1\GeVc.  In this range,
an empirical correction that results in values of $v_2$ that are independent of the track
selection requirements or fraction of fake tracks is applied using
$v_2^{\mathrm{fake}} = 1.3\left\langle {{v_2}} \right\rangle$, where the
yield-weighted average is performed over the transverse momentum range  0.3 to 3 \GeVc, folding in the efficiency-corrected spectra.
This value for $v_2^{\mathrm{fake}}$ is also supported by MC studies using \textsc{hydjet}.

\subsection{Systematic uncertainties}
\label{sec:Systematic Studies}
\subsubsection{Systematic uncertainties in the measurements of  \texorpdfstring{$v_{2}$}{v2}}

The systematic uncertainties in the measurements of $v_{2}$ include those common to all methods, as well as method-specific uncertainties.
They are evaluated as relative uncertainties and are reported as percentages relative to the measured $v_2$ values.
Since we are reporting the results on  $v_{2}$ for nonidentified charged particles, it is important to investigate the
tracking efficiency as a function of particle species. The tracking efficiencies for charged pions, kaons, protons, and
antiprotons are determined using a full simulation of CMS. Subsequently, the value of $v_2(\pt)$ for charged particles is obtained using
different assumptions for the  \pt-dependence of $v_2$  and the transverse momentum spectra of each particle type, taking into
account the corresponding reconstruction efficiencies. The results are compared to those obtained with the assumption of a
particle-species-independent efficiency. The uncertainties in the charged particle  $v_{2}$ results are estimated to be
$\lesssim 0.5\%$, independent of the \pt, $\eta$, and centrality ranges. This uncertainty is listed as ``Part. composition''
in  Tables~\ref{tab:sys_v2pt_eta08_RP}--\ref{tab:sys_v2eta_pt0330}.

Since the  $v_{2}$ value changes with centrality, an uncertainty in the
centrality determination can lead to a shift in the   $v_{2}$ measurement.
This uncertainty is evaluated by varying the value of the minimum-bias trigger efficiency to include ($97\pm3$)\% of the
total inelastic cross section. The resulting uncertainty on  $v_{2}$ is of the order 1\%, independent of the  \pt, $\eta$, and
centrality ranges. This uncertainty is listed as ``Cent. determination'' in
Tables~\ref{tab:sys_v2pt_eta08_RP}--\ref{tab:sys_v2eta_pt0330}.

The kinematic requirements used to select tracks can affect the efficiency of track finding and the relative fraction
of fake tracks in an event. The requirements are varied from their default values to estimate the systematic uncertainty.
For each set of requirements, corresponding corrections are obtained for the fake-track contribution to the
  $v_2$ signal, as described in Section~\ref{sec:corrections}. For a given centrality and \pt range, the systematic
uncertainty is estimated based on the stability of the $v_2$ value after corrections for fake tracks,
independent of the track selection requirements.
The uncertainty
is found to be directly related to the magnitude of the fake-track contribution, with the final results deemed unreliable
when the fake-rate is higher than $\approx$20\%. For the results presented in Section~\ref{sec:Results}, the systematic
uncertainties from this source remain below $4\%$ over the entire range of \pt, $\eta$, and centrality, and are significantly
below this value for  $\pt> 0.5$\GeVc, centrality above $10\%$, and  $|\eta|<1.6$.
The uncertainty in the efficiency corrections is evaluated by determining
the efficiency based on the \textsc{hydjet} model, and by embedding simulated pions into PbPb events in data. Although the two
resulting efficiencies do have differences, the uncertainty on the $v_{2}$ value is small, at most 0.5\%.
Variations in the $v_{2}$ results due to changing detector conditions throughout the data-taking period are studied
by dividing the data into three subgroups and are found to be below 1\% for
all measurements. The combined uncertainties from the efficiency corrections, fake-track corrections, and
variations in detector conditions
are listed under ``Corrections'' in  Tables~\ref{tab:sys_v2pt_eta08_RP}--\ref{tab:sys_v2eta_pt0330}.

Additional studies of the systematic uncertainty are conducted for each method.
In the event-plane method, flattening corrections are obtained using different procedures.
The vertex dependence of the flattening parameters is examined and different subevent $\eta$
gaps are used in obtaining the resolution corrections. The uncertainties from these sources are found to be negligible.
The resolution corrections are measured with the three-subevent and two-subevent methods and are found to be
consistent within the statistical uncertainties. The statistical uncertainties for the resolution correction factor are
less than 1\%, except for the  most peripheral 70--80\% centrality events,
where the statistical uncertainty reaches a value of 2\%. We include the statistical uncertainties associated
with the resolution-correction factors as part of the overall systematic
uncertainty on tracking efficiency and fake-track corrections  for the event-plane method.

In the cumulant method, we examine the numerical stability of the result when the radius parameter  $r_{0}$ used in
the interpolations of the generating function is increased or decreased by half of its central value. The effects
of multiplicity fluctuations are studied for the Lee--Yang zeros and the cumulant methods by analyzing the events in finer 2.5\%
centrality bins, and by using a fixed number of particles chosen at random from each event in a given centrality class.

The systematic uncertainties are smallest for the mid-rapidity region  $|\eta| < 0.8$, $\pt > 0.5$ \GeVc, and in the mid-central
events (10--40\%), and range from 2.0 to 4.5\% for the different methods. At low  \pt for the most central events, and in the forward
pseudorapidity region where the fake-track contributions are larger, the uncertainties increase to 2.4--6.8\%. Similarly, in
the most peripheral events the uncertainties increase mostly due to multiplicity fluctuations and reach up to 3.2--7\%,
depending on the experimental method. A summary of the systematic uncertainties is presented in
Tables~\ref{tab:sys_v2pt_eta08_RP}--\ref{tab:sys_v2eta_pt0330}.

\begin{table}[htb]
\begin{center}
\caption{Systematic uncertainties in the measurement of $v_{2}(\pt)$ for $|\eta| < 0.8$ with the event-plane method
for different \pt  and  centrality ranges.}
\label{tab:sys_v2pt_eta08_RP}
\begin{tabular}{|c|c|c|c|c|}
\hline
 Source & $\pt$ & \multicolumn{3}{c|}{ Centrality } \\ \cline{3-5}
                     &(\GeVc)       & 0 -- 10\% &  10 -- 70\%& 70 -- 80\% \\
\hline
Part. composition & All                   & 0.5\%   & 0.5\%   & 0.5\%   \\
\hline
Cent. determination      & All                   & 1.0\%   & 1.0\%   & 1.0\%     \\
\hline
                    & $<0.3$              &4.0\%    & 2.0\%   & 3.0\%    \\
\cline{2-5}
Corrections         & $0.3$ -- $0.5$           & 2.0\%   & $<1.0\%$& 2.0\%    \\
\cline{2-5}
                     & 0.5 -- 22.0              & $<1.0\%$& $<1.0\%$& $2.0\%$  \\
\hline
Total                &  $<0.3$             & 4.2\%     & 2.3\%     & 3.2\%    \\
\cline{2-5}
                     & 0.3 -- 0.5          & 2.3\%     & 1.5\%     & 2.3\%    \\
\cline{2-5}
                     & 0.5 -- 22.0         & 1.5\%     & 1.5\%     & 2.3\%      \\
\hline
\end{tabular}
\end{center}
\end{table}

\begin{table}[htb]
\begin{center}
\caption{Systematic uncertainties in the measurement of  $v_{2}(\eta)$ for $0.3<\pt<3$\GeVc with the event-plane method for
different $\eta$ and centrality ranges.}
\label{tab:sys_v2eta_pt0330_RP}
\begin{tabular}{|c|c|c|c|c|}
\hline
 Source & $|\eta|$ & \multicolumn{3}{c|}{ Centrality } \\
\cline{3-5}
                     &       & 0 -- 10\% &  10 -- 70\%& 70 -- 80\% \\
\hline
Part. composition     & All                   & 0.5\%      & 0.5\%      & 0.5\%     \\
\hline
Cent. determination          & All                   & 1.0\%      & 1.0\%      & 1.0\%     \\
\hline
Corrections              & 0.0 -- 1.6           & $2.0\%$    & $<1.0\%$   & $2.0\%$    \\
\cline{2-5}
                         & 1.6 -- 2.4              & 4.0\%      & $2.0\%$    & $3.0\%$\\
\hline
 Total              & 0.0 -- 1.6           &3.2\%       & 1.5\%      & $2.3\%$ \\
\cline{2-5}
                         & 1.6 -- 2.4              & 4.2\%      & $ 2.3\%$   & $ 3.2\%$\\

\hline

\end{tabular}
\end{center}
\end{table}

\begin{table*}[htb]
\begin{center}
\caption{Systematic uncertainties in the measurement of $v_{2}(\pt)$ for $|\eta| < 0.8$
with the two-particle cumulant method for different \pt and
centrality ranges.}
\label{tab:sys_v22pt_eta08}
\begin{tabular}{|c|c|c|c|c|c|}
\hline
 Source & $\pt$ & \multicolumn{4}{c|}{ Centrality } \\
\cline{3-6}
                     & (\GeVc)      & 0 -- 5\% & 5 -- 10\% &  10 -- 70\%& 70 -- 80\% \\
\hline
Part. composition & All  & 0.5\%  & 0.5\%  & 0.5\%   & 0.5\%    \\
\hline
Cent. determination      & All  & 1.0\%  & 1.0\%  &  1.0\%  &  1.0\%     \\
\hline
Multiplicity fluct.  & All  & 0.5\%  & 1.5\%  & 4.0\%   & 4.0\%    \\
\hline
$r_{0}$ parameter    & All  & 0.2\%  & 0.2\%   & 0.2\%   & 0.1\%   \\
\hline
Corrections         & 0.3 -- 0.5   & 3.8\%  & 1.7\% & 0.6\%   & 4.0\%    \\
\cline{2-6}
                    & 0.5 -- 22.0  & 2.9\%  & 2.1\% & 0.6\%   & 3.0\%    \\
\hline

 Total   & 0.3 -- 0.5  & 4.0\%  & 2.6\%  & 4.2\%    & 5.8\%   \\
\cline{2-6}
         & 0.5 -- 22.0 & 3.2\%  & 2.9\%  & 4.2\%    & 5.1\%   \\
\hline
\end{tabular}
\end{center}
\end{table*}

\begin{table}[htb]
\begin{center}
\caption{Systematic uncertainties in the measurement of $v_{2}(\eta)$ for the range
\mbox{$0.3<\pt<3$}\GeVc with the two-particle cumulant method
for different $\eta$
and centrality ranges.}
\label{tab:sys_v22eta}
\begin{tabular}{|c|c|c|c|}
\hline
 Source & $|\eta|$ & \multicolumn{2}{c|}{ Centrality } \\
\cline{3-4}
                     &       & 5 -- 10\% &  10 -- 70\%  \\
\hline
Part. composition & All     & 0.5\%  & 0.5\%          \\
\hline
Cent. determination      & All     & 1.0\%  &  1.0\%    \\
\hline
Multiplicity fluct.  & All     & 1.5\%  & 4.0\%     \\
\hline
$r_{0}$ parameter    & All     & 0.2\%  & 0.2\%    \\
\hline

Corrections   & 0.0 -- 1.6    & 0.8\%  & 1\%    \\
\cline{2-4}
             & 1.6 -- 2.4     & 1.5\%  & 1.6\%     \\
\hline

 Total  & 0.0 -- 1.6     & 2.0\% & 4.2\%     \\
\cline{2-4}
 & 1.6 -- 2.4                 & 2.4\% & 4.4\%    \\
\hline

\end{tabular}
\end{center}
\end{table}

\begin{table*}[htb]
\begin{center}
\caption{Systematic uncertainties in the measurement of $v_{2}(\pt)$ for $|\eta| < 0.8$
with the four-particle cumulant method for different \pt  and  centrality ranges.}
\label{tab:sys_v24pt_eta08}
\begin{tabular}{|c|c|c|c|c|c|}
\hline
 Source & $\pt$ & \multicolumn{4}{c|}{ Centrality } \\
\cline{3-6}
                     & (\GeVc)      & 5 -- 10\% &  10 -- 40\%& 40 -- 60\% & 60 -- 70\% \\
\hline
Part. composition & All   & 0.5\%  & 0.5\% & 0.5\%  & 0.5\%     \\
\hline
Cent. determination      & All   & 1.0\%  &  1.0\%  &  1.0\%  &  1.0\%     \\
\hline
Multiplicity fluct.  & All   & 5.0\% & 3.0\%  & 5.0\% & 5.0\%       \\
\hline
$r_{0}$ parameter     & All   & 2.0\% & 3.0\%  & 1.0\%  & 0.1\%     \\
\hline
Corrections & 0.3 -- 0.5          & 4.1\% & 1.4\%  & 3.0\% & 4.5\%     \\
\cline{2-6}
           & 0.5 -- 22.0           & 2.2\% & 1.2\% & 1.1\%  & 3.0\%    \\
\hline

 Total   & 0.3 -- 0.5  & 6.8\%  & 4.6\% & 6.0\%  & 6.8\%    \\
\cline{2-6}
              & 0.5 -- 22.0 & 5.9\%  & 4.5\% & 5.3\%  & 6.0\%   \\
\hline
\end{tabular}
\end{center}
\end{table*}

\begin{table}[htb]
\begin{center}
\caption{Systematic uncertainties in the measurement of $v_{2}(\eta)$ for the range
\mbox{$0.3<\pt<3$}\GeVc with the four-particle cumulant method for different $\eta$
and centrality ranges.}
\label{tab:sys_v24eta}
\begin{tabular}{|c|c|c|c|c|}
\hline
 Source & $|\eta|$ & \multicolumn{3}{c|}{ Centrality } \\
\cline{3-5}
                     &       & 5 -- 10\% &  10 -- 40\%& 40 -- 70\%  \\
\hline
Part. composition & All    & 0.5\%  & 0.5\%    & 0.5\%            \\
\hline
Cent. determination & All         & 1.0\%  &  1.0\%   & 1.0\%   \\
\hline
Multiplicity fluct. & All     & 5.0\%  & 3.0\%    & 5.0\%       \\
\hline
$r_{0}$ parameter   & All     & 2.0\%  & 3.0\%     & 1.0\%     \\
\hline

Corrections & 0.0 -- 1.6    & 1.5\%  & 1.5\%     & 1.5\%     \\
\cline{2-5}
 & 1.6 -- 2.4               & 1.8\%  & 2.1\%     & 1.9\%     \\
\hline

 Total  & 0.0 -- 1.6    & 5.8\%  & 4.8\%    & 5.4\%   \\
\cline{2-5}
 & 1.6 -- 2.4              & 5.8\% & 5.0\%     & 5.6\%   \\
\hline

\end{tabular}
\end{center}
\end{table}

\begin{table}[htb]
\begin{center}
\caption{Systematic uncertainties in the measurement of $v_{2}(\pt)$ for $|\eta| < 0.8$ with the Lee--Yang zeros method for different \pt and centrality ranges.}
\label{tab:sys_v2pt_eta08}
\begin{tabular}{|c|c|c|c|c|}
\hline
 Source & $\pt$ & \multicolumn{3}{c|}{ Centrality } \\
\cline{3-5}
                     & (\GeVc)      & 5 -- 10\% &  10 -- 40\%& 40 -- 50\% \\
\hline
Part. composition & All  & 0.5\% & 0.5\%  & 0.5\% \\
\hline
Cent. determination      & All  & 1.0\%  & 1.0\%  &  1.0\%  \\
\hline
Multiplicity fluct.  & All & 0.1\%  & 0.9\% & 1.9\% \\
\hline

Corrections  & 0.3 -- 0.5 & 2.5\%  & 1.7\% & 0.7\%  \\
\cline{2-5}
             & 0.5 -- 22.0 & 1.5\%  & 1.0\% & 0.6\%  \\
\hline

 Total & 0.3 -- 0.5 & 2.7\%  & 2.2\% & 2.3\%  \\
\cline{2-5}
           & 0.5 -- 22.0 & 1.9\%  & 1.8\% & 2.3\%  \\
\hline

\end{tabular}
\end{center}
\end{table}

\begin{table}[htb]
\begin{center}
\caption{Systematic uncertainties in the measurement of $v_{2}(\eta)$ for \mbox{$0.3<\pt<3$}\GeVc
with the Lee--Yang zeros method for different $\eta$
and centrality ranges.}
\label{tab:sys_v2eta_pt0330}
\begin{tabular}{|c|c|c|c|c|}
\hline
 Source & $|\eta|$ & \multicolumn{3}{c|}{ Centrality } \\
\cline{3-5}
                     &       & 5 -- 10\% &  10 -- 40\%& 40 -- 50\% \\
\hline
Part. composition  & All            & 0.5\%  & 0.5\%  & 0.5\% \\
\hline
Cent. determination       & All            & 1.0\%  & 1.0\%  &  1.0\%  \\
\hline
Multiplicity fluct.   & All            & 0.1\%  & 0.9\% & 1.9\% \\
\hline

Corrections           & 0.0 -- 1.6    & 1.3\%  & 1.0\% & 0.8\%  \\
\cline{2-5}
                      & 1.6 -- 2.4    & 1.5\%  & 1.4\% & 1.3\%  \\
\hline

 Total           & 0.0 -- 1.6    & 1.7\%  & 1.8\% & 2.4\%  \\
\cline{2-5}
                     & 1.6 -- 2.4     & 1.9\%  & 2.0\% & 2.5\%  \\
\hline

\end{tabular}
\end{center}
\end{table}

\subsubsection{Systematic uncertainties in the measurements of the transverse momentum spectra and the mean
transverse momentum \texorpdfstring{$\langle\pt\rangle$}{<pt>} }

Several sources of systematic uncertainty are considered in obtaining the inclusive charged-particle transverse momentum distributions
and their mean values, $\langle\pt\rangle$. These include the uncertainties in the efficiency and fake-track correction factors, the
particle-species-dependent efficiency, and the uncertainty in the minimum-bias trigger efficiency. The effect on the overall
normalization of the spectra is considered separately from the smaller effect on the shape of the spectra as a function of \pt.
To obtain the mean transverse momentum, different functional forms are used to extrapolate the spectra down to $\pt = 0$, and
the \pt range over which the spectra are fitted is varied. The combined point-to-point systematic uncertainties on the spectra
are presented as a function of pseudorapidity in Table~\ref{spectra_pointerror}. The total normalization uncertainty of the
charged-particle spectra measured in each centrality interval is given in Table~\ref{spectra_normerror}. The systematic uncertainties
in the measurement of  $\langle\pt\rangle$ in different pseudorapidity intervals are summarized in Table~\ref{meanpt_pointerror}.

\begin{table}[htb]
\begin{center}
\caption{Point-to-point systematic uncertainties in the measurement of the charged-particle spectra
in different pseudorapidity intervals.}
\label{spectra_pointerror}
\begin{tabular}{|c|c|c|c|}
\hline
Source of Uncertainty & $|\eta|<0.8$ & $0.8<|\eta|<1.2$ & $2.0<|\eta|<2.4$  \\
\hline
Tracking Efficiency &  5\% & 8\% & 13\% \\
\hline
Particle Composition & 1\% & 1\% & 2\% \\
\hline
Trigger Efficiency & 3\% & 3\% & 3\% \\
\hline
Total & 6\% & 9\% & 14\% \\
\hline
\end{tabular}
\end{center}
\end{table}

\begin{table}[htb]
\begin{center}
\caption{Normalization uncertainty in the measurement of the charged-particle spectra
in different centrality intervals resulting from the uncertainty in the minimum-bias trigger
efficiency.
\label{spectra_normerror}}
\begin{tabular}{|c|c|}
\hline
Centrality Range & Normalization Uncertainty \\
\hline
0-5\% & 0.4\% \\
\hline
5-10\% & 1.0\% \\
\hline
10-15\% & 1.7\% \\
\hline
15-20\% & 2.3\% \\
\hline
20-25\% & 3.1\% \\
\hline
25-30\% & 4.1\% \\
\hline
30-35\% & 5.0\% \\
\hline
35-40\% & 6.1\% \\
\hline
40-50\% & 8.0\% \\
\hline
50-60\% & 12\% \\
\hline
60-70\% & 16\% \\
\hline
70-80\% & 21\% \\
\hline
\end{tabular}
\end{center}
\end{table}

\begin{table}[htb]
\begin{center}
\caption{Systematic uncertainty of the mean \pt of charged particles
from each source and in total as a function of pseudorapidity.
\label{meanpt_pointerror}}
\begin{tabular}{|c|c|c|c|}
\hline
Source of Uncertainty & $|\eta|<0.4$ & $0.8<|\eta|<1.2$ & $2.0<|\eta|<2.4$  \\
\hline
Fit function & 3\% & 3\% & 4\% \\
\hline
Trigger Efficiency & 1.5\% & 1.5\% & 1.5\% \\
\hline
Tracking Efficiency & 2\% & 2\% & 2.5\% \\
\hline
Total & 3.9\% & 3.9\% & 4.9\% \\
\hline
\end{tabular}
\end{center}
\end{table}

\section{Results}
\label{sec:Results}
The main results of the analysis using the four methods described above are as follows:

\begin{itemize}
\item{ $v_{2}(\pt)$ at mid-rapidity  $|\eta|<0.8$. }
\item{Integrated  $v_{2}$ at mid-rapidity  $|\eta|<0.8$ and $0.3 < \pt< 3$\GeVc.}
\item{ $v_{2}(\eta)$ for  $0.3 < \pt< 3$\GeVc.}
\end{itemize}
We also measure the charged-particle transverse momentum spectra and their mean \pt for the centrality and
pseudorapidity ranges in which the flow is studied.

The flow studies are performed in the 12 centrality classes listed in  Table~\ref{tbl:glauber}.
Using these results, we examine the  scaling of  the integrated $v_2$ with
the participant eccentricity, as well as perform comparisons to measurements from other experiments. Centrality classes are
regrouped to perform these comparisons, i.e., the results of $v_2(\pt)$, $v_2(\eta)$, or integrated $v_2$ obtained
in the finer bins of centrality are averaged over wider bins, weighted using the corresponding $\rd^{2}N/\rd\pt\, \rd{\eta}$ spectra.
The evolution of the measured elliptic anisotropy
as a function of centrality, center-of-mass energy, and transverse particle density  is studied. The scaling of
$v_2(\eta)$ in the longitudinal dimension is also examined through comparisons to RHIC data.

\subsection{Transverse momentum dependence of \texorpdfstring{$v_{2}$}{v2}}

\begin{figure}[htb]
\begin{center}
\includegraphics[width=\cmsFigWidth]{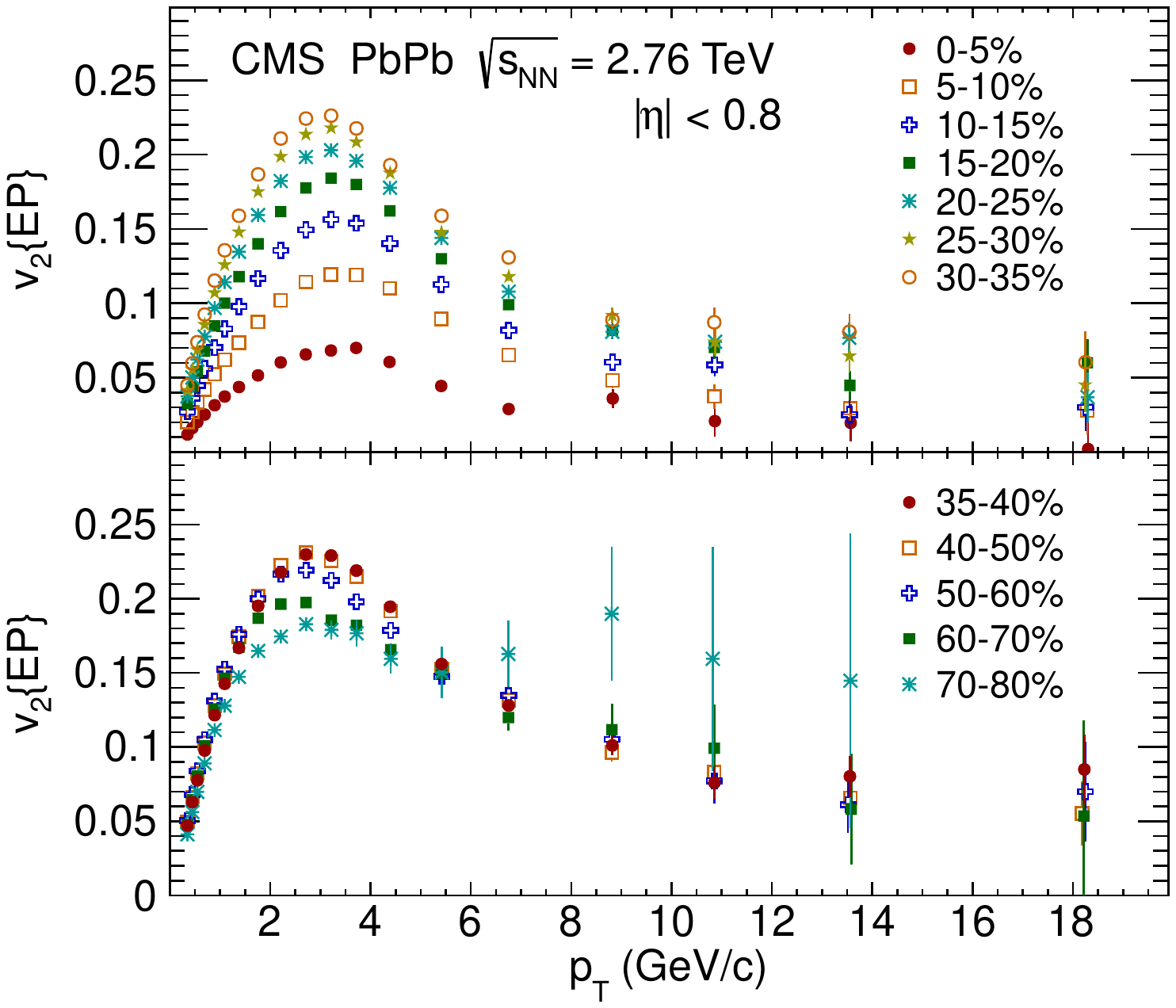}
\caption{\label{fig:v2pt_EP_f}(Color online)
Results from the event-plane (EP) method for $v_{2}$ as a function of \pt at mid-rapidity $|\eta|<0.8$
for the 12 centrality classes given in the legend. The error bars show the statistical uncertainties only.}
\end{center}
\end{figure}

\begin{figure}[htb]
\begin{center}
\includegraphics[width=\cmsFigWidth]{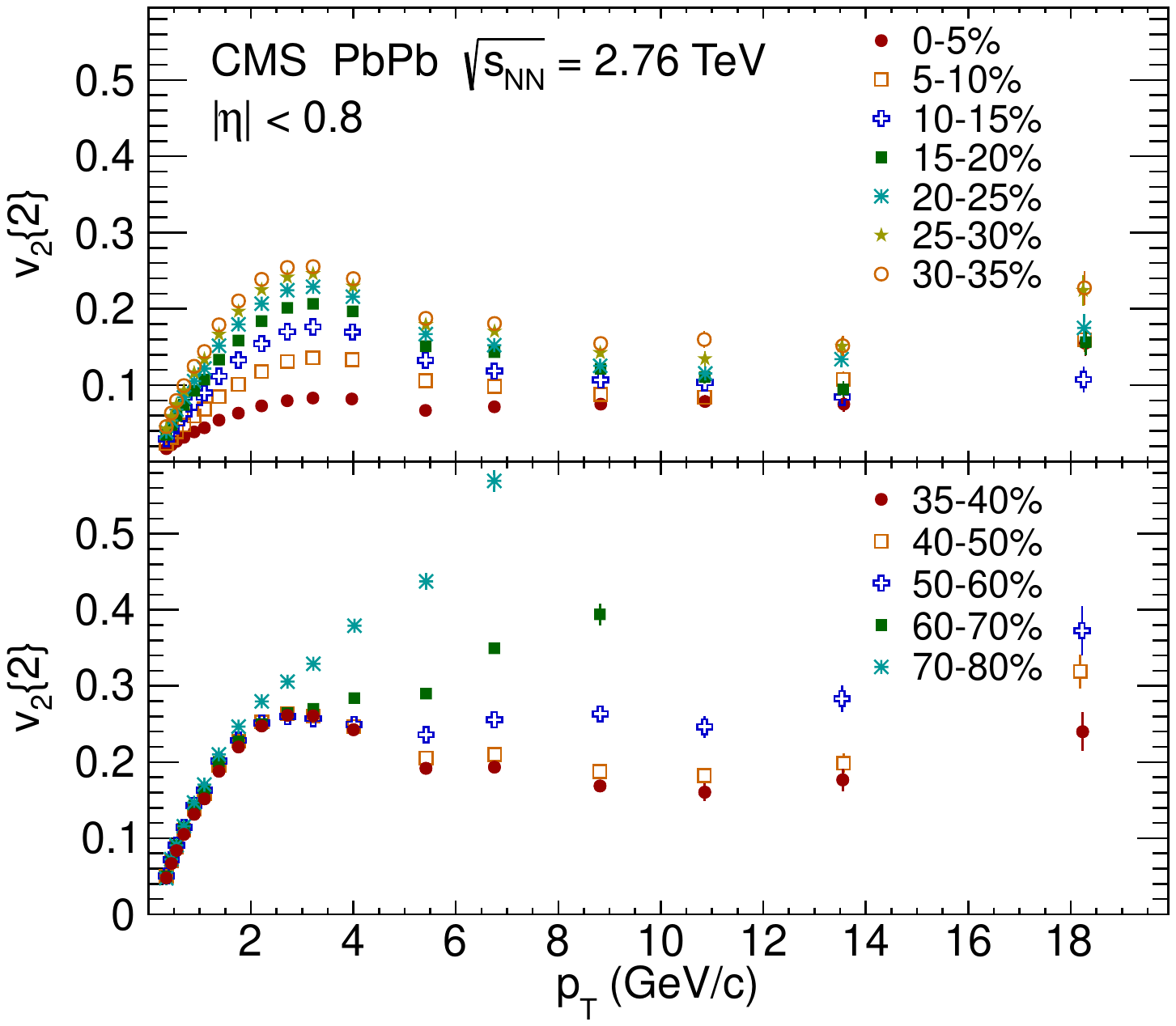}
\caption{\label{fig:v2pt_C2_f}(Color online)
Second-order cumulant results for   $v_{2}$ as a function of \pt at mid-rapidity $|\eta|<0.8$
for the 12 centrality classes given in the legend. The error bars show the statistical uncertainties only.}
\end{center}
\end{figure}

\begin{figure}[htpb]
\begin{center}
\includegraphics[width=\cmsFigWidth]{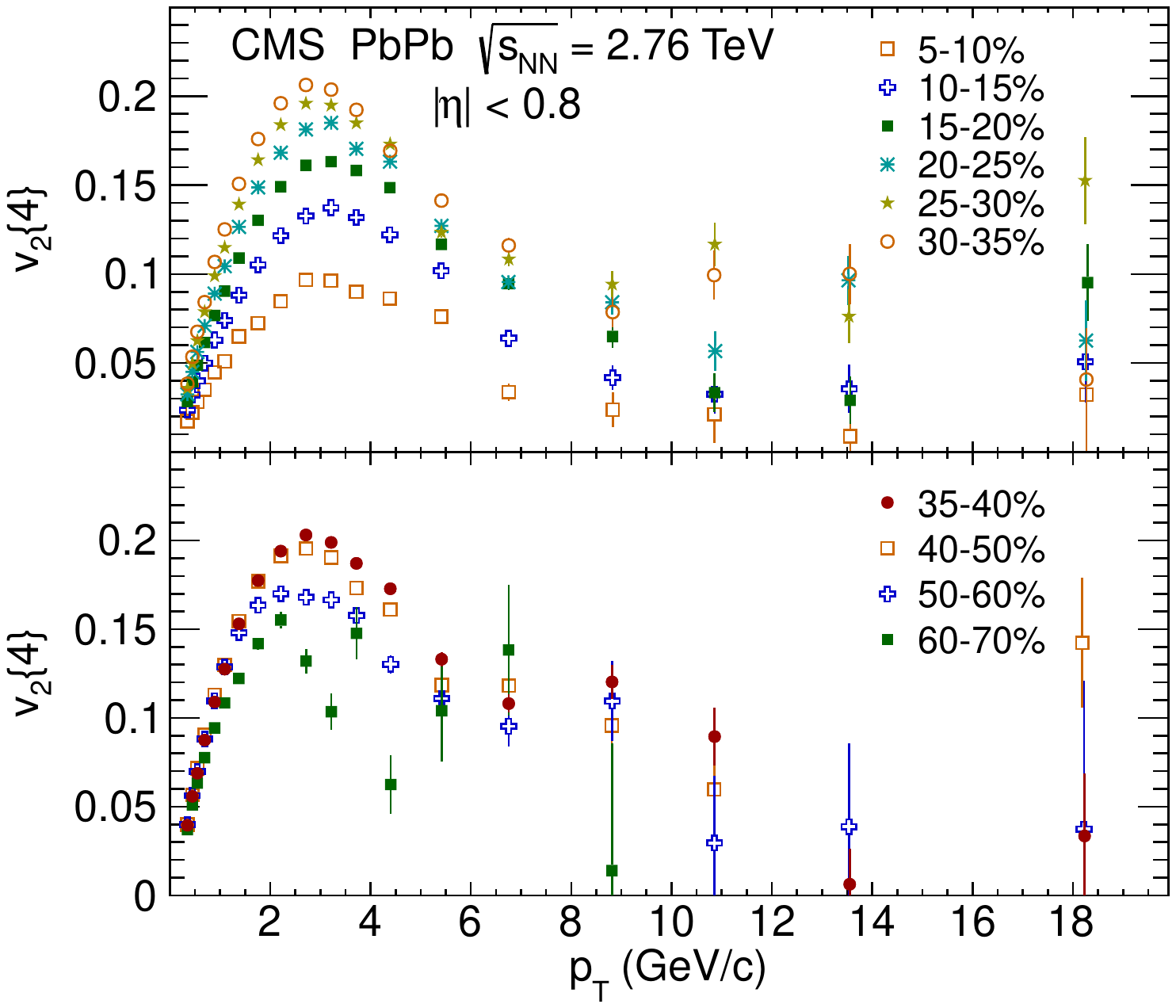}
\caption{\label{fig:v2pt_C4_f}(Color online)
Fourth-order cumulant results for  $v_{2}$ as a function of \pt at mid-rapidity $|\eta|<0.8$
for the 10 centrality classes given in the legend. The error bars show the statistical uncertainties only.}
\end{center}
\end{figure}

\begin{figure}[htpb]
\begin{center}
\includegraphics[width=\cmsFigWidth]{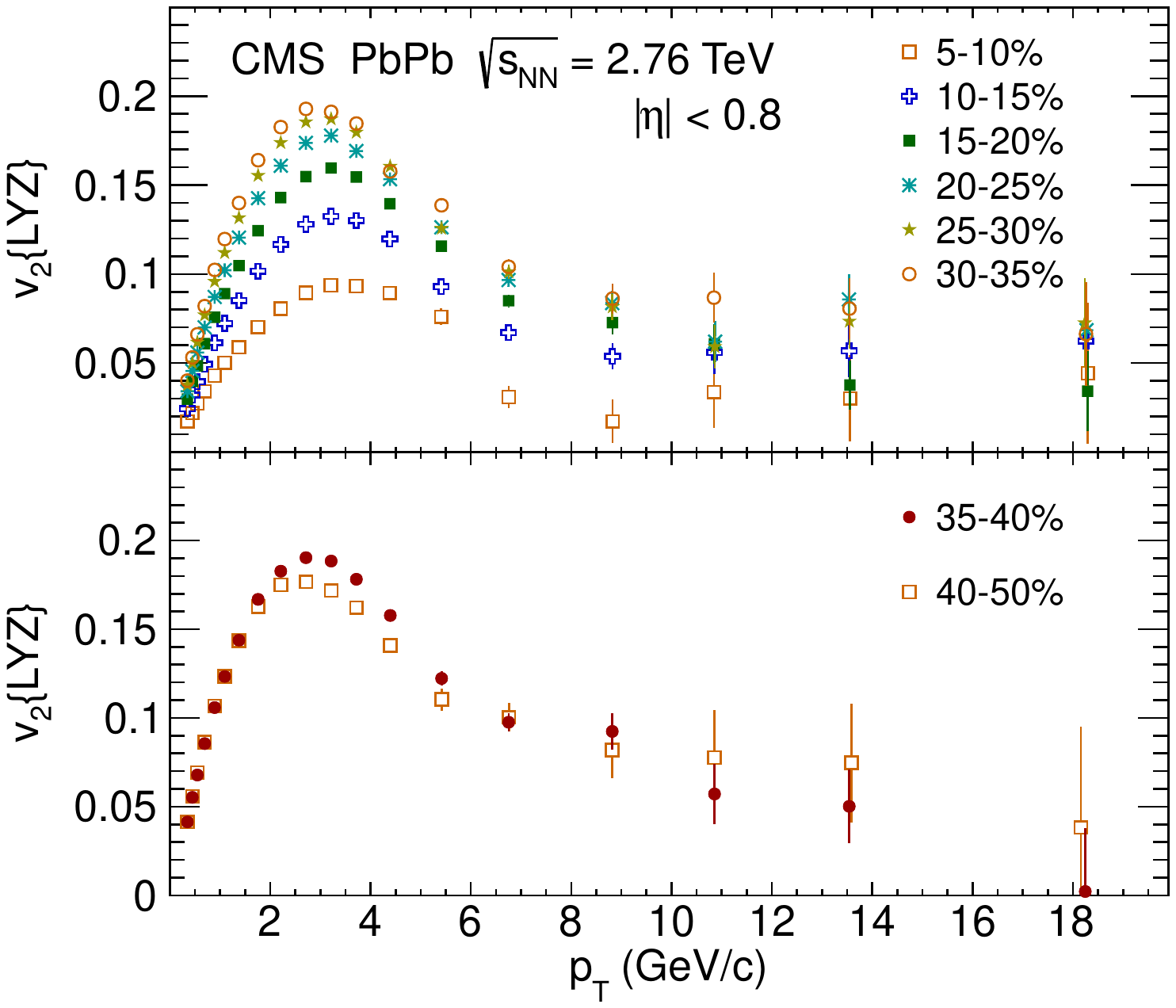}
\caption{\label{fig:v2pt_LYZ_f}(Color online)
Lee--Yang zeros results for   $v_{2}$ as a function of \pt at mid-rapidity $|\eta|<0.8$
for the 8 centrality classes given in the legend. The error bars show the statistical uncertainties only.}
\end{center}
\end{figure}

In Figs.~\ref{fig:v2pt_EP_f}--\ref{fig:v2pt_LYZ_f}, we
present the measurement of $v_{2}$ for charged particles as a function of transverse momentum at
mid-rapidity, obtained by each analysis method.
We use the notation $v_{2}\{\mathrm{EP}\}$ to refer to the measurement of $v_{2}$ using the event-plane
method, and $v_{2}\{2\}$, $v_{2}\{4\}$, $v_{2}\{\mathrm{LYZ}\}$ to refer to those using the two-particle
cumulant, four-particle cumulant, and Lee--Yang zeros methods, respectively.
Several trends can be
observed and related to the physics processes dominating hadron
production in different \pt ranges.  The value of $v_{2}$ increases
from central to peripheral collisions up to 40\% centrality, as expected if the anisotropy is
driven by the spatial anisotropy in the initial state~\cite{Kolb:2003dz,Huovinen:2006jp,Teaney:review}.
The transverse momentum dependence shows a rise of $v_2$ up to $\pt \approx$3\GeVc and
then a decrease.
As a function of centrality, a tendency for the peak position of the $v_2(\pt)$ distribution to move to higher
\pt in more central collisions is observed, with the exception of the results from the most peripheral collisions in
the two-particle cumulant and the event-plane methods.

In ideal hydrodynamics the azimuthal anisotropy continuously increases with increasing
\pt~\cite{Kolb:2003dz,Huovinen:2006jp}. The deviation of the theory from the RHIC data
at $\pt \gtrsim 2$--$3 \GeVc$ has been attributed to incomplete thermalization of the high-\pt
hadrons, and the effects of viscosity. Indeed, viscous hydrodynamic
calculations~\cite{Luzum:2008cw,Romatschke:2009im,Teaney:review} show that the shear viscosity has
the effect of reducing the anisotropy at high \pt. At $\pt \gtrsim 8\GeVc$, where hadron production
is dominated by jet fragmentation, the collective-flow effects are
expected to disappear~\cite{Huovinen:2006jp,Teaney:review}. Instead, an asymmetry in the azimuthal
distribution of hadron emission with respect to the reaction plane
could be generated by path-length-dependent parton energy loss~\cite{gyulassy-2001-86,Bass:2008rv,Adare:2010sp}.
For events with similar charged-particle suppression~\cite{CMS:2012aa}, but different reaction-zone
eccentricity, one might expect that the geometric
information would be imprinted in the elliptic anisotropy signal. The upper panels
of Figs.~\ref{fig:v2pt_EP_f}--\ref{fig:v2pt_LYZ_f} (centrality
0--35\%) show a trend that is consistent with this expectation. In
more central events, where the eccentricity is smaller, the elliptic anisotropy
value is systematically lower. In more peripheral collisions
(centrality 35--80\%), there is a complex interplay between the reduced energy
loss and the increase in eccentricity that influence the $v_2(\pt)$
value in opposite directions. The data presented here provide the
basis for future detailed comparisons to theoretical models.

\begin{figure*}[htb]
\begin{center}
\includegraphics[width=\textwidth]{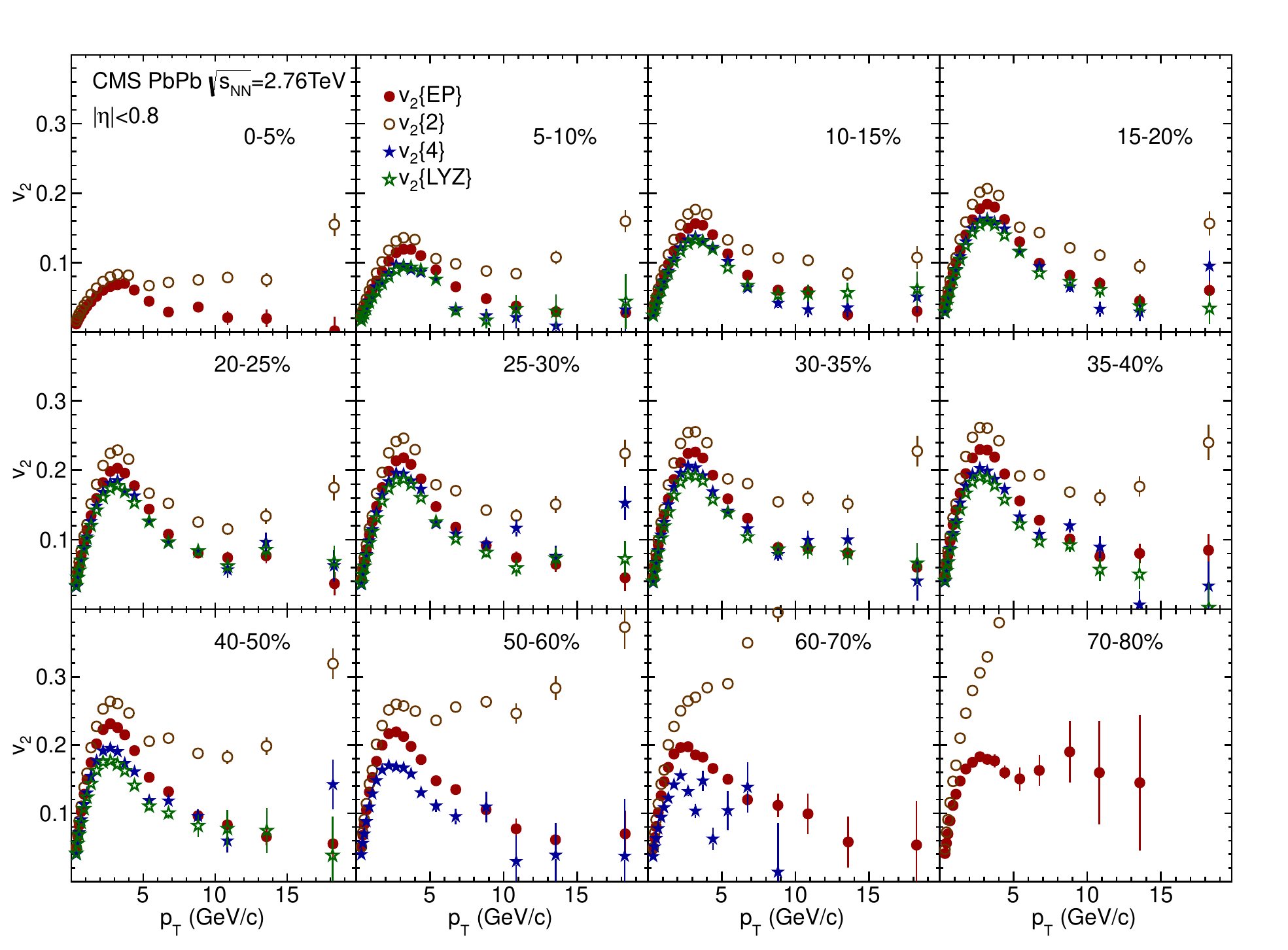}
\caption{\label{fig:v2pt_12cen}(Color online)
Comparison of the four different methods for determining $v_{2}$ as a function of \pt at mid-rapidity ($|\eta|<0.8$)
for the 12 centrality classes given in the figures. The error bars show the statistical uncertainties only.}
\end{center}
\end{figure*}

An important consideration in interpreting the $v_{2}(\pt)$ results
is the contribution from nonflow correlations and initial-state
eccentricity fluctuations.  To aid in assessing the magnitude of these
effects and their evolution with the centrality of the collisions, the results
of $v_{2}(\pt)$ obtained by all methods at mid-rapidity are compared in Fig.~\ref{fig:v2pt_12cen}
for 12 centrality classes. The four methods show differences as expected due to
their sensitivities to nonflow contributions~\cite{Borghini:2001vi,Bhalerao:2003xf,Ollitrault:2009ie}
and eccentricity fluctuations~\cite{Bhalerao:2006tp,PHOBOSeccPART,Qiu:2011iv}.

The method that is most affected by nonflow correlations is the two-particle cumulant, because of the
fact that the reference and the differential flow signals are determined in the
same pseudorapidity range.  The event-plane method is expected to be
similarly affected if dedicated selections are not applied to reduce these
contributions. In our analysis, the particles used in the event-plane determination
and the particles used to measure the flow are at least 3 units of pseudorapidity apart,
which suppresses most nonflow correlations. The differences between the two-particle cumulant
and the event-plane methods are most pronounced at high \pt and in peripheral collisions,
where jet-induced correlations dominate over the collective flow.

In a collision where $M$ particles are produced, direct $k$-particle
correlations are typically of order $1/M^{k-1}$, so that they become smaller as $k$ increases.
Therefore, the fourth-order cumulant and the Lee--Yang zeros methods are expected to be much less affected
by nonflow contributions than the second-order cumulant method~\cite{Borghini:2001vi,Bhalerao:2003xf,Ollitrault:2009ie}.
This trend is seen in our data.

\subsection{Centrality dependence of integrated \texorpdfstring{$v_{2}$}{v2} and eccentricity scaling}
\label{s:integrated v2}
To obtain the integrated
$v_{2}$ values as a function of centrality at mid-rapidity,  $v_2(\pt)$ measurements are averaged over \pt, weighted by
the corresponding charged-particle spectrum. The integration range $0.3 < \pt < 3 $\GeVc is limited to low \pt to
maximize the contribution from soft processes, which facilitates comparisons to hydrodynamic calculations.

\begin{figure}[htb]
\begin{center}
\includegraphics[width=0.5\textwidth]{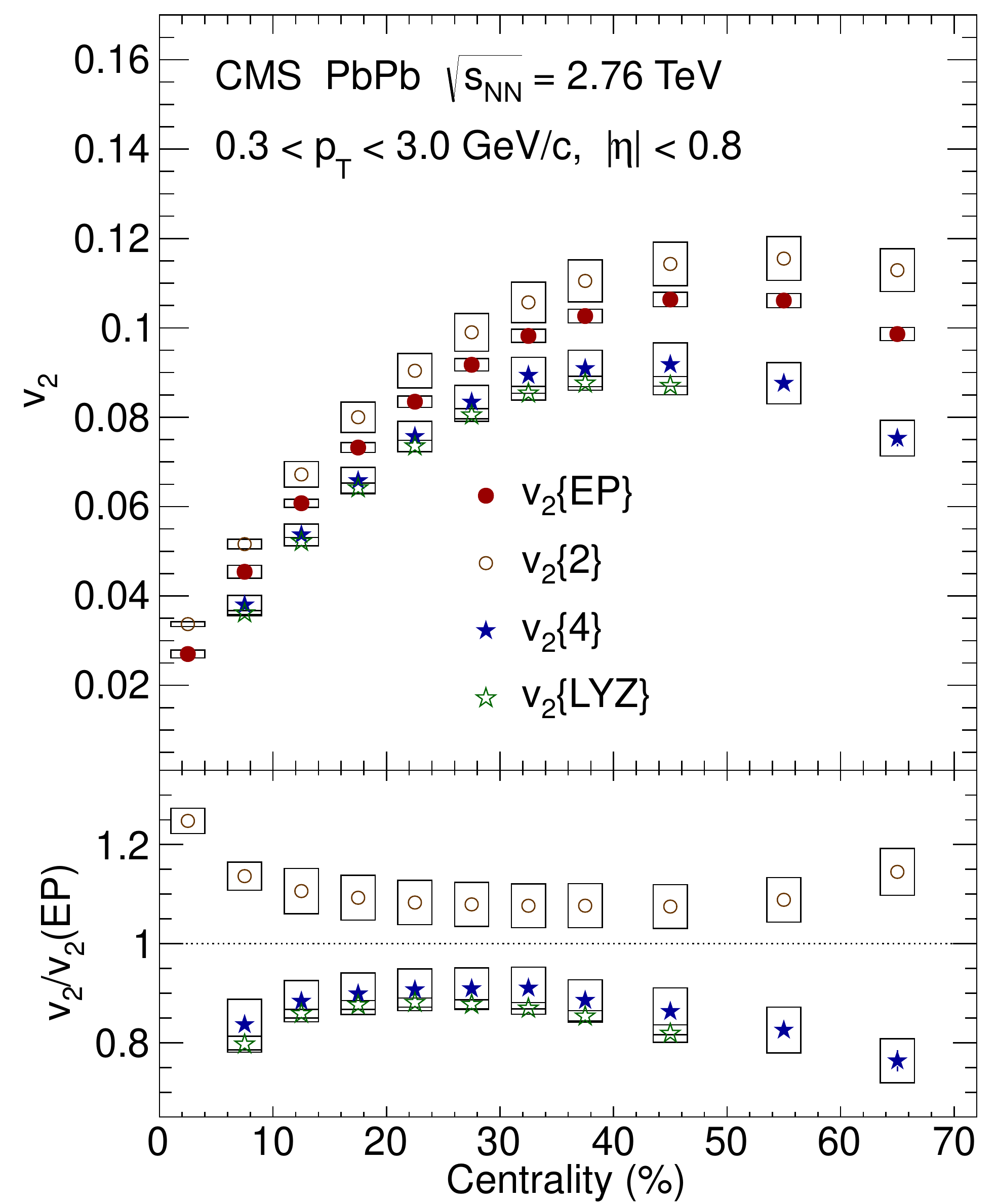}
\caption{\label{fig:v2cen}(Color online)
Top panel: Integrated $v_{2}$ as a function of centrality
at mid-rapidity $|\eta|<0.8$ for the four methods. The boxes represent the
systematic uncertainties. The magnitudes of the statistical uncertainties
are smaller than the size of the symbols. Bottom panel: The values
from three of the methods are divided by the results from the event-plane method. The boxes represent the
systematic uncertainties excluding the sources that are common to all methods. The magnitudes of the
statistical uncertainties are smaller than the size of the symbols.}
\end{center}
\end{figure}

The centrality dependence of the integrated  $v_{2}$ at mid-rapidity $|\eta|<0.8$ is presented in
Fig.~\ref{fig:v2cen}, for the four methods. The  $v_{2}$ values increase from central to peripheral
collisions, reaching a maximum in the 40--50\% centrality range. In the more peripheral collisions, a
decrease in $v_{2}$ is observed in the event-plane and four-particle cumulant measurements, while the
values obtained with the two-particle cumulant method remain constant within their uncertainties. The results for
$v_{2}\{2\}$  are larger than those for  $v_{2}\{\mathrm{EP}\}$,
while the  $v_{2}\{4\}$ and  $v_{2}\{\mathrm{LYZ}\}$ values are
smaller. To facilitate a quantitative comparison between the methods,
including their respective systematic uncertainties, the bottom panel of Fig.~\ref{fig:v2cen} shows the results
from the cumulant and the Lee--Yang zeros methods divided by those obtained from the event-plane method. The boxes
represent the systematic uncertainty in the ratios, excluding sources of uncertainty common to all methods. The ratios
are relatively constant in the 10-60\% centrality range, but the differences between the methods increase for the most
central and the most peripheral collisions. These findings are similar to results obtained by the STAR experiment at
RHIC~\cite{STAR:2008ed}. Below, we further investigate the differences in the  $v_{2}$ values returned by each method.

The collective motion of the system, and therefore the anisotropy
parameter, depend on the initial shape of the nucleus-nucleus
collision area and the fluctuations in the positions of the
interacting nucleons.  By dividing $v_2$ by the participant
eccentricity, one may potentially remove this dependence across
centralities, colliding species, and center-of-mass energies, enabling
a comparison of results
in terms of the underlying physics driving the flow.

\begin{figure*}[htb]
\begin{center}
\includegraphics[width=\textwidth]{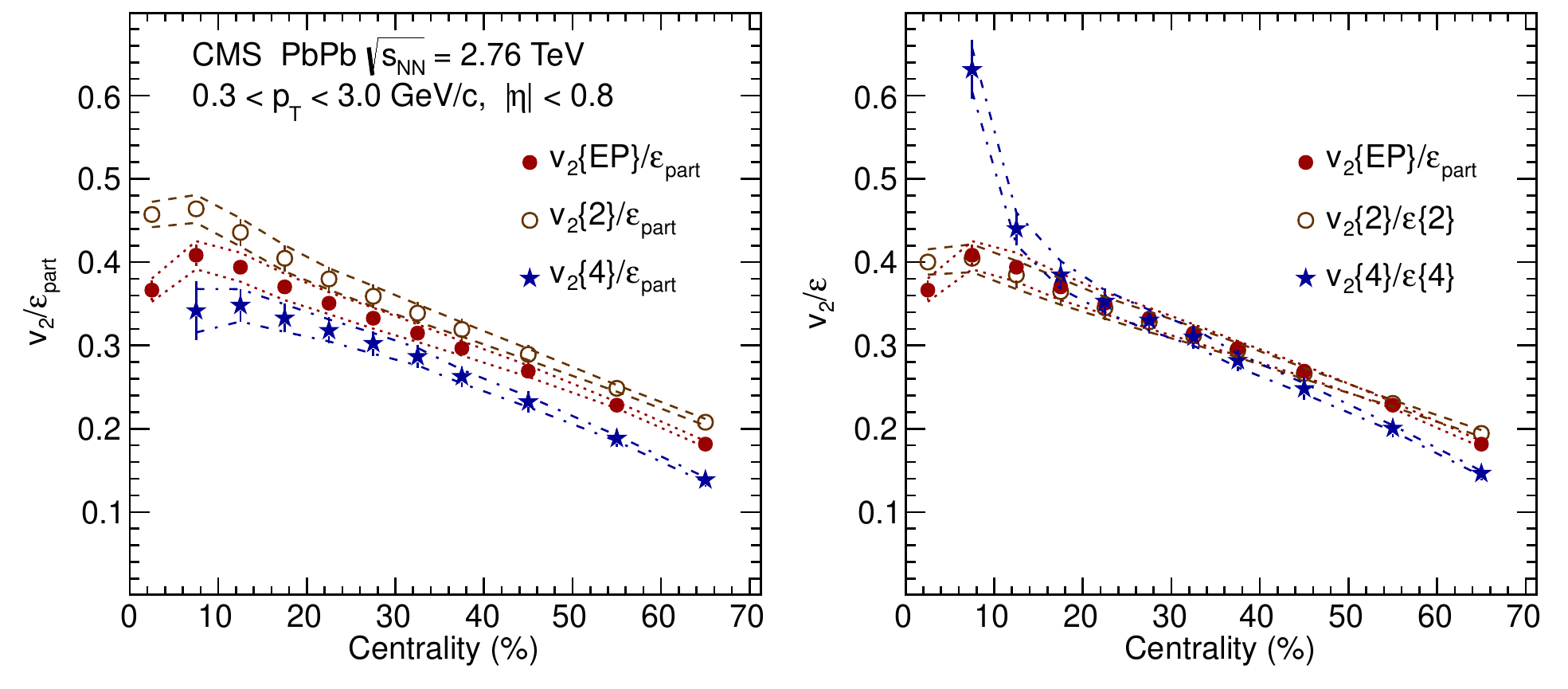}
\caption{\label{f:V2overecc} (Color online)
Left panel: Centrality dependence of the integrated $v_{2}$ divided by
the participant eccentricity, $\epsilon_{\mathrm{part}}$, obtained at
mid-rapidity $|\eta|<0.8$ for the event-plane and two- and four-particle cumulant
methods.
Right panel: The measurements of $v_{2}$ as a function of centrality
are the same as in the left panel, but here the two-particle and four-particle
cumulant results are divided by their corresponding moments of the
participant eccentricity, $\epsilon\{2\}$ and $\epsilon\{4\}$.
In both panels, the error bars show the sum in quadrature of the statistical
and systematic uncertainties in the $v_{2}$ measurement, and the lines
represent the systematic uncertainties in the eccentricity
determination.}

\end{center}
\end{figure*}

In Fig.~\ref{f:V2overecc}, we examine the centrality dependence of
the eccentricity-scaled anisotropy parameter obtained with the event-plane
and cumulant methods at mid-rapidity, $|\eta|<0.8$. The
participant eccentricity and its cumulant moments are obtained from a
Glauber-model simulation, as discussed in Section~\ref{sec:Centrality
Determination}. The statistical and systematic uncertainties in
the integrated $v_{2}$ measurements are added in quadrature and represented
by the error bars.  The dashed lines show the systematic uncertainties
in the eccentricity determination. In the left panel of
Fig.~\ref{f:V2overecc}, the results from each method are divided by
the participant eccentricity, $\epsilon_{\mathrm{part}}$. The data show a
near-linear decrease from central to peripheral collisions, with
differences between methods that were already observed in
Fig.~\ref{fig:v2cen}. In the right panel, the $v_{2}$ values for the
cumulant measurements are scaled by their respective moments of the
participant eccentricity, thus taking into account the corresponding
eccentricity fluctuations~\cite{Bhalerao:2006tp,PHOBOSeccPART,Voloshin:2007pc,Ollitrault:2009ie,Qiu:2011iv}.
With this scaling, the two-particle cumulant and the event-plane results
become nearly identical, except for the most central and the most
peripheral collisions, where the cumulant results are more affected by
nonflow contributions. This is expected~\cite{Ollitrault:2009ie} because in our
application of the method there is no separation in rapidity between
the particles used for the reference flow and those used in the
differential flow measurement.  In the centrality range of 15--40\%,
the four-particle cumulant measurement of $v_{2}\{4\}/\epsilon\{4\}$ is also
in better agreement with the other two methods. This indicates that
the main difference in the results from the different methods could be
attributed to their sensitivity to eccentricity fluctuations.
In the most central events, where the eccentricity  $\epsilon\{4\}$ is very small,
and in the most peripheral events, where the fluctuations are large, $v_{2}\{4\}/\epsilon_2\{4\}$ deviates
from the common scaling behavior.
For centralities above 50\% the differences between the four-particle
cumulant method and the other two methods do not seem to be accounted
for by the initial-state fluctuations, as described in our implementation of the Glauber model.
In this centrality range, the event-by-event fluctuations in the eccentricity are non-Gaussian due to the underlying Poisson distributions
 from discrete nucleons~\cite{PHOBOSeccPART,Qiu:2011iv} and
are more difficult to model. It has also been suggested~\cite{Ollitrault:2009ie,PHOBOSeccPART} that when the event-plane
resolution is smaller than $\approx$0.6, as is the case for the peripheral collisions studied in CMS, the results from the
event-plane method should be evaluated using the two-particle cumulant eccentricity $\epsilon\{2\}$, rather
than the participant eccentricity  $\epsilon_{\mathrm{part}}$. We have used a common definition of eccentricity ($\epsilon_{\mathrm{part}}$)
for all centrality classes studied in our event-plane analysis. This would lower the measurements of
 $v_{2}\{\mathrm{EP}\}/\epsilon$  by about 10\% in the most peripheral collisions, which is not sufficient to reconcile
the differences between the event-plane and the four-particle cumulant results.

Another model of the initial state that has been used in the
literature~\cite{Hirano:2005xf,Hirano:2010jg,Song:2010mg,Nagle:2011uz,Shen:2011eg,Qiu:2011iv}, but has not been explored here,
is the color glass condensate (CGC) model~\cite{PhysRevC.76.041903}, which takes into account that  at
very high energies or small values of Bjorken $x$, the gluon density becomes very large and saturates. The CGC model
predicts eccentricities that exceed the Glauber-model eccentricities by an approximately constant factor of around
1.2, with some  deviation from this behavior in the most central and most peripheral collisions~\cite{Nagle:2011uz,Qiu:2011iv}.
The results presented here may give further insight into the nature of the initial-state fluctuations, especially in the regions
where the eccentricity fluctuations become non-Gaussian.

\subsection{Pseudorapidity dependence of \texorpdfstring{$v_{2}$}{v2}}
\label{s:v2(eta)}
The pseudorapidity dependence of the anisotropy parameter provides
additional constraints on the system evolution in the longitudinal
direction.  To obtain the $v_{2}(\eta)$ distribution with the event-plane method, we
first measure $v_{2}(\pt)$ in pseudorapidity bins
of $\Delta\eta = 0.4$, and then average the results over the range
$0.3 < \pt < 3 $\GeVc, weighting with the efficiency and
fake-rate-corrected spectrum.

For the cumulant and Lee--Yang zeros methods, the measurements are
done using all particles in the range $|\eta|<2.4$,  and either $0.3 < \pt< 3 $\GeVc
or $0.3 < \pt< 12$\GeVc  in the generating function, to obtain the reference flow,
and then extracting the pseudorapidity dependence in small  pseudorapidity intervals of $\Delta\eta = 0.4$. Tracking efficiency and
fake-rate corrections are applied using a track-by-track weight in forming the differential
generating functions. As a crosscheck, we have confirmed that at mid-rapidity the values obtained with this method agree
with the ones obtained from a direct yield-weighted average of the $v_{2}(\pt)$ results from
Figs.~\ref{fig:v2pt_C2_f}--\ref{fig:v2pt_LYZ_f}, within the stated systematic uncertainties.

\begin{figure*}[htb]
\begin{center}
\includegraphics[width=\textwidth]{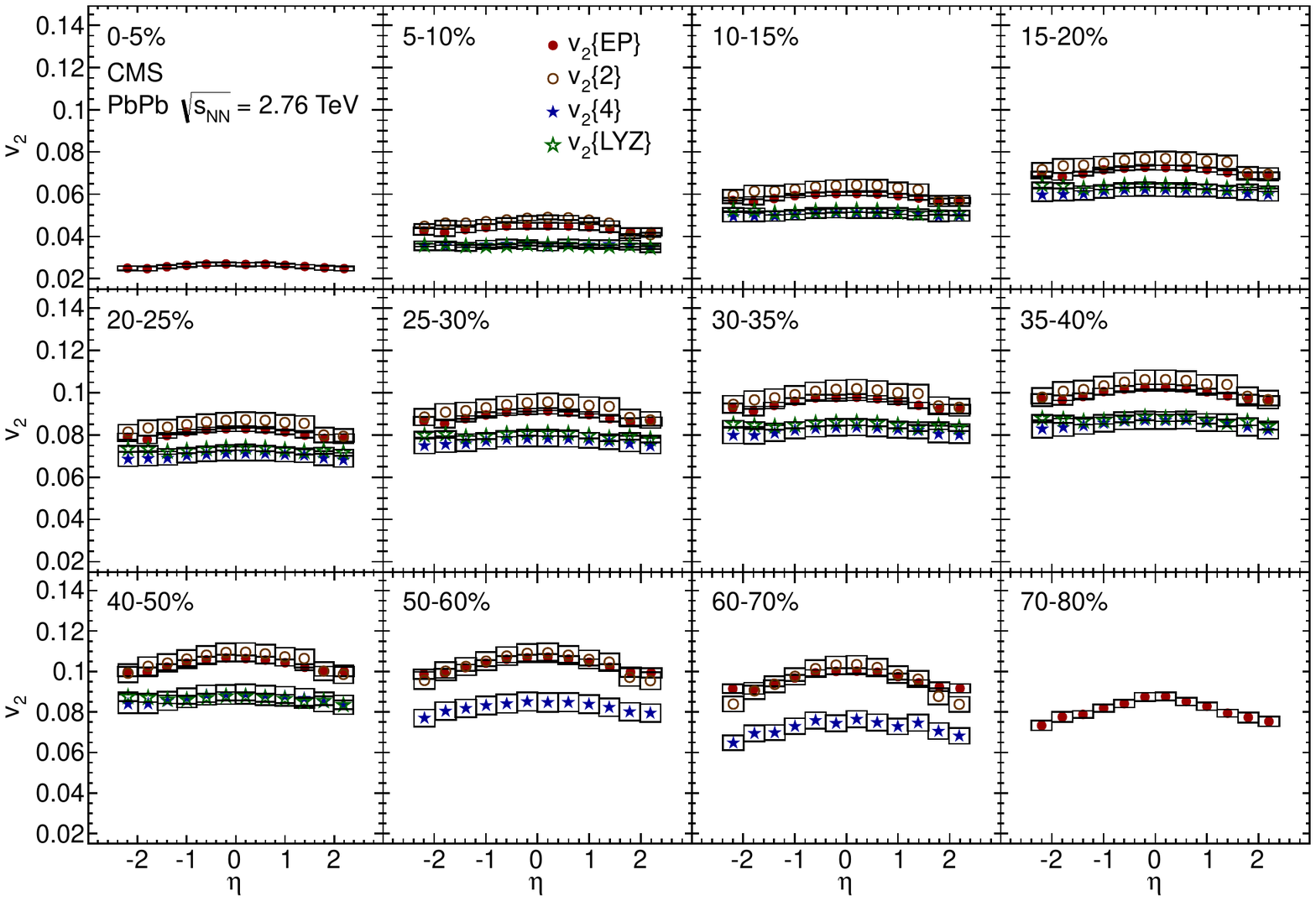}
\caption{\label{fig:v2eta} (Color online)
Pseudorapidity dependence of $v_{2}$ for $0.3 < \pt< 3$\GeVc with all four methods in 12 centrality classes.
The boxes give the systematic uncertainties. The magnitudes of the statistical uncertainties
are smaller than the size of the symbols.}
\end{center}
\end{figure*}

As observed at mid-rapidity ($|\eta|<0.8$) in Fig.~\ref{fig:v2cen},
the values of $v_{2}\{4\}$ and $v_{2}\{\mathrm{LYZ}\}$ are in agreement and are smaller than
 $v_{2}\{2\}$ and $v_{2}\{\mathrm{EP}\}$.  This behavior
persists at larger pseudorapidity, as shown in Fig.~\ref{fig:v2eta},
which suggests that similar nonflow correlations and eccentricity fluctuations
affect the results over the full measured pseudorapidity range.  The
results show that the value of $v_2(\eta)$ is greatest at mid-rapidity
and is constant or decreases very slowly at larger values of $|\eta|$. This behavior is most
pronounced in peripheral collisions and for the
two-particle cumulant method, which is most affected by nonflow
contributions.

\begin{figure*}[htb]
\begin{center}
\includegraphics[width=\textwidth]{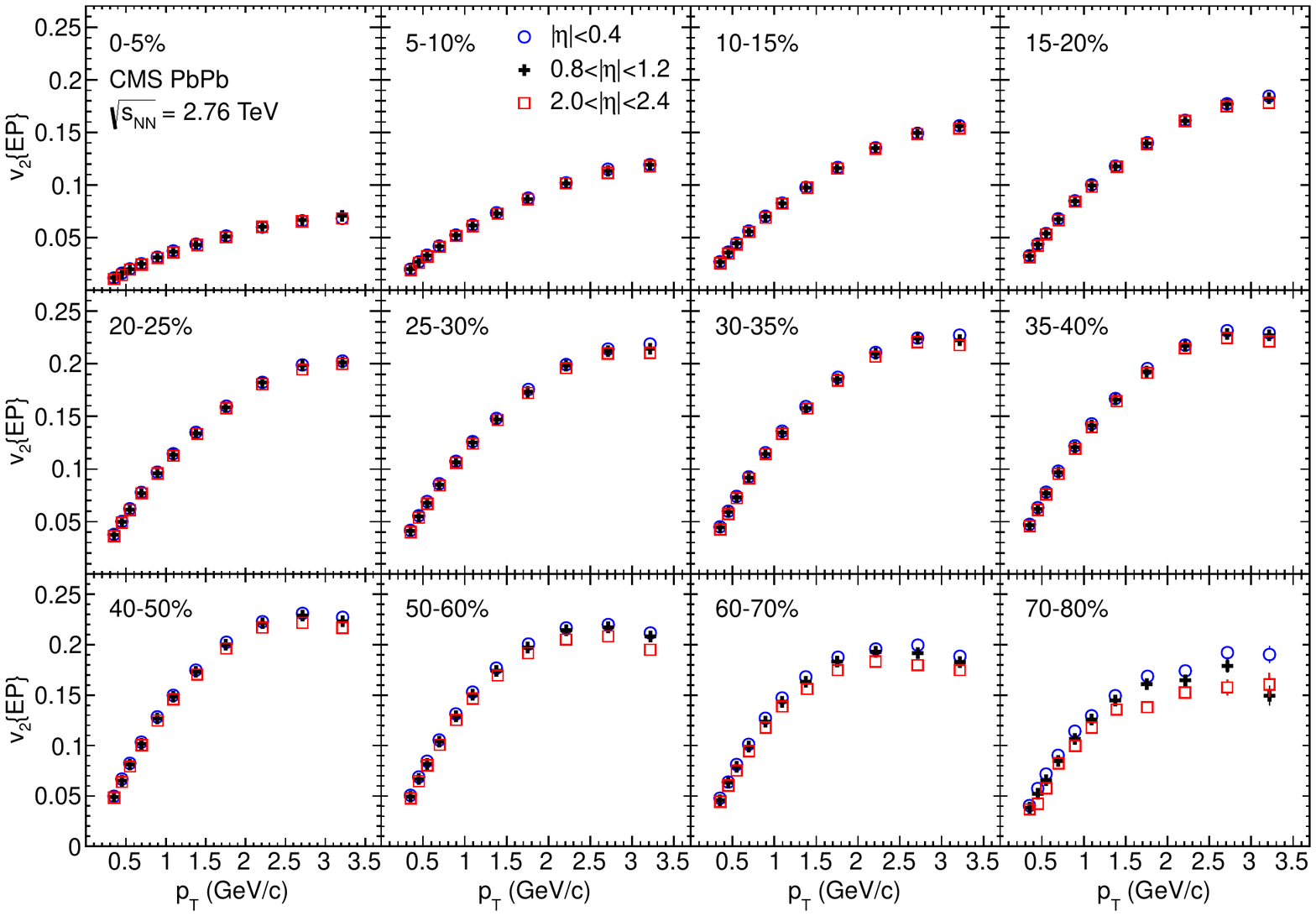}
\caption{\label{fig:v2pteta} (Color online)
Results from the event-plane method for $v_{2}(\pt)$ in three pseudorapidity regions and in 12 centrality
classes. The error bars show the statistical uncertainties that in most cases have magnitudes smaller than the size of the symbol. }
\end{center}
\end{figure*}

To assess whether the observed decrease in $v_{2}(\eta)$ in the forward pseudorapidity region in peripheral
PbPb collisions is due to a pseudorapidity dependence in the $v_{2}(\pt)$ distributions or in the
underlying charged-particle spectra, in Fig.~\ref{fig:v2pteta} we examine the values of $v_{2}(\pt)$ obtained with the
event-plane method for several pseudorapidity intervals in each of the 12 centrality classes shown in
Fig.~\ref{fig:v2eta}. From the most central events up to 35--40\% centrality there is no change in the
$v_{2}(\pt)$ distributions with pseudorapidity within the statistical uncertainties. Therefore, any change
in the $v_{2}(\eta)$ distribution can be attributed to changes in the underlying charged-particle transverse momentum spectra. A
gradual decrease is observed in the  $v_{2}(\pt)$ values at forward pseudorapidity ($2.0<|\eta|<2.4$) in
more peripheral events. For the 70--80\% centrality class  the values of $v_{2}(\pt)$ decrease by approximately
10\% between the central pseudorapidity region  $|\eta|<0.4$ and the forward region $2.0<|\eta|<2.4$. Thus,
the pseudorapidity dependence in $v_{2}(\eta)$ for peripheral collisions observed in Fig.~\ref{fig:v2eta}
is caused by changes in the $v_{2}(\pt)$ distributions with pseudorapidity, as well as changes in the underlying
transverse momentum spectra presented in Section~\ref{s:spectra}.

\subsection{Centrality and pseudorapidity dependence of the transverse momentum distributions }
\label{s:spectra}
Elliptic flow measures the azimuthal anisotropy in the invariant yield of the final-state particles. Therefore,
the charged-particle transverse momentum distributions influence the observed results.
The soft-particle-production mechanism and the evolution of the expanding nuclear medium are reflected in the
low-\pt range of the transverse momentum spectra. In the hydrodynamics calculations, measurements of the pseudorapidity density of
charged particles produced in collisions with different centrality constrain the description of the initial entropy
and the energy density distribution in the collision zone, while the mean transverse momentum of the
particle spectra constrains the final temperature and the radial-flow velocity of the system. With additional input on the
equation-of-state that is typically provided by lattice QCD, the hydrodynamics calculations then provide a description of
the system evolution from some initial time, when local thermal equilibrium is achieved, to the ``freeze-out'', when
the particle interactions cease.

\begin{figure}[htb]
\begin{center}
\includegraphics[width=\linewidth]{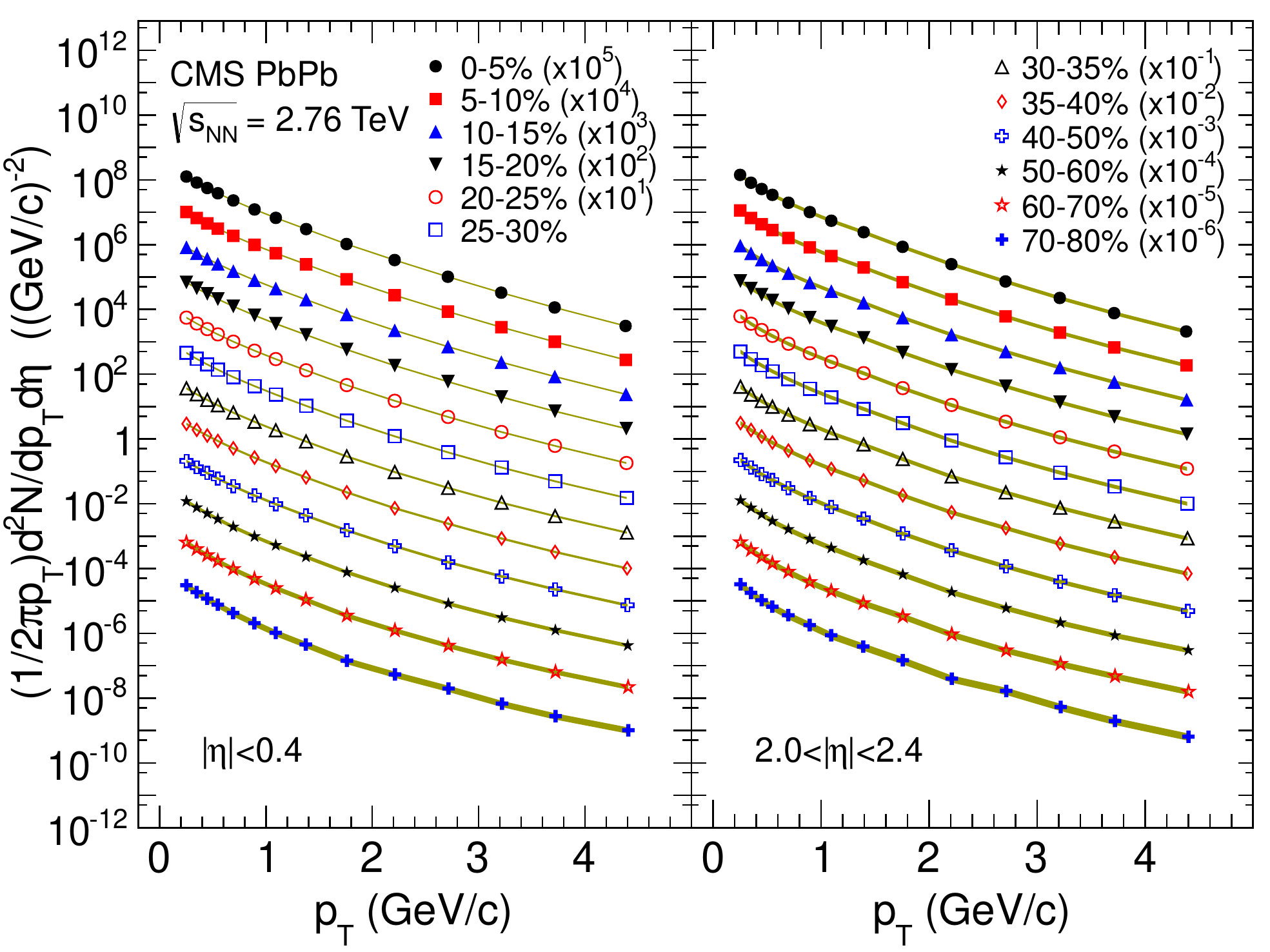}
\caption{\label{fig:spectra_final}
(Color online) Inclusive charged-particle spectra at
mid-rapidity (left) and forward rapidity (right), for the 12 centrality classes given in the legend.
The distributions are offset by arbitrary factors given in the legend for clarity. The shaded bands represent the
statistical and systematic uncertainties added in quadrature, including the overall normalization uncertainties from the trigger efficiency
estimation.
}
 \end{center}
 \end{figure}

We have measured the charged-particle transverse momentum spectra for 12 centrality classes over the pseudorapidity range $|\eta| < 2.4$
in bins of $\Delta \eta = 0.4$. Examples of these distributions for the mid-rapidity ($|\eta| < 0.4$) and
forward-rapidity ($2.0 <|\eta| < 2.4$) regions are shown in Fig.~\ref{fig:spectra_final}. These results extend the measurements of
charged-particle spectra previously reported by CMS~\cite{CMS:2012aa} down to  $\pt = 0.3$\GeVc and to forward pseudorapidity.
The measurements presented here are in good agreement with the results in  Ref.~\cite{CMS:2012aa} in their common ranges of \pt,
pseudorapidity, and centrality.

\begin{figure}[htb]
\begin{center}
\includegraphics[width=\linewidth]{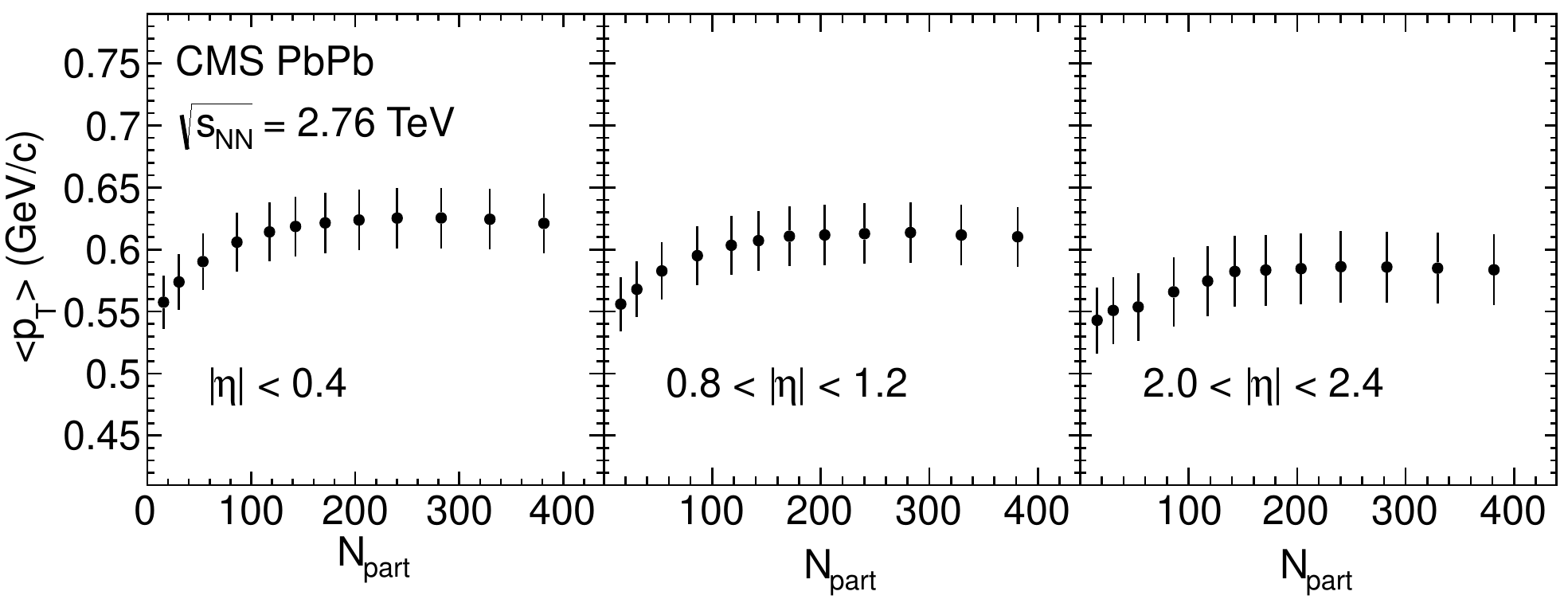}
\caption{\label{fig:meanpt_final}
Mean transverse momentum of the charged-particle spectra as a function of $N_{\mathrm{part}}$ in three pseudorapidity intervals marked in the figure.
The error bars represent the quadratic sum of the statistical and systematic uncertainties.}
\end{center}
\end{figure}

The evolution of the charged-particle spectra with centrality and pseudorapidity can be quantified in terms of the mean \pt  of the
transverse momentum distributions. The values of  $\langle\pt\rangle$   as a function of $N_{\mathrm{part}}$, which is derived
from the centrality of the event, are shown in Fig.~\ref{fig:meanpt_final}. In each pseudorapidity interval, the  values of
$\langle\pt\rangle$ increase with $N_{\mathrm{part}}$ up to  $N_{\mathrm{part}}\approx$150 and then saturate, indicating that the
freeze-out conditions of the produced system are similar over a broad range of collision centralities (0--35\%). This behavior is in contrast
to the centrality dependence in the integrated $v_{2}$ values at mid-rapidity shown in Fig.~\ref{fig:v2cen} that vary strongly in this
centrality range. On the other hand, in more peripheral collisions (centrality greater than 35\%) the  $v_{2}(\pt)$ values do not vary
much with centrality at low \pt, as shown in the
bottom panels of Figs.~\ref{fig:v2pt_EP_f}--\ref{fig:v2pt_LYZ_f}, while the spectral shapes change, as indicated by the $\langle\pt\rangle$ measurement.
This behavior is qualitatively similar at forward pseudorapidities, as shown in Figs.~\ref{fig:v2pteta} and~\ref{fig:meanpt_final}. These
measurements taken together will help in understanding the early-time dynamics in
the system evolution, which is reflected in the elliptic flow, and the
overall evolution through the hadronic stage, which is reflected in the charged-particle spectra.

\subsection{Comparison with other measurements of   \texorpdfstring{$v_{2}$}{v2} at the LHC}

\begin{figure}[hbtp]
  \begin{center}
    \includegraphics[width=0.5\textwidth]{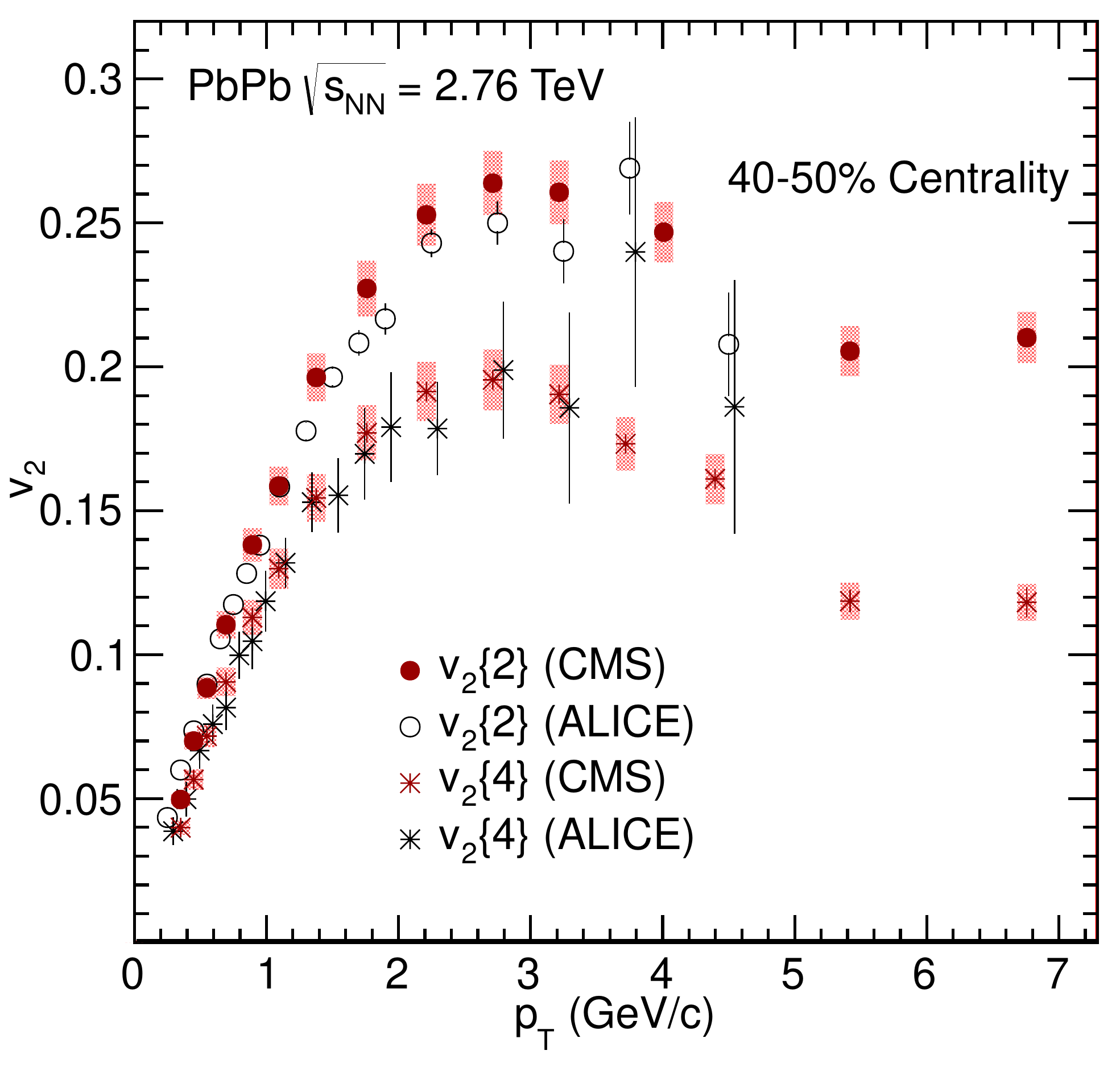}
    \caption{ (Color online) The values of $v_{2}\{2\}$ and $v_{2}\{4\}$  obtained with the cumulant method, as a function of \pt from CMS (closed
      symbols) and ALICE (open symbols)~\cite{ALICE:PhysRevLett.105.252302}, measured in the range $|\eta| < 0.8$
      for the  40--50\% centrality class. The error bars show the statistical uncertainties, and the boxes give the systematic
uncertainties in the CMS measurement.}
    \label{fig:v2ptcumulant_cms_alice_4050}
  \end{center}
\end{figure}

Results on the elliptic anisotropy measured in PbPb collisions in $\sqrt{s_{NN}}$ = 2.76 TeV
have previously been reported by the ALICE~\cite{ALICE:PhysRevLett.105.252302} and ATLAS~\cite{:2011yk} experiments.
A comparison of $v_{2}\{2\}$ and $v_{2}\{4\}$ as a function of \pt
in the 40--50\% centrality class for $|\eta| < 0.8$ from CMS and
ALICE~\cite{ALICE:PhysRevLett.105.252302} is shown in
Fig.~\ref{fig:v2ptcumulant_cms_alice_4050}. The error bars give the statistical uncertainties, and the boxes
represent the systematic uncertainties in the CMS measurements. The two measurements are in
good agreement over their common \pt range.

\begin{figure*}[hbtp]
  \begin{center}
    \includegraphics[width=\textwidth]{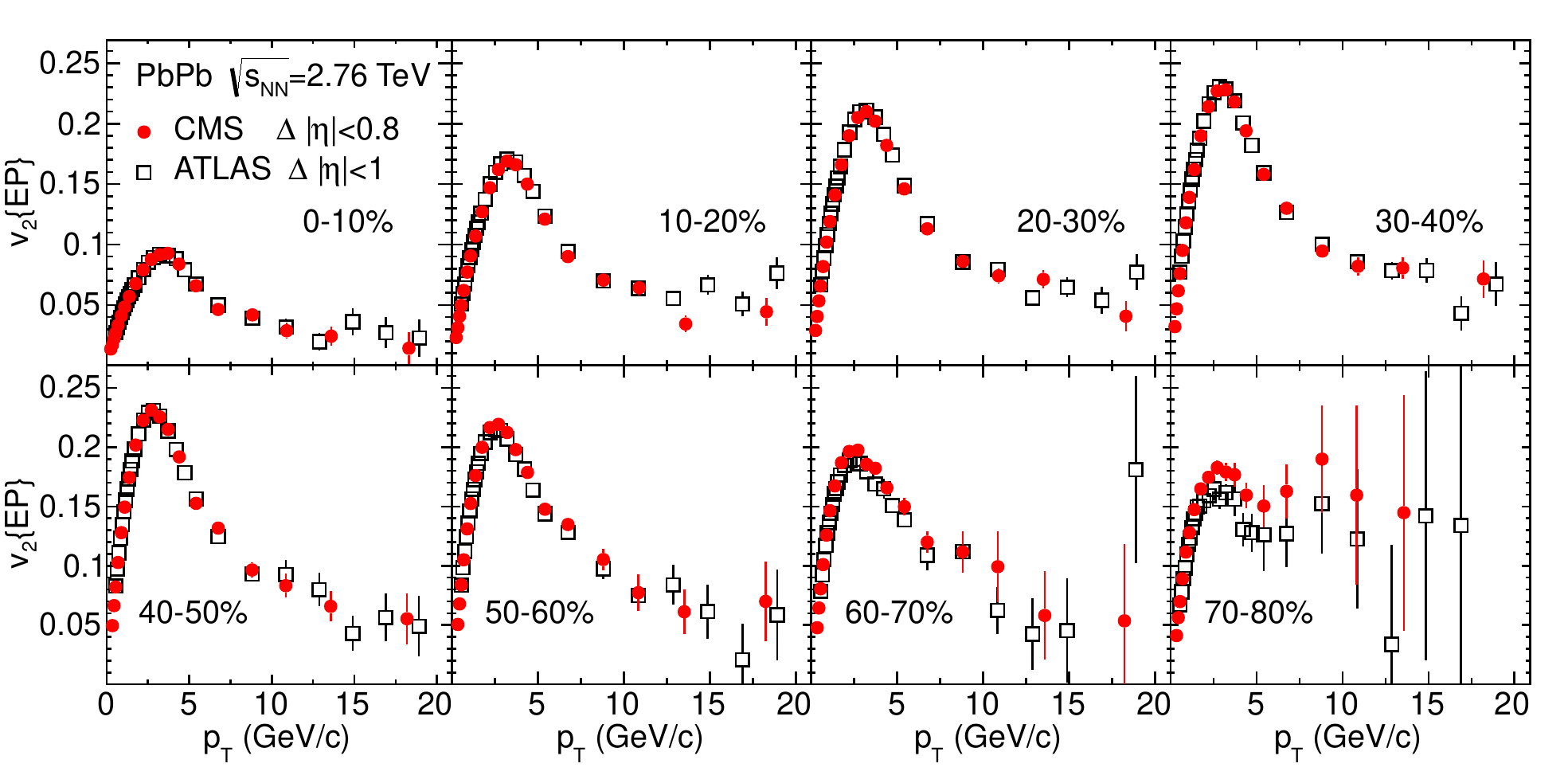}
    \caption{(Color online) Comparison of results for $v_{2}(\pt)$ obtained with the event-plane method from CMS (closed symbols) and ATLAS (open symbols) for the centrality classes marked in the figure. The error bars show the statistical and systematic uncertainties added in quadrature.}
    \label{fig:ATLAS}
\end{center}
\end{figure*}

A comparison of  $v_{2}(\pt)$ obtained with the event-plane method at mid-rapidity from CMS and
ATLAS~\cite{:2011yk} is presented in Fig.~\ref{fig:ATLAS} for the centrality ranges of the ATLAS measurement.
The error bars show the statistical and systematic uncertainties  added in quadrature.
The results are in good agreement within the
statistical and systematic uncertainties.

\section{Discussion}
\label{sec:Discussion}

\begin{figure*}[htb]
\begin{center}
\includegraphics[width=\textwidth]{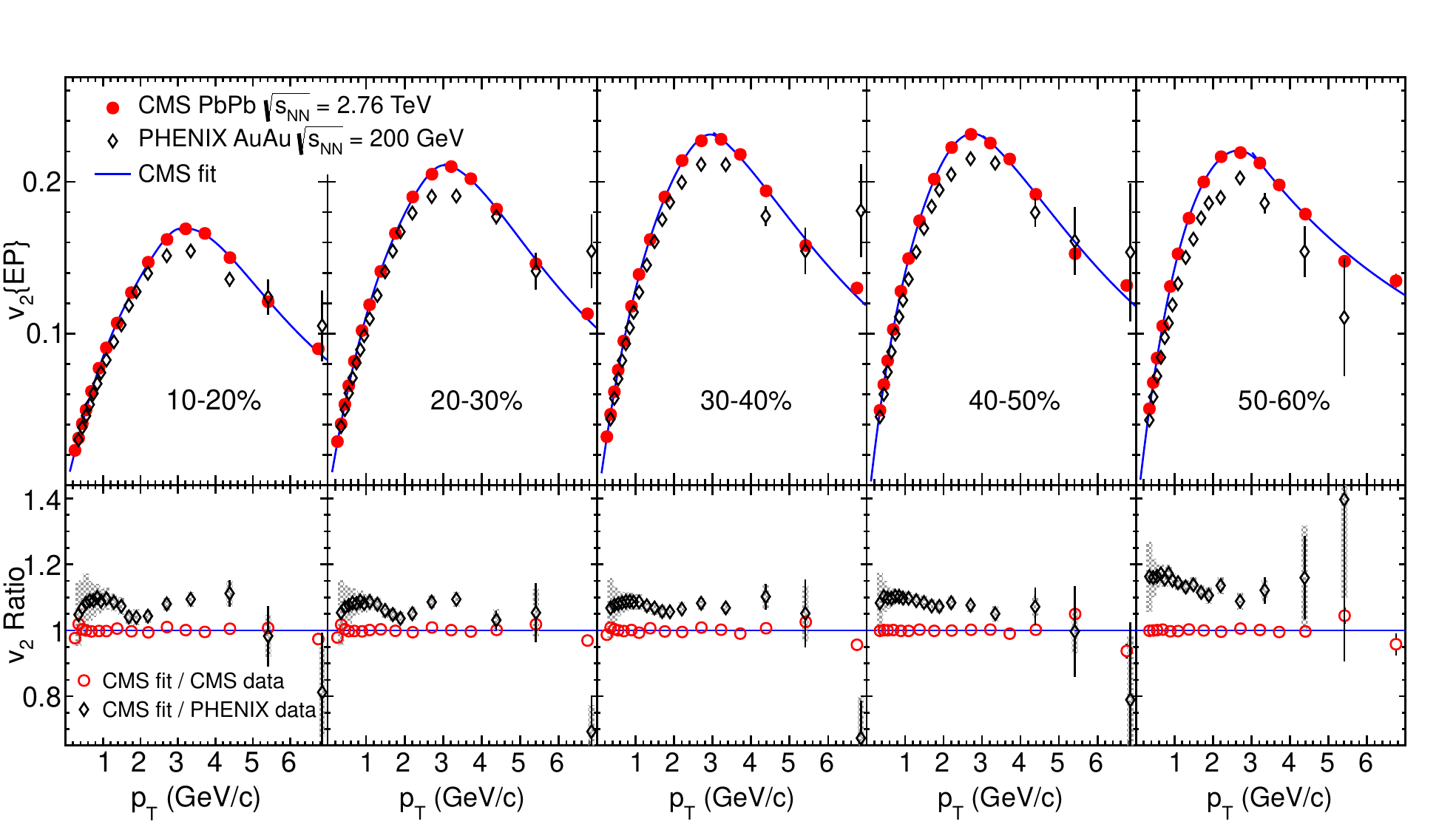}
\caption{\label{fig:EP_CMS_PHENIX} (Color online)
Top panels: Comparison of $v_2(\pt)$ using the event-plane method as
measured by CMS (solid circles) at $\sqrt{s_{NN}}= 2.76$\TeV, and
PHENIX~\cite{Afanasiev:2009wq} (open diamonds) at $\sqrt{s_{NN}}= 200$\GeV
for mid-rapidity ($|\eta|<0.8$ and $|\eta|<0.35$,
respectively). The error bars represent the statistical uncertainties. The
solid line is a fit to the CMS data. Bottom panels: Ratios of the CMS
fit to the PHENIX data (open diamonds) and to the CMS data (open
circles).  The error bars show the statistical uncertainties, while
the shaded boxes give the quadrature sum of the CMS and PHENIX systematic uncertainties.}
\end{center}
\end{figure*}

We compare the CMS elliptic flow  measurements presented in
Section~\ref{sec:Results} with results obtained at RHIC by the PHENIX,
STAR, and PHOBOS experiments. Since each method for measuring $v_2$
has a different sensitivity to nonflow correlations and initial-state
fluctuations, we make these comparisons for measurements conducted
with the same method and with similar kinematic requirements in the method's
implementation. In Fig.~\ref{fig:EP_CMS_PHENIX} the mid-rapidity
measurement of $v_2(\pt)$ with the event-plane method in CMS is
compared to results from PHENIX~\cite{Afanasiev:2009wq} for
$\sqrt{s_{NN}} = 200$\GeV AuAu collisions. For the PHENIX measurement,
the event plane was determined at forward pseudorapidities, $|\eta| =$ 3.1--3.9, while the $v_2(\pt)$
measurement was performed in the pseudorapidity interval $|\eta|<
0.35$, thus providing a separation of at least 2.75 units of
pseudorapidity between the charged particles used for the $v_2(\pt)$ analysis
and the particles used in the event-plane determination. This
procedure is comparable to the CMS approach, where a separation of at
least 3 units of pseudorapidity is used. These large pseudorapidity
gaps are expected~\cite{Ollitrault:2009ie} to suppress nonflow contributions in both
measurements. The pseudorapidity interval for the CMS measurement is
wider, $|\eta|< 0.8$, but since the pseudorapidity dependence of $v_2(\eta)$ was shown to be weak (see Fig.~\ref{fig:v2eta}),
this difference should not influence the comparison of the
results. The top panels in Fig.~\ref{fig:EP_CMS_PHENIX} show the
measurements of $v_2(\pt)$ from CMS (closed symbols) and PHENIX (open
symbols) for several centrality classes. The
error bars represent the statistical uncertainties only. The shape of the
$v_2(\pt)$ distributions and the magnitude of the signals are
similar, in spite of the factor of $\approx$14 increase in
the center-of-mass energy.  To facilitate a quantitative comparison of
these results, the CMS measurements are fitted with a combination of a
fifth-order polynomial function (for $\pt < 3.2$\GeVc) and a Landau
distribution (for $ 3 < \pt < 7 $\GeVc).  There is no
physical significance attributed to these functional choices, other than
an attempt to analytically describe the CMS $v_2(\pt)$ distributions, so
that the value of $v_2$ can be easily compared to results from other
experiments, which have been obtained with different \pt binning.
The results from the fits are plotted as solid lines in the top panels of
Fig.~\ref{fig:EP_CMS_PHENIX}. In the bottom panels, the fit function
is used to evaluate $v_2(\pt)$ at the \pt values for each data
set, and then to form the ratios between the CMS fit values and the PHENIX data,
and the CMS fit to the CMS measurements. The
error bars represent the statistical uncertainties. The systematic
uncertainties from the CMS and  PHENIX measurements are added in
quadrature and plotted as shaded boxes. The $v_2(\pt)$ values measured by CMS are
systematically higher than those from PHENIX in all centrality classes and
over the entire transverse momentum range measured by PHENIX. The relative deviations of are the order
10\%, except for the most peripheral collisions where they reach 15\%.

\begin{figure*}[htb]
\begin{center}
\includegraphics[width=\textwidth]{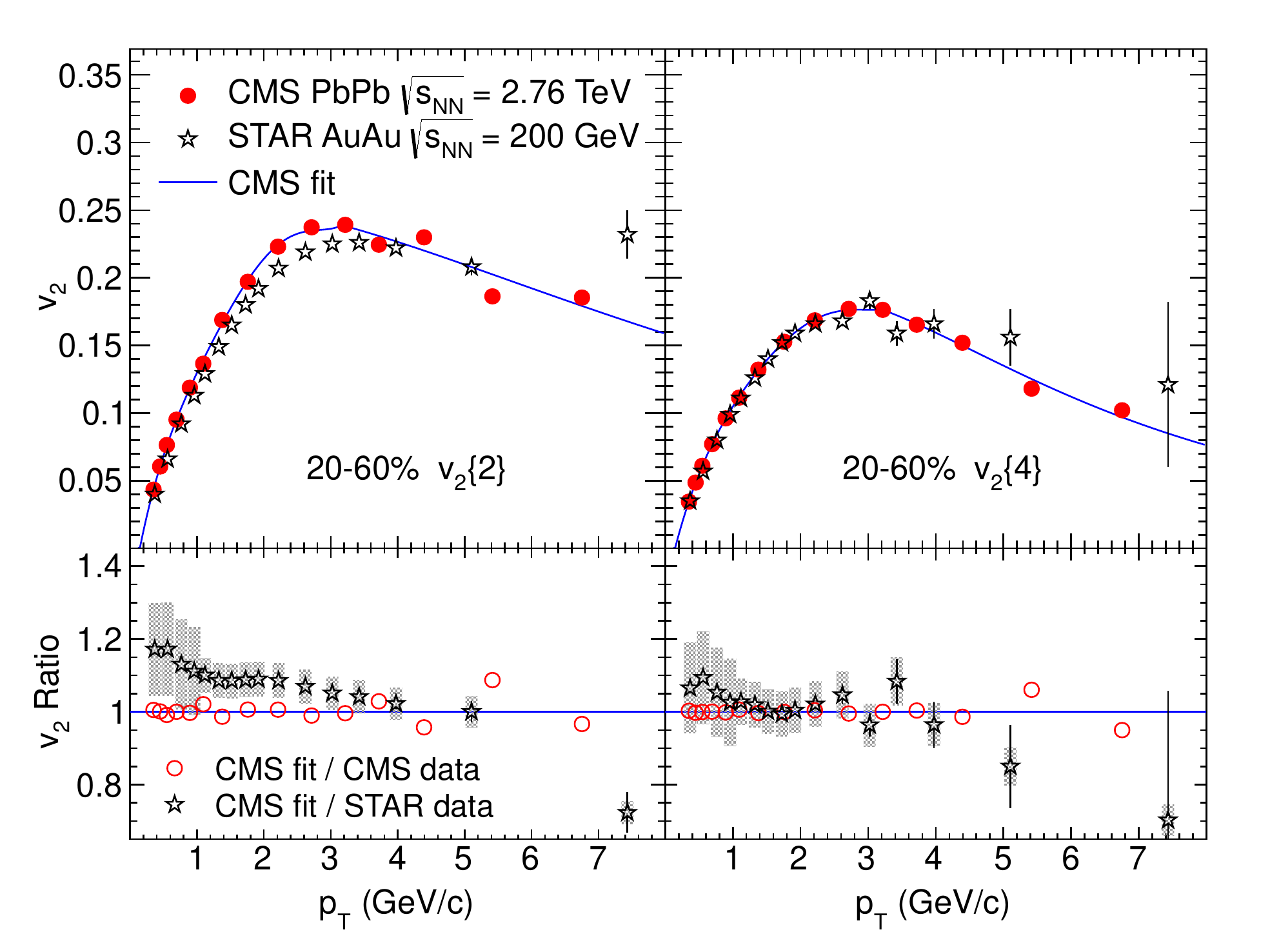}
\caption{\label{fig:C2C4_CMS_STAR}(Color online)
Top panels: Comparison of $v_2(\pt)$ using the two-particle (left)
and the four-particle (right) cumulant method as measured
by CMS (solid circles) at $\sqrt{s_{NN}}= 2.76$\TeV, and
STAR~\cite{Adams:2004wz} (open stars) at $\sqrt{s_{NN}}= 200$\GeV at
mid-rapidity ($|\eta|<0.8$ and $|\eta|<1.3$, respectively). The error
bars represent the statistical uncertainties. The  line is a fit to
the CMS data. Bottom panels: Ratios of the CMS fit values to the STAR data
(open diamonds) and to the CMS data (open circles). The error bars show
the statistical uncertainties, while the shaded boxes give the quadrature sum of the CMS and STAR  systematic
uncertainties.
}
\end{center}
\end{figure*}

A similar comparison is carried out for the two-particle and four-particle cumulant methods.
In Fig.~\ref{fig:C2C4_CMS_STAR}, the
results from the STAR experiment~\cite{Adams:2004wz} for  AuAu collisions at $\sqrt{s_{NN}} = 200$\GeV
in the 20--60\% centrality range are compared with the CMS measurements.
The pseudorapidity interval for the STAR measurement is $|\eta|< 1.3$, compared to
$|\eta|< 0.8$ for CMS. These $\eta$ ranges are within a pseudorapidity
region in which the $v_2(\eta)$ values only weakly depend on the
pseudorapidity. The kinematic selections imposed on the charged particles
used in determining the reference flow are also similar in the CMS and
STAR measurements. The top panels of
Fig.~\ref{fig:C2C4_CMS_STAR} show the $v_2(\pt)$ distributions for
the two-particle (left) and  four-particle (right) cumulant
method  from both experiments, along with fits to the CMS data
(lines). The functional form used for the fit of the CMS
$v_2(\pt)$ distributions is the same as the one used in
Fig.~\ref{fig:EP_CMS_PHENIX}. The bottom panels in
Fig.~\ref{fig:C2C4_CMS_STAR} show the ratios of the fits to the CMS
data to the actual measurements from CMS and STAR. The error bars represent the statistical uncertainties. The
systematic uncertainties from the CMS and STAR measurements are
added in quadrature and plotted as shaded boxes. At low \pt, the $v_2(\pt)$
values measured by CMS are larger than in the STAR data, but
the relative deviations are smaller than 5\% for the four-particle cumulant method, and are of the order
10--15\% for the two-particle cumulant method.
Taken together, the comparisons to the RHIC results in
Figs.~\ref{fig:EP_CMS_PHENIX} and~\ref{fig:C2C4_CMS_STAR} indicate only a moderate increase
in $v_2(\pt)$ at low \pt from the highest RHIC energy to the LHC, despite the
large increase in the center-of-mass energy.

\begin{figure}[hbtp]
  \begin{center}
   \includegraphics[width=0.5\textwidth]{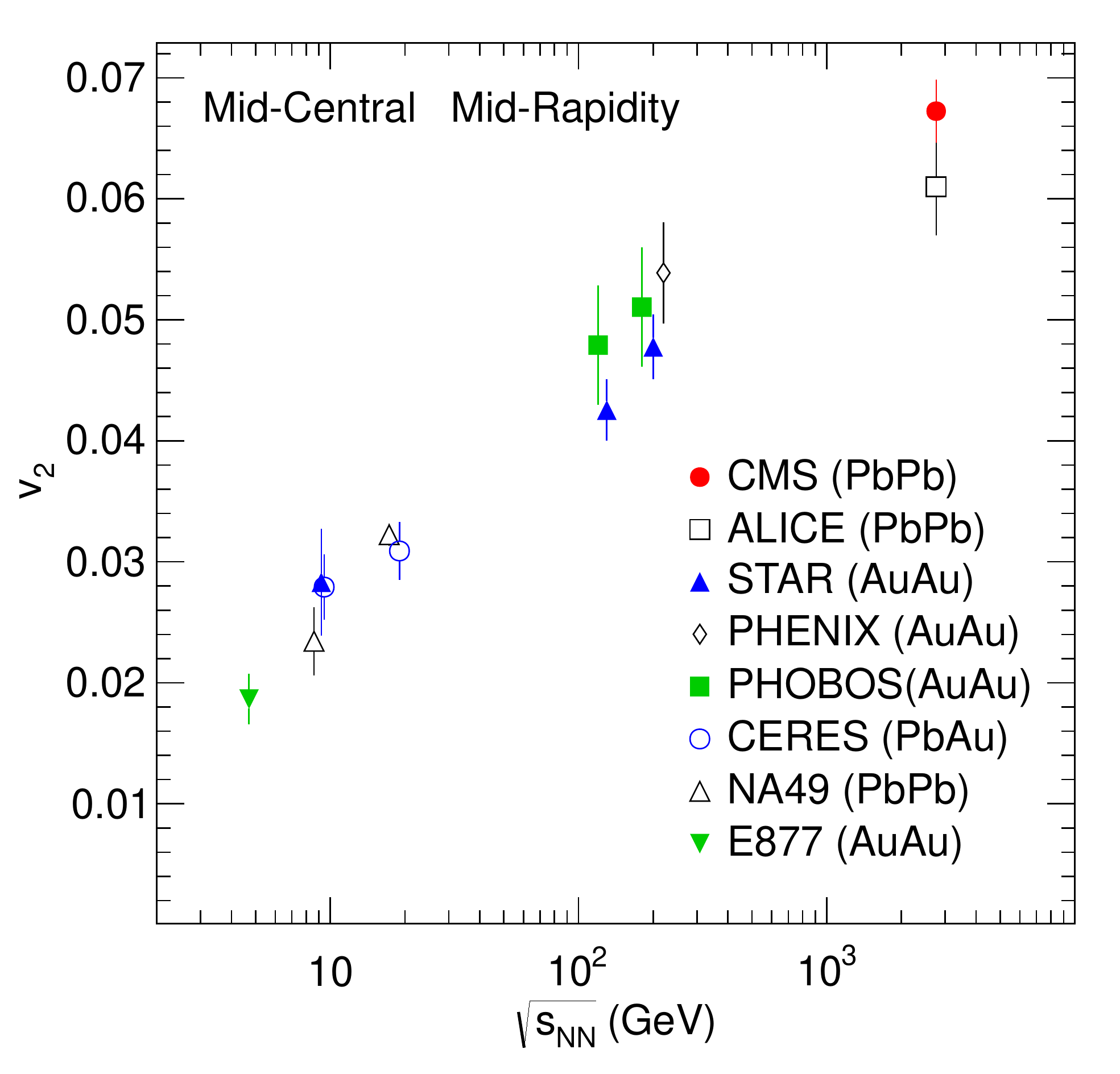}
    \caption{(Color online)
The CMS integrated $v_2$ values  for 20--30\% centrality from the range $|\eta| < 0.8$ and $0< \pt < 3 \GeVc$ obtained using the event-plane method
is compared as a function of  $\sqrt{s_{NN}}$ to results at mid-rapidity and similar centrality from ALICE~\cite{ALICE:PhysRevLett.105.252302},
STAR~\cite{PhysRevC.81.024911},
PHENIX~\cite{PHENIX_scaling},
PHOBOS~\cite{PhysRevC.72.051901, PhysRevLett.89.222301, PhysRevLett.98.242302},
NA49~\cite{PhysRevC.68.034903},
E877~\cite{PhysRevC.55.1420},
and CERES~\cite{CERES:2002253}.
The error bars for the lower-energy results represent statistical
uncertainties; for the  CMS and ALICE measurements the statistical and systematic uncertainties are added in quadrature.}
    \label{fig:ExcitationFun}
\end{center}
\end{figure}

\begin{figure}[hbtp]
  \begin{center}
   \includegraphics[width=0.5\textwidth]{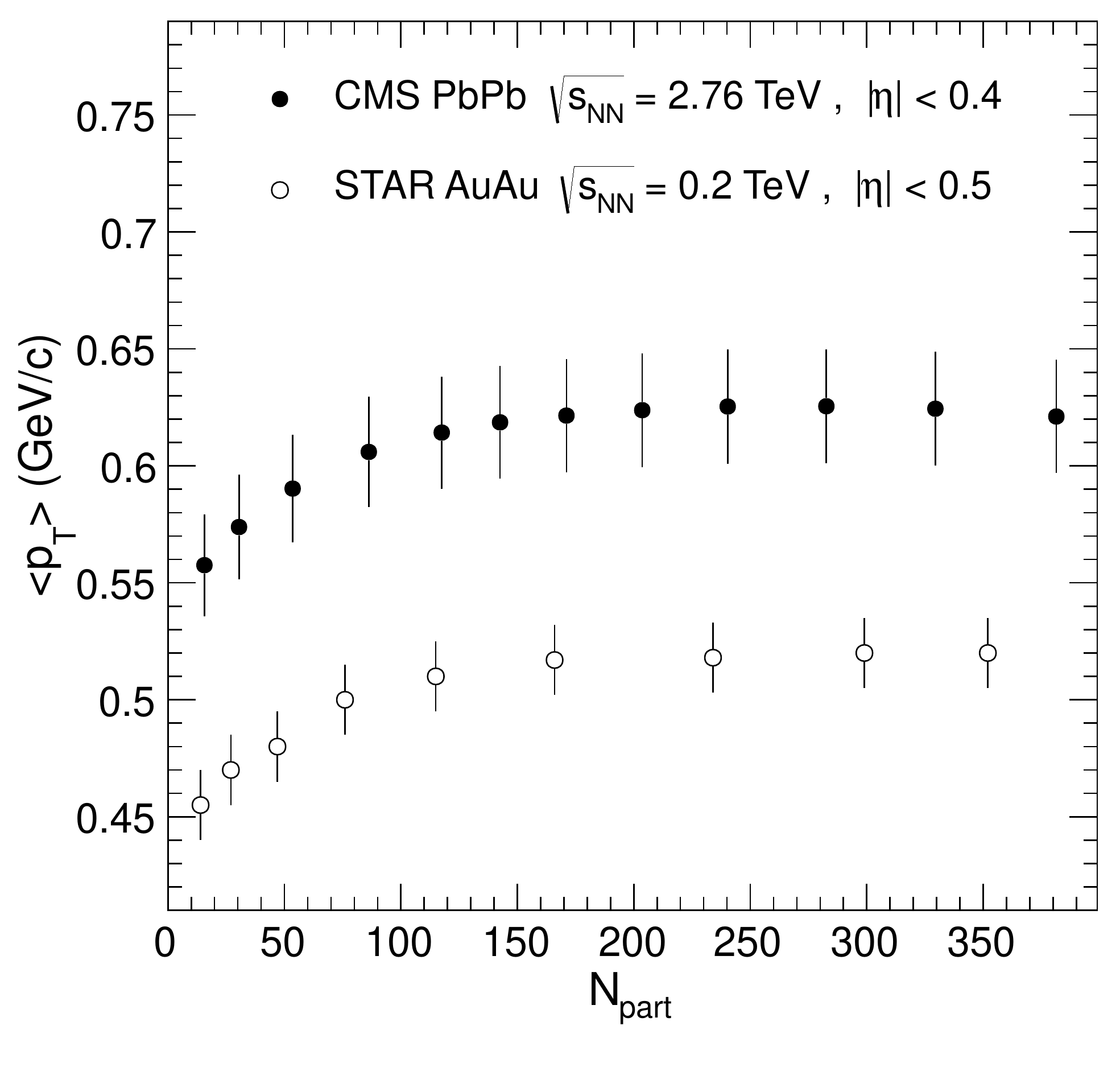}
    \caption{
Mean transverse momentum of the charged-particle spectra as a function of $N_{\mathrm{part}}$ measured by CMS in PbPb collisions at
 $\sqrt{s_{NN}} = 2.76$\TeV (closed circles) and by STAR~\cite{Adams:2004cb} in AuAu  collisions at  $\sqrt{s_{NN}} = 200$\GeV (open circles). The error bars represent
the quadratic sum of statistical and systematic uncertainties. }
    \label{fig:meanptCMS-STAR}
\end{center}
\end{figure}

\begin{figure}[hbtp]
  \begin{center}
   \includegraphics[width=0.5\textwidth]{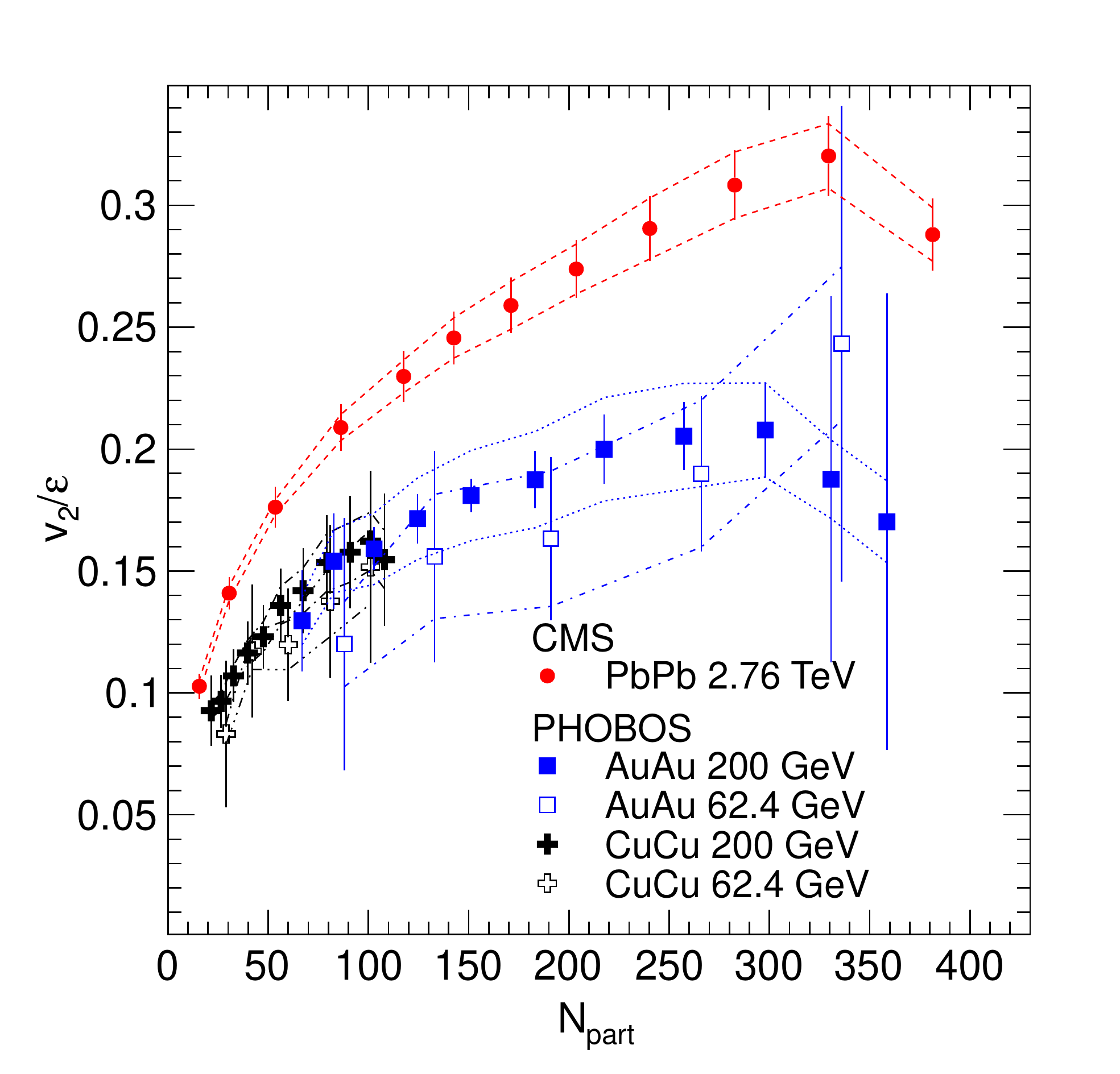}
    \caption{(Color online) The CMS integrated $v_{2}$ values from the event-plane method divided by
the participant eccentricity as a function of $N_{\mathrm{part}}$ with
$|\eta| < 0.8$ and $0< \pt < 3 \GeVc$. These results are compared
with those from  PHOBOS~\cite{PHOBOSeccPART} for different nuclear species
and collision energies. The PHOBOS  $v_{2}$  values are divided by the cumulant eccentricity $\epsilon\{2\}$
(see text). The error bars give the statistical and
systematic uncertainties in the $v_{2}$ measurements added in quadrature. The dashed lines represent the
systematic uncertainties in the eccentricity determination. }
    \label{fig:v2eps_Npart}
\end{center}
\end{figure}

\begin{figure}[hbtp]
  \begin{center}
    \includegraphics[width=0.5\textwidth]{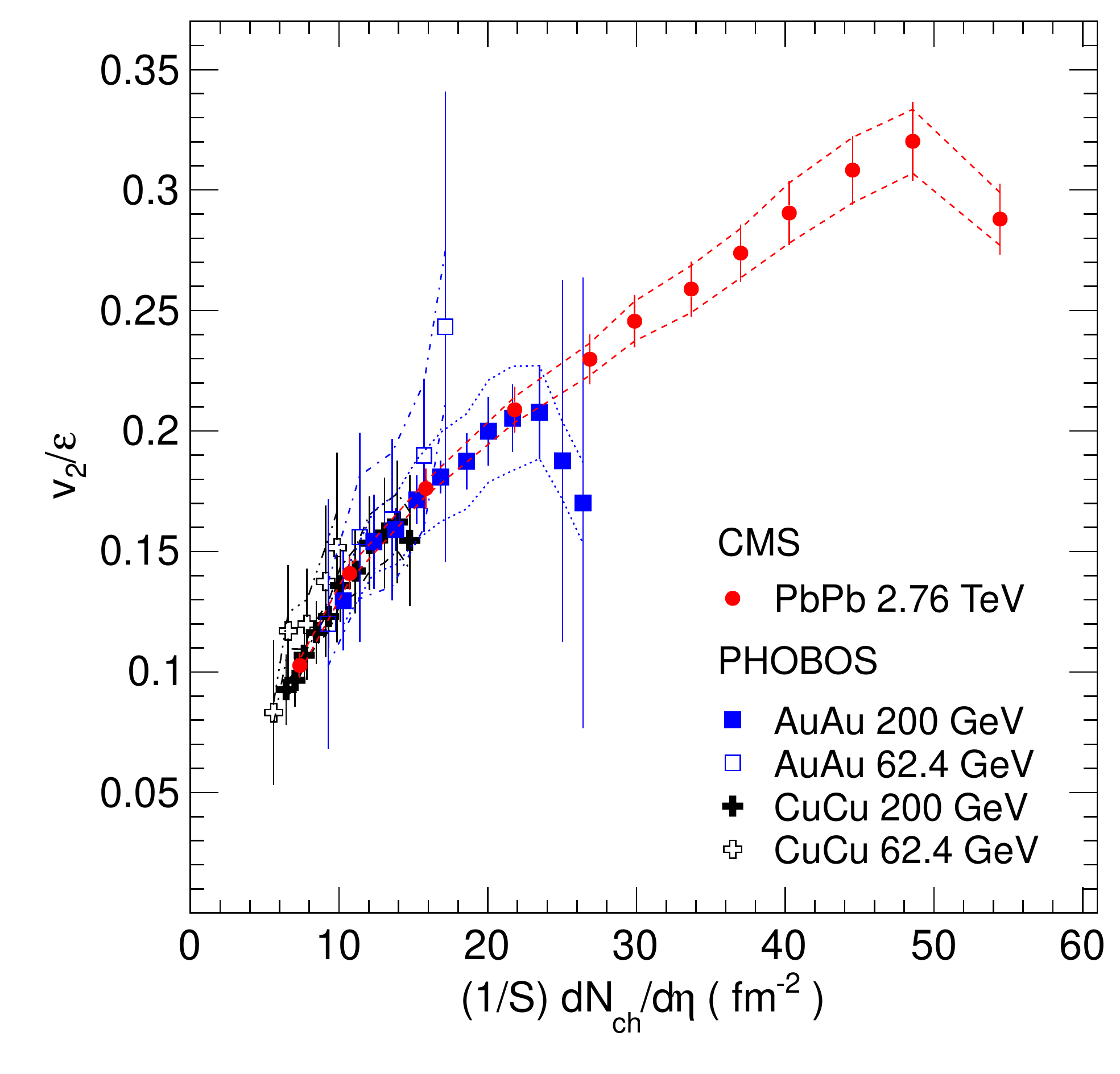}
    \caption{(Color online) Eccentricity-scaled $v_{2}$ as a function of the transverse charged-particle density from CMS and
PHOBOS~\cite{PHOBOSeccPART}. The error bars include both statistical and systematic uncertainties in $v_{2}$.
The dashed lines represent the systematic uncertainties in the eccentricity determination. }
    \label{fig:v2ecc_SdNdeta_0030}
\end{center}
\end{figure}

\begin{figure*}[hbtp]
  \begin{center}
    \includegraphics[width=\textwidth]{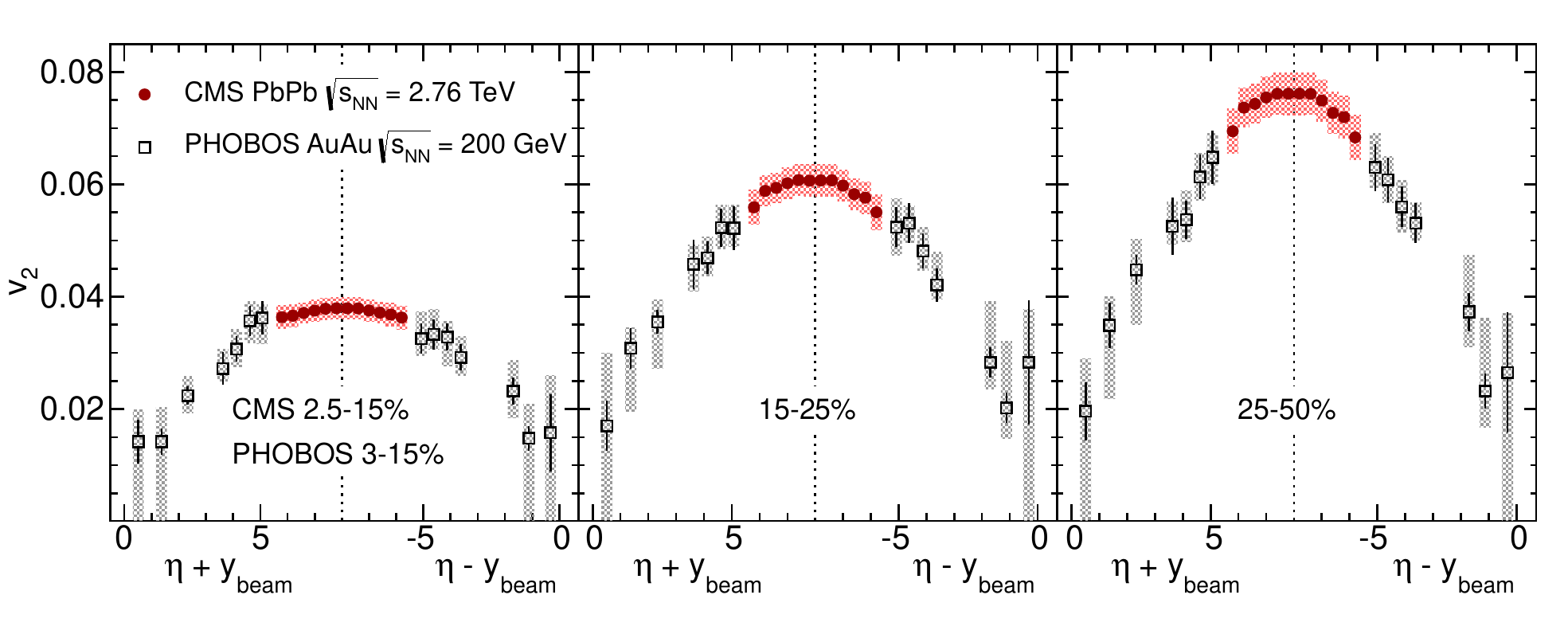}
    \caption{(Color online) Measurements of $v_2$ as a function of the pseudorapidity of particles in the rest frame of
the colliding nuclei $\eta \pm y_{\mathrm{beam}}$ from CMS (closed symbols) and PHOBOS (open symbols)~\cite{PhysRevC.72.051901}
in three centrality intervals. The error bars show the statistical uncertainties and the boxes give the systematic ones.}
    \label{fig:eta_long}
  \end{center}
\end{figure*}

In Fig.~\ref{fig:ExcitationFun}, we examine the  $\sqrt{s_{NN}}$ dependence
of the integrated $v_2$ from  mid-central collisions
spanning $\sqrt{s_{NN}}= 4.7$ GeV to $\sqrt{s_{NN}}= 2.76$ TeV.
The CMS measurement is obtained with the event-plane method in the 20--30\%
centrality class by extrapolating the
$v_2(\pt)$ and the charged-particle spectra down to $\pt = 0$.
In the extrapolation it is assumed that  $v_2(0) = 0 $, and the charged-particle yield is constrained to
match the  $\rd N_{\mathrm{ch}}/\rd\eta$ values  measured by CMS~\cite{Chatrchyan:2011pb}.
The low-energy data are from
Refs.~\cite{Abelev:2007qg,PhysRevC.81.024911,PHENIX_scaling,PhysRevC.72.051901,
PhysRevLett.89.222301,PhysRevLett.98.242302,PhysRevC.68.034903,PhysRevC.55.1420,CERES:2002253},
as compiled in Ref.~\cite{PhysRevC.68.034903} and tabulated in Ref.~\cite{PhysRevC.81.024911}. The error bars for
the low-energy data represent the statistical uncertainties. For the
CMS data the error bar is the quadrature sum of the statistical and
systematic uncertainties. The integrated $v_2$ values  increase
approximately logarithmically with $\sqrt{s_{NN}}$ over the full energy range, with a 20--30\% increase from the highest
RHIC energy to that of the LHC.
This has contributions from the
increase in the mean \pt of the charged-particle spectra with
$\sqrt{s_{NN}}$, shown in Fig.~\ref{fig:meanptCMS-STAR},  and from the moderate increase in the
$v_2(\pt)$ distributions at low \pt, shown in
Fig.~\ref{fig:EP_CMS_PHENIX}. We note that the centrality selections, the collision species, and the methods employed in
the integrated $v_2$ measurements are not identical in all experiments, so the comparison presented in Fig.~\ref{fig:ExcitationFun}
is only approximate. Further comparisons to results from lower energies are presented in
Figs.~\ref{fig:v2eps_Npart}--\ref{fig:eta_long}.

In ideal hydrodynamics, the eccentricity-scaled elliptic flow is
constant over a broad range of impact parameters;
deviations from this behavior are expected in peripheral collisions, in
which the system freezes out before the elliptic flow fully builds up
and saturates~\cite{Kolb:2000sd}. A weak centrality and beam-energy dependence is
expected through variations in the equation-of-state. In addition, the
system is also affected by viscosity, both in the sQGP and
hadronic stages~\cite{Hirano:2010jg,Luzum:2009sb,Shen:2011kn,Heinz:2009xj} of
its evolution. Therefore, the centrality and $\sqrt{s_{NN}}$ dependence of
$v_{2}/\epsilon$ can be used to extract the ratio of the shear viscosity  to the entropy density
of the system.

In Fig.~\ref{fig:v2eps_Npart}, the integrated $v_2$ obtained from
the event-plane method is divided by the eccentricity of the
collisions and plotted as a function of $N_{\mathrm{part}}$, which is derived
from the centrality of the event. The result is compared to lower-energy AuAu
and CuCu measurements from the PHOBOS experiment~\cite{PHOBOSeccPART}.
For the CMS measurement, the value of $v_2$ is divided by the participant eccentricity $\epsilon_{\mathrm{part}}$
since the event-plane resolution factor shown in Fig.~\ref{fig:Rescor} is
greater than 0.6 for all but the most central and  most peripheral
event selections in our analysis. It has been argued~\cite{PHOBOSeccPART,Ollitrault:2009ie}
 that for lower-resolution parameters, the event-plane method measures the r.m.s. of
the azimuthal anisotropy, rather than the mean, and therefore,
the relevant eccentricity parameter in this case should be the
second-order cumulant eccentricity $\epsilon\{2\} \equiv
\sqrt{\langle \epsilon_\mathrm{part}^{2} \rangle}$. Thus, the
comparison with the PHOBOS $v_2$ results, which were obtained with low event-plane resolution,
is done by implementing this scaling using the data from Ref.~\cite{PHOBOSeccPART}.  An approximately 25\%
increase  in the integrated $v_2$ scaled by the eccentricity
between RHIC and LHC energies is observed, and with a similar $N_{\mathrm{part}}$ dependence.

It was previously observed~\cite{Adler:2002pu,PhysRevC.68.034903,PHOBOSeccPART} that the
$v_2/\epsilon$ values obtained in different collision systems and
varying beam energies scale with the charged-particle rapidity density per unit
transverse overlap area $(1/S) (\rd N_{\mathrm{ch}}/\rd y)$, which is proportional to the initial
entropy density. In addition, it has been pointed out~\cite{Song:2010mg} that in this
representation the sensitivity to the modeling of the initial conditions of the heavy-ion collisions
is largely removed, thus enabling the extraction of the shear viscosity  to the entropy density ratio from the data through the
comparison with viscous hydrodynamics calculations. With the factor of 2.1  increase in the
charged-particle pseudorapidity density per participant pair,
$(\rd N_{\mathrm{ch}}/\rd\eta)/(N_{\mathrm{part}}/2)$, from the highest RHIC energy to the
LHC~\cite{Aamodt:2010cz,Chatrchyan:2011pb}, this scaling behavior can
be tested over a much broader range of initial entropy densities. In
Fig.~\ref{fig:v2ecc_SdNdeta_0030}, we compare the CMS results for
$v_{2}/\epsilon$ from the event-plane method to results from the
PHOBOS experiment~\cite{PHOBOSeccPART} for CuCu and AuAu collisions
with $\sqrt{s_{NN}} = 62.4$\GeV and 200\GeV.

At lower energies, the scaling has been examined using the charged-particle
rapidity density $\rd N_{\mathrm{ch}}/\rd y$~\cite{Adler:2002pu,PhysRevC.68.034903,PHOBOSeccPART}.
However, since we do not identify the species of charged particles in this analysis, we
perform the comparison using $(1/S) (\rd N_{\mathrm{ch}}/\rd\eta$) to avoid introducing
uncertainties related to assumptions about the detailed behavior of the
identified particle transverse momentum spectra that are
needed to perform this conversion. In Fig.~\ref{fig:v2ecc_SdNdeta_0030},
the charged-particle pseudorapidity density $\rd N_{\mathrm{ch}}/\rd\eta$ measured by
CMS~\cite{Chatrchyan:2011pb} is used, and the value of the integrated $v_{2}$ for the
ranges $0< \pt < 3\GeVc$ and $|\eta| < 0.8$.  The transverse nuclear-overlap
area $S$ and the participant eccentricity are
listed in Table~\ref{tbl:glauber}.
The PHOBOS results from Ref.~\cite{PHOBOSeccPART}
used $\rd N_{\mathrm{ch}}/\rd y$, and applied two factors to
perform the Jacobian transformation from $\rd N_{\mathrm{ch}}/\rd\eta$: the $x$ axis was scaled by
a factor 1.15, and the $y$ axis by 0.9. Both of these factors are reversed in
order to compare the CMS and  PHOBOS measurements in Fig.~\ref{fig:v2ecc_SdNdeta_0030}.
As in Fig.~\ref{fig:v2eps_Npart}, the PHOBOS data are scaled
by $\epsilon\{2\}$, while the CMS data are  scaled by the
participant eccentricity $\epsilon_{\mathrm{part}}$, taking into account the
event-plane resolution factors in the two measurements. The CMS
result extends to very peripheral collisions (70--80\%
centrality), which allows for a significant overlap in the transverse
charged-particle density measured at RHIC and LHC. Despite the
large systematic uncertainties quoted for the PHOBOS measurements, the
data are in good agreement over the common $(1/S) (\rd N_{\mathrm{ch}}/\rd\eta)$
range. A smooth increase in $v_{2}/\epsilon$  proportional to the
transverse particle density is observed over the entire measured range, except for
a small decrease in the most central collisions in both the
RHIC and LHC data.
The theoretical predictions~\cite{Hirano:2010jg,Luzum:2009sb,Shen:2011eg} for
the $\sqrt{s_{NN}}$ dependence of the transverse-particle-density scaling
of $v_{2}/\epsilon$ differ, and do not generally predict a universal behavior.
The data presented here provide
constraints on the model descriptions of the dynamical evolution of the
system, and thus should aid the reliable extraction of the transport properties of the hot QCD medium
from data.

The  $v_{2}(\eta)$ results can be used to test theoretical descriptions of the  longitudinal dynamics in the
expanding system, as they have been shown to be sensitive to the choice of the initial conditions, the
event-by-event fluctuations in the eccentricity, and the viscosity in the sQGP and the hadronic stages of the
system evolution~\cite{Hirano:2005xf,Andrade:2006yh}.
The PHOBOS experiment  observed~\cite{Back:2004zg} that the
elliptic flow measured over a broad range of collision energies ($\sqrt{s_{NN}}=19.6$, 62.4, 130, and 200\GeV)
exhibited longitudinal scaling extending over several units of pseudorapidity when viewed in the rest frame of
one of the incident nuclei.
A similar phenomenon for soft-particle yields and spectra is known as
limiting fragmentation~\cite{PhysRev.188.2159}.
Furthermore, with increasing $\sqrt{s_{NN}}$, this beam-energy independence
of $v_2(\eta)$ was found to extend over an increasingly wider pseudorapidity
range~\cite{Back:2004zg}.
To investigate the potential continuation of extended
longitudinal scaling of the elliptic flow to LHC energies, in Fig.~\ref{fig:eta_long} we
compare the pseudorapidity dependence of $v_2(\eta)$ measured by CMS with that
measured by PHOBOS at $\sqrt{s_{NN}} = 200$\GeV in three centrality intervals.
Neither the CMS nor the PHOBOS measurements
are performed using identified particles.
The pseudorapidity $\eta^+$ ($\eta^-$) of the
particles in the rest frame of the nuclei moving in the positive
(negative) direction is approximated by $\eta^\pm = \eta \pm y_{\mathrm{beam}}$, where
$\eta$ is  the pseudorapidity of the particles in the center-of-mass
frame, and
$y_{\mathrm{beam}} = \textrm{arccosh} (E_{\mathrm{lab}}/Am_N c^2) \approx \ln(\sqrt{s_{NN}}~\mathrm{[GeV]})$,
with $E_{\mathrm{lab}}$ denoting the energy of the beam in the
laboratory frame, $A$ the nuclear mass number, and  $m_N$ the nucleon mass.
In Fig.~\ref{fig:eta_long}, the left (right) half of each plot depicts $v_2$ in the rest frame of the
beam moving in  the positive (negative) lab direction.
The PHOBOS  $v_2(\eta)$ results are from the hit-based analysis from Fig. 4 in Ref.~\cite{PhysRevC.72.051901}.
The CMS results are obtained with the event-plane method in 0.4-unit-wide bins of pseudorapidity
by averaging the corresponding $v_{2}(\pt)$ distributions over the  range $0<\pt<3$\GeVc and folding in the
efficiency-corrected charged-particle spectra. A comparable centrality interval (2.5--15\% in CMS, and 3--15\% in PHOBOS)
is analyzed for the
most central collisions, while for mid-central (15--25\%) and more peripheral (25--50\%) collisions, the centrality
selections are the same in both experiments. In Fig.~\ref{fig:eta_long}, the statistical uncertainties
are shown as error bars, while the systematic ones are represented by the shaded boxes surrounding
the points. The CMS data cover 4.8 units of pseudorapidity, but do not overlap in pseudorapidity with the
PHOBOS data when plotted in the  rest frames of the colliding nuclei.
The CMS results show  weaker pseudorapidity dependence than observed in the PHOBOS measurement.
The data suggest a nearly boost-invariant region that is several units wide for
central events but considerably smaller for peripheral ones.
It has been noted~\cite{Borghini:2007ub} that if the QCD matter produced at mid-rapidity at RHIC is in
local equilibrium, then deviations from the triangular shape of  $v_{2}(\eta)$ observed in the PHOBOS measurements~\cite{Back:2004zg}
would be expected around mid-rapidity at LHC energies.
Detailed comparisons of theoretical calculations to the results presented here can give new insights into the
nature of the matter produced at both RHIC and LHC energies.

\section{Summary}
\label{sec:Summary}
Detailed measurements of the charged-particle azimuthal anisotropies in
$\sqrt{s_{NN}} = 2.76\TeV$ PbPb collisions and comparisons to lower collision energy results
have been presented. The results cover a broad kinematic range: $0.3 < \pt < 20$\GeVc,
$|\eta| < 2.4$, and 12 centrality classes from 0 to 80\%. The measurements employ four different
methods that have different sensitivities to fluctuations in the initial conditions and nonflow
correlations. The systematic comparison between the methods provides the
possibility to explore the underlying physics processes that cause these
differences.

The elliptic anisotropy parameter $v_2(\pt)$ for $|\eta|<0.8$ is found to increase with \pt up to  $\pt\approx3\GeVc$,
and then to  decrease  in the range  $3 <\pt < 10\GeVc $.
For transverse momenta of $10 <\pt < 20\GeVc $, no strong dependence of $v_2$ on \pt is observed.
The study of the high-\pt azimuthal
anisotropy in charged-particle production may constrain the theoretical descriptions of
parton energy loss and its dependence on the path-length traveled through the medium.

The shapes of the $v_{2}(\pt)$
distributions are found to be similar to those measured at RHIC. At low \pt, only a moderate increase (5--15\%)
is observed in the comparison between results obtained at the highest RHIC energy and the LHC,
despite the large increase in the center-of-mass energy. The integrated $v_{2}$ at mid-rapidity and in
mid-central collisions (20--30\% centrality) increases approximately logarithmically with $\sqrt{s_{NN}}$.
An increase by 20--30\% from the highest RHIC energy to that of the LHC is observed, which is mostly due to the increase
in the mean \pt of the underlying charged-particle spectra.  The integrated $v_{2}$ signal increases from the most central
collisions to the 40--50\% centrality range, after which a decrease is observed. Conversely, the values
of $\langle\pt\rangle$ increase with $N_{\mathrm{part}}$ up to  $N_{\mathrm{part}}\approx$150 (from the most peripheral collisions
up to centrality$\approx$35\%) and then saturate, indicating similar freeze-out conditions in the more central collisions.
The different methods of measuring $v_{2}$ give consistent results over a broad range
of centrality, when scaled by their respective participant eccentricity moments. Deviations from this scaling are
observed in the most central collisions and in peripheral (centrality above 50\%) collisions.
The eccentricity-scaled  $v_{2}$ at mid-rapidity is measured to be approximately linear in the transverse particle density,
and a universal scaling is observed in the comparison of results from different collision systems and
center-of-mass energies measured at RHIC and the LHC. The value of $v_{2}(\eta)$ is found to be weakly
dependent on pseudorapidity in central collisions; for peripheral collisions the values of $v_{2}(\eta)$ gradually decrease
 as the pseudorapidity increases. The results presented here provide further input to the theoretical models of relativistic
nucleus-nucleus collisions and will aid in determining the initial conditions of the system, the degree of
equilibration, and the transport properties of hot QCD matter produced in heavy-ion collisions.

\section*{Acknowledgments}
\hyphenation{Bundes-ministerium Forschungs-gemeinschaft Forschungs-zentren} We congratulate our colleagues in the CERN accelerator departments for the excellent performance of the LHC and thank the technical and administrative staffs at CERN and at other CMS institutes for their contributions to the success of the CMS effort. In addition, we gratefully acknowledge the computing centres and personnel of the Worldwide LHC Computing Grid for delivering so effectively the computing infrastructure essential to our analyses. Finally, we acknowledge the enduring support for the construction and operation of the LHC and the CMS detector provided by the following funding agencies: the Austrian Federal Ministry of Science and Research; the Belgian Fonds de la Recherche Scientifique, and Fonds voor Wetenschappelijk Onderzoek; the Brazilian Funding Agencies (CNPq, CAPES, FAPERJ, and FAPESP); the Bulgarian Ministry of Education, Youth and Science; CERN; the Chinese Academy of Sciences, Ministry of Science and Technology, and National Natural Science Foundation of China; the Colombian Funding Agency (COLCIENCIAS); the Croatian Ministry of Science, Education and Sport; the Research Promotion Foundation, Cyprus; the Ministry of Education and Research, Recurrent financing contract SF0690030s09 and European Regional Development Fund, Estonia; the Academy of Finland, Finnish Ministry of Education and Culture, and Helsinki Institute of Physics; the Institut National de Physique Nucl\'eaire et de Physique des Particules~/~CNRS, and Commissariat \`a l'\'Energie Atomique et aux \'Energies Alternatives~/~CEA, France; the Bundesministerium f\"ur Bildung und Forschung, Deutsche Forschungsgemeinschaft, and Helmholtz-Gemeinschaft Deutscher Forschungszentren, Germany; the General Secretariat for Research and Technology, Greece; the National Scientific Research Foundation, and National Office for Research and Technology, Hungary; the Department of Atomic Energy and the Department of Science and Technology, India; the Institute for Studies in Theoretical Physics and Mathematics, Iran; the Science Foundation, Ireland; the Istituto Nazionale di Fisica Nucleare, Italy; the Korean Ministry of Education, Science and Technology and the World Class University program of NRF, Korea; the Lithuanian Academy of Sciences; the Mexican Funding Agencies (CINVESTAV, CONACYT, SEP, and UASLP-FAI); the Ministry of Science and Innovation, New Zealand; the Pakistan Atomic Energy Commission; the Ministry of Science and Higher Education and the National Science Centre, Poland; the Funda\c{c}\~ao para a Ci\^encia e a Tecnologia, Portugal; JINR (Armenia, Belarus, Georgia, Ukraine, Uzbekistan); the Ministry of Education and Science of the Russian Federation, the Federal Agency of Atomic Energy of the Russian Federation, Russian Academy of Sciences, and the Russian Foundation for Basic Research; the Ministry of Science and Technological Development of Serbia; the Secretar\'{\i}a de Estado de Investigaci\'on, Desarrollo e Innovaci\'on and Programa Consolider-Ingenio 2010, Spain; the Swiss Funding Agencies (ETH Board, ETH Zurich, PSI, SNF, UniZH, Canton Zurich, and SER); the National Science Council, Taipei; the Thailand Center of Excellence in Physics, the Institute for the Promotion of Teaching Science and Technology and National Electronics and Computer Technology Center; the Scientific and Technical Research Council of Turkey, and Turkish Atomic Energy Authority; the Science and Technology Facilities Council, UK; the US Department of Energy, and the US National Science Foundation.

Individuals have received support from the Marie-Curie programme and the European Research Council (European Union); the Leventis Foundation; the A. P. Sloan Foundation; the Alexander von Humboldt Foundation; the Belgian Federal Science Policy Office; the Fonds pour la Formation \`a la Recherche dans l'Industrie et dans l'Agriculture (FRIA-Belgium); the Agentschap voor Innovatie door Wetenschap en Technologie (IWT-Belgium); the Ministry of Education, Youth and Sports (MEYS) of Czech Republic; the Council of Science and Industrial Research, India; the Compagnia di San Paolo (Torino); and the HOMING PLUS programme of Foundation for Polish Science, cofinanced from European Union, Regional Development Fund.

\bibliography{auto_generated}   

\cleardoublepage \appendix\section{The CMS Collaboration \label{app:collab}}\begin{sloppypar}\hyphenpenalty=5000\widowpenalty=500\clubpenalty=5000\textbf{Yerevan Physics Institute,  Yerevan,  Armenia}\\*[0pt]
S.~Chatrchyan, V.~Khachatryan, A.M.~Sirunyan, A.~Tumasyan
\vskip\cmsinstskip
\textbf{Institut f\"{u}r Hochenergiephysik der OeAW,  Wien,  Austria}\\*[0pt]
W.~Adam, T.~Bergauer, M.~Dragicevic, J.~Er\"{o}, C.~Fabjan, M.~Friedl, R.~Fr\"{u}hwirth, V.M.~Ghete, J.~Hammer\cmsAuthorMark{1}, N.~H\"{o}rmann, J.~Hrubec, M.~Jeitler, W.~Kiesenhofer, M.~Krammer, D.~Liko, I.~Mikulec, M.~Pernicka$^{\textrm{\dag}}$, B.~Rahbaran, C.~Rohringer, H.~Rohringer, R.~Sch\"{o}fbeck, J.~Strauss, A.~Taurok, F.~Teischinger, P.~Wagner, W.~Waltenberger, G.~Walzel, E.~Widl, C.-E.~Wulz
% ...
\vskip\cmsinstskip
\dag:~Deceased\\
1:~~Also at CERN, European Organization for Nuclear Research, Geneva, Switzerland\\
% ...

\end{sloppypar}
\end{document}